\pdfoutput=1
\documentclass[12pt]{article}
\usepackage[T1]{fontenc}
\usepackage[utf8]{inputenc}
\usepackage[a4paper]{geometry}
\usepackage{color}
\usepackage{longtable}
\usepackage{float}
\usepackage{calc}
\usepackage{textcomp}
\usepackage{url}
\usepackage{amsmath}
\usepackage{amssymb}
\usepackage{graphicx}
\PassOptionsToPackage{version=3}{mhchem}
\usepackage{mhchem}
\usepackage{subscript}
\usepackage[unicode=true,
 bookmarks=false,
 breaklinks=false,pdfborder={0 0 0},pdfborderstyle={},backref=section,colorlinks=true]
 {hyperref}
\hypersetup{pdftitle={GlassNet: a multitask deep neural network for predicting many glass properties},
 pdfauthor={Daniel R. Cassar},
 pdfcreator={Emacs 27.1 (Org mode 9.6)},pdflang={English},pdfborderstyle={},urlcolor=DodgerBlue4,citecolor=DodgerBlue4,linkcolor=DodgerBlue4}

\makeatletter

\providecommand{\tabularnewline}{\\}
\newcommand{\lyxdot}{.}

\pdfoutput=1
\usepackage{wrapfig}
\usepackage{rotating}
\usepackage[normalem]{ulem}
\usepackage{capt-of}
\author{Daniel R.~Cassar}
\usepackage[x11names]{xcolor}
\usepackage{indentfirst}
\usepackage{subcaption}

\usepackage{pdflscape}
\usepackage{placeins}
\usepackage{xurl}
\author{Daniel R. Cassar}
\date{\small{Ilum School of Science, Brazilian Center for Research in Energy and Materials (CNPEM), Zip Code 13083-970, Campinas, Sao Paulo, Brazil.}\\daniel.cassar@ilum.cnpem.br}
\title{GlassNet: a multitask deep neural network for predicting many glass properties}

\newcommand{\citeprocitem}[2]{\hyper@linkstart{cite}{citeproc_bib_item_#1}#2\hyper@linkend}

\usepackage[notquote]{hanging}

\makeatother

\begin{document}
\maketitle \vspace{0.5cm}

\noindent %
\noindent\fbox{\begin{minipage}[t]{1\linewidth - 2\fboxsep - 2\fboxrule}%
\begin{center}
For the publisher's version, please
visit \url{https://doi.org/10.1016/j.ceramint.2023.08.281} \\
~ \\
\textbf{Cite as}: D. R. Cassar, “GlassNet: A multitask deep neural network for predicting many glass properties,” Ceramics International, vol. 49, no. 22, Part B, pp. 36013–36024, Nov. 2023, doi: 10.1016/j.ceramint.2023.08.281.
\par\end{center}%
\end{minipage}}

\vspace{1cm}

\section*{Abstract}

\label{sec:orgf39b299} A multitask deep neural network model was
trained on more than 218k different glass compositions. This model,
called GlassNet, can predict 85 different properties (such as optical,
electrical, dielectric, mechanical, and thermal properties, as well
as density, viscosity/relaxation, crystallization, surface tension,
and liquidus temperature) of glasses and glass-forming liquids of
different chemistries (such as oxides, chalcogenides, halides, and
others). The model and the data used to train it are available in
the \texttt{GlassPy} Python module as free and open source software
for the community to use and build upon. As a proof of concept, GlassNet
was used with the MYEGA viscosity equation to predict the temperature
dependence of viscosity and outperformed another general purpose viscosity
model available in the literature (ViscNet) on unseen data. An explainable
AI algorithm (SHAP) was used to extract knowledge correlating the
input (physicochemical information) and output (glass properties)
of the model, providing valuable insights for glass manufacturing
and design. It is hoped that GlassNet, with its free and open source
nature, can be used to enable faster and better computer-aided design
of new technological glasses.

\vspace{0.5cm}
 \texttt{Keywords}: non-metallic glasses, artificial neural networks,
property prediction

\newpage{}

\section{Introduction}

\label{sec:orgb52ddf0}

Glasses are particularly interesting materials for data-driven modeling.
First, there is no need for crystal structure descriptors, because
these materials are noncrystalline. Second, commercial glass properties
are highly dependent on glass chemical composition, because the vast
majority of these glasses are manufactured by the same process, the
melt and quench technique \citeprocitem{1}{{[}1{]}}. These facts
are often exploited in the literature with good to great results \citeprocitem{2}{{[}2{]}}--\citeprocitem{9}{{[}9{]}}.

Recently, Le Losq et al. \citeprocitem{10}{{[}10{]}} reported a
new model called i-Melt, which is a multitask deep neural network
capable of predicting 18 different properties of melts and glasses
in the K\textsubscript{2}O--Na\textsubscript{2}O--Al\textsubscript{2}O\textsubscript{3}--SiO\textsubscript{2}
system. A multitask model is a single model that has more than one
output \citeprocitem{11}{{[}11{]}}. The expectation is that learning
how to predict one output can help in predicting another related output.

This work is inspired by i-Melt. The goal is to test whether multitask
learning can be beneficial to glass modeling when we increase the
scope of the model, both the input (more glass chemistries) and the
output (more properties). In the literature, single-output neural
networks are viable algorithms for glass modeling \citeprocitem{12}{{[}12{]}}--\citeprocitem{22}{{[}22{]}}.
However, multitask models have not been explored except for the work
of Le Losq et al. \citeprocitem{10}{{[}10{]}}.

The main hypothesis investigated in this work was \textbf{H}\textsubscript{1}:
multitask learning improves the performance of predictive models of
glass properties.

\section{Materials and methods}

\label{sec:org04b33db} 

\subsection{Data acquisition and processing}

\label{sec:org1755a8e}

The SciGlass database, which is openly available to the community
in a GitHub repository \citeprocitem{23}{{[}23{]}}, provides the
data used in this paper. This database consists of several dataframes.
For this work, all the data in the main dataframe were collected.
The processing steps for the raw data were as follows:
\begin{enumerate}
\item Only glasses with a sum of the molar fractions of the compounds between
0.99 and 1.01 were taken into account. Sum of molar fractions that
are too far away from unity are probably due to typing errors.
\item Glasses with non-zero amounts of any compound in \{``Al2O3+Fe2O3'',
``MoO3+WO3'', ``CaO+MgO'', ``FeO+Fe2O3'', ``Li2O+Na2O+K2O'',
``Na2O+K2O'', ``F2O-1'', ``FemOn'', \linebreak{}
 ``HF+H2O'', ``R2O'', ``R2O3'', ``R2O3'', ``RO'', ``RmOn''\}
were removed. These compounds cannot be converted to atomic mass (see
next step).
\item The chemical features were converted from compound molar fraction
to atomic molar fraction. The atomic fractions were then rebalanced
so that their sum is 1 for all examples.
\item Only the elements between atomic numbers 1 and 83 (hydrogen and bismuth
included) were considered, excluding promethium and the noble gases.
Glasses containing non-zero amounts of excluded elements were removed.
This strategy will become clear in the next section when discussing
feature extraction.
\item A total of 85 target properties were considered. Glasses without values
for at least one property were removed. These properties and related
information will be presented later (see Table \ref{tab:descript_metrics}
for the complete list of targets and Table \ref{tab:symbols} for
the meaning of the symbols).
\item Some of the properties were processed by setting an acceptable minimum
or maximum value. These limits are shown in the Supplementary Material. 
\end{enumerate}
After these steps, the processed dataset had 281,093 examples with
919,164 filled target cells. Many target cells are unfilled, as expected.
This is not an issue for inducing a neural network model, because
the unfilled cells do not contribute in the backpropagation step \citeprocitem{24}{{[}24{]}}.

The next step was deduplication, the removal of duplicate entries
to avoid data leakage \citeprocitem{25}{{[}25{]}}. For this step,
the atomic fraction was first rounded to the third decimal place,
and then glasses with the same composition were grouped together and
collapsed into a single entry. The resulting target value was the
median of the targets from the duplicate groups, ignoring the unfilled
cells. If all cells for a given target were unfilled, then the final
collapsed cell was also unfilled. The median was chosen instead of
the mean because it is robust to outliers in small datasets.

After deduplication, the final dataset had 218,533 examples of different
glasses, with 795,298 filled target cells. Finally, the data were
shuffled and 10\% were randomly selected to be part of the holdout
dataset. These data were not used for anything else, except at the
very end to test the predictive ability of the selected models (simulating
what happens when the model sees new data). Hereafter, any reference
to the dataset refers to the 90\% of the data that was \emph{not}
selected for the holdout dataset, unless otherwise stated.

\subsection{Feature extraction, feature selection, and data scaling}

\label{sec:org6ed90f0}

Inspired by the works of Ward et al. \citeprocitem{26}{{[}26{]}},
Hu et al. \citeprocitem{27}{{[}27{]}}, and Nakamura et al. \citeprocitem{28}{{[}28{]}},
\citeprocitem{29}{{[}29{]}}, we extracted physicochemical features
from the chemical information of the glasses, following a similar
procedure reported in a previous communication \citeprocitem{19}{{[}19{]}}.
This procedure involved three steps.

The first step was to collect the physicochemical properties of the
elements considered in this work (see step 4 of data processing in
the previous section). A total of 55 elemental properties were collected
using the Python modules \texttt{mendeleev} \citeprocitem{30}{{[}30{]}}
and \texttt{matminer} \citeprocitem{31}{{[}31{]}}. The only restriction
at this stage was that the property had to be available for all elements
studied (justifying why promethium and noble gases were not considered
in the previous section). See Table \ref{tab:symbols} for 25 of those
55 properties that were selected. The Supplementary Material lists
the physicochemical properties that were considered but not selected.

The second step was using the physicochemical properties to compute
new features. Let us consider one glass as an example. Let $\boldsymbol{C}=[x_{\mathrm{H}},x_{\mathrm{Li}},\cdots,x_{\mathrm{Bi}}]$
be a vector of the atomic mole fractions of the chemical elements
that make this glass. Let $\boldsymbol{S}=[s_{\mathrm{H}},s_{\mathrm{Li}},\cdots,s_{\mathrm{Bi}}]$
be a vector of a certain physicochemical property $s$ (atomic radius,
for example). Each vector element of $\boldsymbol{C}$ and $\boldsymbol{S}$
correspond to the atomic mole fraction or physicochemical property
of a chemical element. Having these vectors, we can compute the weighted
features with

\begin{equation}
w=f(\boldsymbol{C}\circ\boldsymbol{S})\label{eq:feat_w}
\end{equation}

and the absolute features with

\begin{equation}
a=f(\left\lceil \boldsymbol{C}\right\rceil \circ\boldsymbol{S}).\label{eq:feat_a}
\end{equation}

In the equations above, $f$ is an aggregator function and $\circ$
is the Hadamard product, also known as the element-wise product. The
aggregator functions considered in this paper are $\left\{ \mathrm{sum},\mathrm{min},\mathrm{max},\mathrm{mean},\mathrm{std}\right\} $.
Note that $w$ is a feature that changes continuously as the amount
of each chemical element in the glass changes. In contrast, $a$ is
a feature that is only sensitive to the presence or absence of the
chemical elements (the ceil operator will keep any zero-valued element
of the vector $C$ as zero, but change any non-zero-valued element
to unity). The expectation is that the $a$ features will help the
model cluster different glass chemistries, while the $w$ features
will help the model fine-tune its prediction for specific glasses.

The third step was feature selection. After the previous step, the
dataset had 627 features: 77 representing the atomic mole fraction
of the chemical elements, 275 representing the weighted physicochemical
features (55 physicochemical properties times 5 aggregator functions),
and 275 representing the absolute physicochemical features. The first
procedure was to remove features with extremely low variance, defined
as those with a standard deviation less than 10\textsuperscript{-3}.
This threshold value was used in a previous paper \citeprocitem{19}{{[}19{]}}
and, while arbitrary, it is a sufficiently low value to remove features
that contribute little to the training process. The second procedure
was to remove features with high multicollinearity using the Variance
Inflation Factor (VIF) \citeprocitem{32}{{[}32{]}}. The VIF is
calculated by attempting to predict each available feature by a linear
regression of all other features. The higher the VIF value, the easier
it is to predict the selected feature, thus the higher the multicollinearity
of that feature. The steps of the procedure were:
\begin{enumerate}
\item Compute the VIF for all remaining features;
\item If all values of VIF are below 5, then stop;
\item Otherwise, remove the feature with the highest VIF and return to step
1. 
\end{enumerate}
The rationale of feature extraction is to add more relevant information
to the problem to be leveraged by the algorithm when inducing the
predictive model. The benefit of feature selection is twofold: reduce
the computation cost to induce the model \citeprocitem{33}{{[}33{]}},
and reduce multicollinearity, which improves convergence of the model.
See Table \ref{tab:selected_features} for a list of the selected
features.

Finally, the dataset was scaled using a min-max scaler (all features
and targets were scaled). This is a linear transformation that converts
all features and targets to values between zero and one. This strategy
is often used when training neural networks to improve convergence
and reduce the difference in magnitude between features and targets.
The advantages of using the min-max scaler instead of the standard
scaler is that the former is robust against small standard deviations
of the features and it preserves zero entries in sparse non-negative datasets (which
is the case of compositional features of glasses) \citeprocitem{34}{{[}34{]}}.

\subsection{Designing and training a multitask neural network}

\label{sec:orgf1c21d5}

An artificial neural network (NN) algorithm was used to induce the
predictive models of this work. More specifically, it was a multitask
feedforward neural network (also known as a multilayer perceptron,
MLP). This is a well-known algorithm in the machine learning field,
and the formalism can be found elsewhere \citeprocitem{35}{{[}35{]}}.
All NN models in this work were trained with \texttt{pytorch} \citeprocitem{36}{{[}36{]}}
using \texttt{lightning} \citeprocitem{37}{{[}37{]}}.

Backpropagation to adjust the NN weights and biases was performed
using the weighted Adam optimizer. This optimizer was chosen based
on the results of a previous publication \citeprocitem{19}{{[}19{]}}.
At the start of training, 10\% of the training data was randomly selected
and reserved as the validation dataset. These data were not used to
change the parameters (weights and biases) of the NN, but instead
were used for the early stopping routine. This routine reduces overfitting
of the NN by stopping its training after a certain number of epochs
without improvement in the validation loss transpire. This number
of epochs without improvement is a tunable hyperparameter called \emph{patience}.
An epoch is when all the training data ``passes'' through the neural
network (forward pass) and then through backpropagation, and can be
considered as one cycle of the training process.

Some of the hyperparameters (HP) of the algorithm were not part of
the HP tuning; the loss function is one of them. We used the multitask
loss function of Liebel and Körner \citeprocitem{38}{{[}38{]}},
which combines the single-task loss (mean squared error) for each
predicted property into a single number. This multitask loss function
includes a regularization term with individual weights for each target.
These weights are learnable parameters and their purpose is to avoid
giving too much weight to those properties with much more data than
others (appropriate for the unbalanced database studied).

Other hyperparameters of the algorithm were tuned and a total of 1000
different HP sets were tested. Table \ref{tab:search_space} shows
the search space. The search was performed with the \texttt{ray{[}tune{]}}
Python module \citeprocitem{39}{{[}39{]}}, using an Asynchronous
Successive Halving Algorithm (ASHA) scheduler \citeprocitem{40}{{[}40{]}}
with a grace period of 20 and a reduction factor of 4. Search space
navigation was performed with suggestions from a Tree-structured Parzen
Estimator algorithm \citeprocitem{41}{{[}41{]}}. Each layer of
the NN was allowed to have its own activation function (that is, the
activation function of the hidden layers could be different). The
activation functions considered were hyperbolic tangent, sigmoid,
ReLU, Leaky ReLU, Softplus, GELU, ELU, PReLU, SiLU, SELU, and Mish.
The maximum number of epochs for HP tuning was set to 1000.

After this initial search, 10 HP sets were selected for the next step,
which was a 10-fold cross-validation. These 10 HP sets were manually
selected from the best scoring sets (using multitask loss as the score),
taking care to select \emph{sufficiently different} network architectures.
The rationale is that the top positions of these experiments commonly
consist of neural networks that are far too similar; therefore, manually
selecting different architectures increases the diversity of architectures
to be tested. After cross-validation, the selected HP set was the
one with the lowest mean loss considering the 10 local test datasets.
This selected architecture will be discussed in the results section,
but the reader can already check its hyperparameters in the far right
column of Table \ref{tab:search_space}.

The NN obtained is a multilayer perceptron. One interesting characteristic
of the network that inspired this work (i-Melt) is that it is a multi-headed
feedforward NN: it has a shared multilayer perceptron and two different
heads (a ``head'' here means a set of artificial neurons), one for
property prediction and another for predicting Raman spectra.
The specialized heads can leverage the overall training (shared neurons)
and then focus on predicting its specific task (isolated neurons).
This strategy may be interesting in this work as well. The main idea
is that a shared network could learn the overall patterns of the glasses
and each head could learn the specific characteristic of each property.
This gives us the opportunity to test a secondary hypothesis of this
work \textbf{H}\textsubscript{2}: for the induction of multitask
predictive models of glass properties, multi-headed feedforward neural
networks have a better performance than multilayer perceptrons.

To test H\textsubscript{2}, we used a trained NN with the selected
HP set and converted it into a multi-headed NN. We did so by replacing
the last layer (output layer) with 85 new layers of 10 neurons each
(ReLU activation function) in parallel, one layer for each property.
These new layers are relatively small and highly specialized for the
prediction of their respective property.

After these procedures, we had two models: a multitask MLP (MT-MLP)
and a multitask multi-headed neural network (MT-MH). The performance
of these two models was tested on the holdout dataset to see how they
predicted new data. As will be discussed in detail later, the MT-MH
performed better in this test, supporting H\textsubscript{2}.

To test H\textsubscript{1}, we need to train single-task models.
To keep it simple, we trained single-task NNs (ST-NN) with the same
architecture as the MT-MH model, but with only one output neuron (instead
of 85). Tests of H\textsubscript{1} using a single-task random forest
algorithm and the eXtreme Gradient Boosting (XGBoost) algorithm are
discussed in the Supplementary Material.

After observing a reasonable performance of the NNs (will be discussed
in the Results section), we proceeded to train the final models using
the chosen architecture and considering \emph{all} the dataset (training
\emph{and} holdout) \citeprocitem{42}{{[}42{]}}. This final collection
of models was given the name GlassNet.

Finally, relevant information for the glass community was obtained
by computing the SHapley Additive exPlanation (SHAP) values using
the \texttt{shap} Python module \citeprocitem{43}{{[}43{]}}--\citeprocitem{45}{{[}45{]}}.
This is an analysis that allows the trained model to be interpreted
in search of patterns that can then be used by glass scientists and
engineers when designing new glasses. Unfortunately, it was not possible
to obtain the interaction SHAP values recently used by the community
\citeprocitem{9}{{[}9{]}}, because this calculation cannot be performed
for neural network-based models.

\subsection{Modeling the temperature-dependence of viscosity}

\label{sec:org8daa1e5}

Knowledge of the temperature-dependence of shear viscosity is essential
for glass manufacturing, because it is used to adjust process variables
such as melting, working, and annealing temperatures. Recent publications
have already reported on data-driven models for predicting this property,
ViscNet \citeprocitem{19}{{[}19{]}} and i-Melt \citeprocitem{10}{{[}10{]}}
being two examples of free and openly available data-driven viscosity
models.

Many of the properties investigated in this work are directly or indirectly
related to shear viscosity. Here, we exploited this fact by using
the trained models to generate (temperature, viscosity) data tuples,
and then using these data points to perform a non-linear regression
of the MYEGA equation (Eq. \ref{eq:myega}), which is a physical model
of viscosity \citeprocitem{46}{{[}46{]}}.

\begin{equation}
\log_{10}(\eta(T))=\log_{10}(\eta_{\infty})+\frac{T_{12,\mathrm{M}}}{T}[12-\log_{10}(\eta_{\infty})]\exp\left(\left[\frac{m}{12-\log_{10}(\eta_{\infty})}-1\right]\left[\frac{T_{12,\mathrm{M}}}{T}-1\right]\right)\label{eq:myega}
\end{equation}

In the previous equation, $\eta$ is the equilibrium shear viscosity,
$\eta_{\infty}$ is the asymptotic viscosity ($\eta_{\infty}\equiv\lim_{T\rightarrow\infty}\eta(T)$),
$m$ is the liquid fragility index \citeprocitem{47}{{[}47{]}},
and $T_{12,\mathrm{M}}$ has the same definition as $T_{12}$ ($\eta(T_{12})\equiv10^{12}\,\textrm{Pa. s}$),
but it is written with a different notation to indicate that it comes
from a non-linear regression of the MYEGA equation. Using this approach,
GlassNet can predict three additional properties.

Knowing that the prediction of material properties by machine learning
models is much more susceptible to noise (compared to measured experimental data), using a robust non-linear regression of Eq.
(\ref{eq:myega}) is a good strategy to compensate for this disadvantage.
In this work, we used a Cauchy loss for the least squares regression.
The mathematical equation for this loss is $\rho(z)=\ln(1+z)$, where
$z$ is the standard least squares loss.

To test the viscosity prediction capabilities of GlassNet, a set of
147,007 data points of composition, temperature, and measured viscosity
were collected from the SciGlass database. Entries containing thorium
or uranium were removed, because GlassNet cannot predict glasses with
either of these elements. Entries with $\log_{10}(\eta)$ outside
the $[-5,12]$ range, measured at temperatures above 3000 K, or measured
at temperatures below the glass transition (predicted by GlassNet)
were also removed, as these data are prone to higher measurement errors.
Composition (in atomic fraction) was rounded to the third decimal
place, and temperature (in Kelvin) was rounded to the first decimal
place before duplicate entries were merged into one with the median
value of $\log_{10}(\eta)$. Finally, entries with the same chemical
composition as one of the glasses used to train GlassNet were removed,
to make this experiment another test of GlassNet's predictive power
for unseen compositions.

\subsection{Model and data availability}

\label{sec:org1b54d8c}

The model reported here is called GlassNet and is available to the
community, along with the training data, in the \texttt{GlassPy} Python
module, a free and open source module for researchers working with
glass materials. See the official repository at \url{https://github.com/drcassar/glasspy}
for instructions on installing and using this module. In \texttt{GlassPy},
the user has the option to load the MT-MLP or MT-MH version of GlassNet.
\texttt{GlassPy} can also use the single-task models for those properties
where they outperformed the multitask models (see Table \ref{tab:descript_metrics};
note that when doing this, the inference is slower). GlassNet can
also predict the temperature-dependence of viscosity and the MYEGA
parameters.

\texttt{GlassPy} also provides an easy way to load SciGlass data into
a \texttt{pandas} DataFrame \citeprocitem{48}{{[}48{]}}, including
(but not limited to) the GlassNet training dataset. These \texttt{pandas}
DataFrames are state of the art Python objects for data analysis.
Two advantages of using \texttt{GlassPy} as a frontend for exploring
SciGlass data are
\begin{enumerate}
\item \texttt{GlassPy} already translates the SciGlass data into a ready-to-use
DataFrame with an intuitive naming scheme;
\item \texttt{GlassPy} does not require installing any legacy non-free software
(the official SciGlass repository only provides the database files
in a legacy proprietary Microsoft Access format). 
\end{enumerate}
The dataset used to train GlassNet and the viscosity dataset used
to test the viscosity prediction of GlassNet are also available as
Supplementary Data to this communication.

\section{Results and discussion}

\label{sec:orge391ca2} 

\subsection{Data analysis and selected physicochemical features}

\label{sec:org9cac298}

Table \ref{tab:symbols} lists the symbols used in this work, their
meaning, and their units. Table \ref{tab:descript_metrics} shows
the descriptive statistics (count, minimum, mean, and maximum) of
the 85 targets of GlassNet. These numbers reflect the whole dataset
after data processing, just before the holdout split.

Of the 85 properties, 26 had more than 10k instances when all data
were considered. In this group we find the glass transition temperature,
density at ambient temperature, refractive index, Abbe number, Young's
modulus, microhardness, linear coefficient of thermal expansion below
$T_{g}$, crystallization peak, crystallization onset, and others.

In contrast, 21 out of 85 properties had less than 1k examples. This
group includes maximum crystal growth velocity, density and surface
tension at temperatures above ambient, heat capacity at constant pressure,
and others.

This representation problem regarding the number of examples of different
targets can be an issue if left unattended. For example, the NN might
favor properties with more data (by the simple fact that they can
have a greater impact on the loss value) over those with less data,
defeating the purpose of having a multitask model. This problem is
minimized by using the multitask loss weights of Liebel and Körner
\citeprocitem{38}{{[}38{]}}, as discussed in the Materials and
Methods section. A test and discussion of training a multitask model
without the loss weights of Liebel and Körner can be found in the
Supplementary Material.

Table \ref{tab:elements} shows the descriptive statistics (count,
mean, and standard deviation) and elemental mole fraction information
(first quartile, median, third quartile, and maximum) for the chemical
elements in the glasses used to train GlassNet. These values reflect
the full dataset just before the holdout split, similar to the data
shown in Table \ref{tab:descript_metrics}.

Considering all the data, 25 of the 72 elements were present in more
than 10k examples. In this group, we find the most common chemical
elements of inorganic glasses such as oxygen, silicon, boron, sodium,
aluminum, calcium, potassium, and others. Similarly, 14 of the 72
elements were present in fewer than 1k examples. In this group we
find dysprosium, hafnium, europium, terbium, and others.

Remarkably, some elements in Table \ref{tab:elements} have a maximum
elemental mole fraction of 1, in other words, these are monoatomic
examples present in the dataset, and it is known that most of them
are quite resistant to glass formation \citeprocitem{1}{{[}1{]}}.
While the original authors' rationale for including these particular
examples in the SciGlass database is not clear, the information in
these entries is typically related to liquidus temperature, which
(by definition) is a property of crystalline materials and not glasses.
Given the small number of these data points and the importance of
the liquidus temperature for glass making \citeprocitem{1}{{[}1{]}},
these data points were not excluded.

The induced models will inevitably be better at predicting the properties
of glasses with chemistries that are more represented in the training
dataset. The expectation is that the strategy of physicochemical feature
extraction from the data (see methods) will reduce this problem.

Finally, Table \ref{tab:selected_features} shows the 98 selected
features out of the 627 considered. Of these 98 features, 64 are elemental
mole fractions, 12 are weighted physicochemical features (see Eq.
\ref{eq:feat_w}), and 22 are absolute physicochemical features (see
Eq. \ref{eq:feat_a}). Some remarks on these selected features: the
aggregator function ``mean'' is not present; out of a total of 77
elemental mole fraction features, 13 were not selected (e.g., silicon
and oxygen are in this group of 13); out of a total of 55 physicochemical
properties considered, only 25 are part of the selected features.

\subsection{Neural network}

\label{sec:orgb989650}

The last column of Table \ref{tab:search_space} shows the hyperparameters
selected after HP tuning. This is a deep neural network with four
hidden layers. Each hidden layer has a different activation function.
All hidden layers have a dropout probability that decreases from the
first to the second layer and then keeps increasing until the last
layer. Batch normalization is only present in the first two hidden
layers.

The root mean squared error (RMSE) metrics of the NN models (MT-MLP,
MT-MH, and ST-NN) are shown in Table \ref{tab:descript_metrics},
one metric for each property. The values are the prediction metrics
for the holdout dataset (\emph{not} used to train the model) and the
standard deviation is that obtained in a 10-fold cross-validation
experiment. The RMSE gives an estimate of the prediction error in
the same units and magnitude as the target. A t-test (95\% confidence)
was used to compare how the three models performed; these results
are also reported in Table \ref{tab:descript_metrics}.

The MT-MH model outperforms the MT-MLP model for 13 properties, 12
of which are properties with more than 10k examples. There was only
one property for which the MT-MH model had a worse prediction, and
that was $d(1073\,\mathrm{K})$. This result supports H\textsubscript{2};
a multi-headed NN improves performance for targets with a large number
of examples, without losing generalization power for targets with
fewer examples.

Comparing the multitask models with the single-task models produced
mixed results: for 29 properties, there was no statistical difference
between their performances. The MT-MLP outperformed the ST-NN in 24
properties and was outperformed in 28 properties. The MT-MH produced
slightly better results: it outperformed the ST-NN in 24 properties
and was outperformed in 24 properties.

The properties that the multitask models predicted better than the
single-task models were mostly targets with more than one measurement
at different temperatures; most of these properties are targets related
to viscosity. Properties predicted better by single-task models than multitask models were mostly targets with more than 10k examples.

Overall, it is not possible to support or reject hypothesis H\textsubscript{1}.
On the one hand, a multitask model can improve the prediction of glass
properties for targets with measurements in different experimental
conditions (e.g., different temperatures) compared to specialized
single-task models. The reason for such improvement is attributed
to the shared hidden layers of a multitask NN, because they can exploit
the connections between the targets. On the other hand, the specialized
single-task models can perform better when more data are available.

\subsection{Interpreting the trained model}

\label{sec:org8647c2c}

\begin{figure}
\centering \begin{subfigure}[h]{0.4\textwidth} \includegraphics[width=1\textwidth,keepaspectratio]{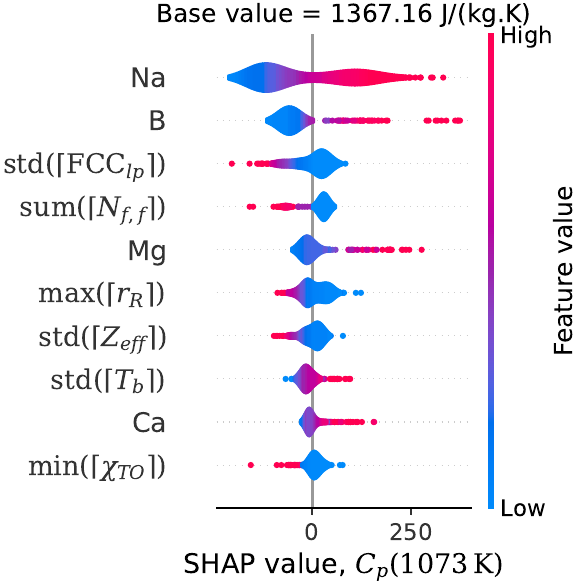}
\caption{}
\end{subfigure}

\begin{subfigure}[h]{0.4\textwidth} \includegraphics[width=1\textwidth,keepaspectratio]{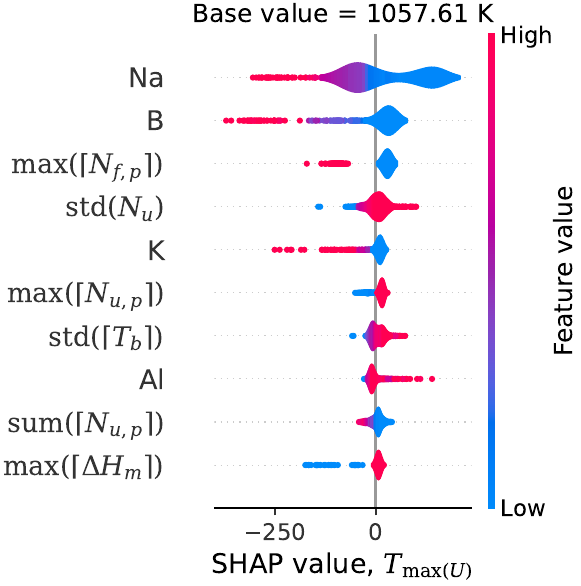}
\caption{}
\end{subfigure}

\begin{subfigure}[h]{0.4\textwidth} \includegraphics[width=1\textwidth,keepaspectratio]{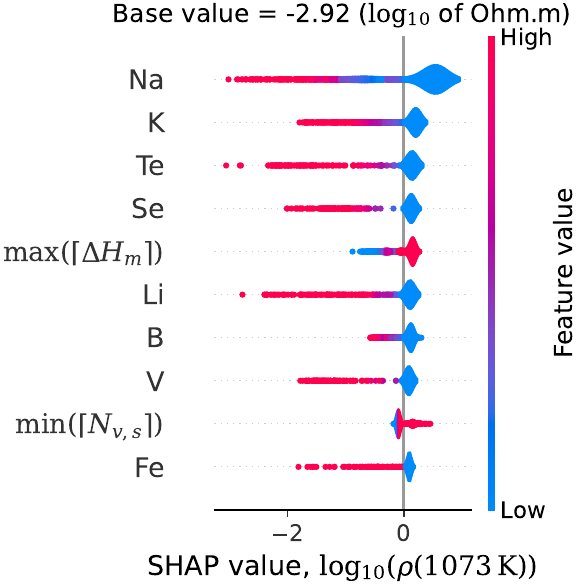}
\caption{}
\end{subfigure} \caption{\label{fig:shap_glassnet}Violin plots of the SHAP values for (a)
$C_{p}(1073\,\mathrm{K})$, (b) $T_{\mathrm{max}(U)}$, and (c) $\log_{10}(\rho(1073\,\mathrm{K}))$.}
\end{figure}

Figure \ref{fig:shap_glassnet} shows the SHAP value violin plots
of three properties: $C_{p}(1073\,\mathrm{K})$, $T_{\mathrm{max}(U)}$,
and $\log_{10}(\rho(1073\,\mathrm{K}))$. These plots show the 10
most important features (those with the highest mean absolute SHAP value)
on the y-axis, with SHAP values on the x-axis. The base value (marked
as a vertical gray line) is the mean value of the property (considering
the entire dataset) and has a SHAP value of zero. The width of the
violins represents the number of examples with the same SHAP value,
while the color within the violins represents the value of the feature.
See the Supplementary Material for the violin plots for the other
properties.

As shown in Figure \ref{fig:shap_glassnet}a, $C_{p}(1073\,\mathrm{K})$
increases with increasing amounts of sodium, boron, magnesium, or
calcium. Sodium is the element with the greatest impact on this property.
Increasing the standard deviation of the boiling point of the elements
present in the glass also increases this property. One can decrease
this property by increasing the standard deviation of the FCC lattice
parameter, the maximum atomic radius, the standard deviation of the
effective nuclear charge, and the sum of the number of filled f valence
orbitals.

Figure \ref{fig:shap_glassnet}b shows that sodium and boron are the
most important features for modeling $T_{\mathrm{max}(U)}$. The modeling
of this rather complex property involves many physicochemical features.
With the exception of the melting enthalpy, the other physicochemical
features shown in Figure \ref{fig:shap_glassnet}b are not often discussed
in the crystal growth literature. The temperature of the maximum crystal
growth velocity is related to the glass-forming ability \citeprocitem{49}{{[}49{]}},
and is one of key properties for glass-ceramics design, together with
the glass transition temperature and the maximum crystal growth velocity
\citeprocitem{50}{{[}50{]}}.

Finally, Figure \ref{fig:shap_glassnet}c shows sodium once again
as the most important feature, this time for $\log_{10}(\rho(1073\,\mathrm{K}))$.
Notably, 8 of the 10 most important features are elemental features
for this property, the opposite of what was observed for $C_{p}(1073\,\mathrm{K})$
and $T_{\mathrm{max}(U)}$. This list includes elements that are well
known in the electrical property community, like potassium, lithium,
and vanadium.

The SHAP analysis allows us to answer other questions: What are the
most relevant features overall (considering all the properties studied
here)? And what are the most relevant features for each group of properties?
Examining the frequency of the 10 most important features for each
property is one strategy for answering these questions.

Table \ref{tab:shap_analysis} shows the most frequent features overall
and for property groups. Overall, we can see that some of the chemical
elements that are most commonly used in glass making are also the
most important ones in our analysis. Sodium, boron, lead, lithium,
aluminum, potassium, and calcium are on this list. One might think
that this is just a list of the most abundant chemical elements in
the dataset, but this is not the case, as barium, magnesium, phosphorus,
zinc, titanium, and fluor are not on this list (and they are all more
abundant than lead in the dataset). Melting enthalpy, number of unfilled
valence orbitals, and number of filled d orbitals are physicochemical
features that are also collectively relevant to glass properties.

The most relevant features for viscosity and relaxation, optical properties,
electrical and dielectric properties, mechanical properties, density,
thermal properties, crystallization, and surface tension are also
listed in Table \ref{tab:shap_analysis}. Again, sodium and boron
are relevant features for most of these property groups. Lithium and
lead are also relevant features, appearing many times in this table.
Interestingly, the model understood that the optical properties are
highly dependent on the filled valence orbitals (where the absorption
process takes place) and on lead, bismuth, titanium, niobium, lanthanum,
and germanium, known elements used for optical glasses.

\subsection{Viscosity modeling}

\label{sec:orged5c213}

The final viscosity dataset (after all the procedures mentioned in
the Materials and Methods section) consisted of 31,976 examples. For
each of these examples, the viscosity at the measurement temperature
was predicted using GlassNet. To evaluate this prediction, the RMSE
between the reported and predicted values of $\log_{10}(\eta)$ was
calculated (with $\eta$ in units of Pa.s).

GlassNet had an overall RMSE of 0.9 in this test, which was better
than ViscNet \citeprocitem{19}{{[}19{]}} (another deep learning
model used to predict viscosity), which had an RMSE of 1.1 for unseen
data. The performance of GlassNet considering only silicates (RMSE:
0.56), aluminosilicates (RMSE: 0.45), or borosilicates (RMSE: 0.75)
was better than the overall performance, supporting that GlassNet
could learn reasonable patterns for these chemistries to be able to
predict viscosity for unseen data. The performance for silicates and
aluminosilicates is comparable to that reported by i-Melt (RMSE: 0.4)
\citeprocitem{10}{{[}10{]}} and obtained using thermodynamic models
(RMSE in the range of 0.2--0.4) \citeprocitem{51}{{[}51{]}}--\citeprocitem{53}{{[}53{]}}.

However, for other chemistries the performance of GlassNet was not
great, with an RMSE in the range of 1.2--1.4 for germanate, borate,
germanosilicate, and chalcogenide liquids, and in the range of 1.5--1.7
for phosphosilicate, halide, and chalcohalide liquids. For these chemistries,
GlassNet can give a general trend for the temperature-dependence of
viscosity that may be useful as a rough estimate, but the reader is
advised to use it with extra caution. Performance was poor for phosphate
liquids with an RMSE of 2.3. The viscosity data set is available as
supplementary data to this communication. More information on how
the different liquid chemistries were defined is available in the
Supplementary Material.

\section{Summary and conclusion}

\label{sec:org85777b2}

In this work, we collected more than 218k different glass compositions
with more than 795k data points on 85 properties. Neural network models
were designed and trained with this rich dataset. We observed that
the multitask models outperformed the specialized single-task models
in predicting some targets of the same property measured under different
conditions (e.g., different temperatures). However, the specialized
models outperformed the multitask models for some properties with
many data points (more than 10k).

An advantage of the proposed model, compared to other models reported
in the literature, is that it is not limited to a specific inorganic
glass chemistry and can be used to predict properties of oxides, chalcogenides,
halides, and other types of glasses.

We show how the purely data-driven predictions of the proposed model
can be used together with a physical model to predict the temperature
dependence of viscosity. This approach was tested with about 32k viscosity
data points and yielded better results than another general viscosity
model (ViscNet \citeprocitem{19}{{[}19{]}}) and comparable results
to specialized models for silicate and aluminosilicate liquids. Additionally,
useful insights for glass manufacturing and design were obtained from
the trained model using SHAP analysis.

The obtained model was named GlassNet and is free software, available
to the community (together with the training data) in the Python module
\texttt{GlassPy}. This free and open source suite for inorganic glass
data and property prediction is expected to benefit the community,
improve the reproducibility of data-driven publications, and accelerate
the development of new and exciting glasses and glass-ceramics (e.g.,
by using GlassNet with inverse design tools such as GLAS \citeprocitem{20}{{[}20{]}}).
In addition, the open nature of GlassNet allows the community to build
on it to solve other property prediction problems using transfer learning
\citeprocitem{35}{{[}35{]}}, \citeprocitem{54}{{[}54{]}}, \citeprocitem{55}{{[}55{]}}.

\section*{Acknowledgements}

\label{sec:org6b4858e} The author acknowledges funding from the CNPq
/ INCT - Materials Informatics project. The author also thanks Carolina
B. Zanelli for the English revision.

\section*{Declaration of Generative AI and AI-assisted technologies in the
writing process}

\label{sec:org08bcb11} During the preparation of this work the author
used DeepL and Grammarly in order to improve readability and grammar.
After using these tools, the author reviewed and edited the content
as needed and takes full responsibility for the content of the publication.

\section*{References}

\label{sec:orge2de473} \begingroup \leftskip=17pt \parindent=-\leftskip
\hypertarget{citeproc_bib_item_1}{{[}1{]} A. K. Varshneya and J.
C. Mauro, \textit{Fundamentals of Inorganic Glasses}, 3 edition. Elsevier,
2019.}

\hypertarget{citeproc_bib_item_2}{{[}2{]} S. Bishnoi \textit{et al.},
“Predicting Young’s modulus of oxide glasses with sparse datasets
using machine learning,” \textit{Journal of non-crystalline solids},
vol. 524, p. 119643, 2019, doi: \href{https://doi.org/10.1016/j.jnoncrysol.2019.119643}{10.1016/j.jnoncrysol.2019.119643}.}

\hypertarget{citeproc_bib_item_3}{{[}3{]} A. Tandia, M. C. Onbasli,
and J. C. Mauro, “Machine learning for glass modeling,” in \textit{Springer
Handbook of Glass}, J. D. Musgraves, J. Hu, and L. Calvez, Eds., in
Springer Handbooks. Cham: Springer International Publishing, 2019,
pp. 1157--1192.}

\hypertarget{citeproc_bib_item_4}{{[}4{]} E. Alcobaça \textit{et
al.}, “Explainable machine learning algorithms for predicting glass
transition temperatures,” \textit{Acta materialia}, vol. 188, pp.
92--100, 2020, doi: \href{https://doi.org/10.1016/j.actamat.2020.01.047}{10.1016/j.actamat.2020.01.047}.}

\hypertarget{citeproc_bib_item_5}{{[}5{]} B. Deng, “Machine learning
on density and elastic property of oxide glasses driven by large dataset,”
\textit{Journal of non-crystalline solids}, vol. 529, p. 119768, 2020,
doi: \href{https://doi.org/10.1016/j.jnoncrysol.2019.119768}{10.1016/j.jnoncrysol.2019.119768}.}

\hypertarget{citeproc_bib_item_6}{{[}6{]} Y. Tokuda, M. Fujisawa,
D. M. Packwood, M. Kambayashi, and Y. Ueda, “Data-driven design of
glasses with desirable optical properties using statistical regression,”
\textit{Aip advances}, vol. 10, no. 10, p. 105110, 2020, doi: \href{https://doi.org/10.1063/5.0022451}{10.1063/5.0022451}.}

\hypertarget{citeproc_bib_item_7}{{[}7{]} S. Bishnoi, R. Ravinder,
H. S. Grover, H. Kodamana, and N. M. A. Krishnan, “Scalable Gaussian
processes for predicting the optical, physical, thermal, and mechanical
properties of inorganic glasses with large datasets,” \textit{Materials
advances}, vol. 2, no. 1, pp. 477--487, 2021, doi: \href{https://doi.org/10.1039/D0MA00764A}{10.1039/D0MA00764A}.}

\hypertarget{citeproc_bib_item_8}{{[}8{]} D. R. Cassar, S. M. Mastelini,
T. Botari, E. Alcobaça, A. C. P. L. F. de Carvalho, and E. D. Zanotto,
“Predicting and interpreting oxide glass properties by machine learning
using large datasets,” \textit{Ceramics international}, vol. 47, no.
17, pp. 23958--23972, 2021, doi: \href{https://doi.org/10.1016/j.ceramint.2021.05.105}{10.1016/j.ceramint.2021.05.105}.}

\hypertarget{citeproc_bib_item_9}{{[}9{]} R. Bhattoo, S. Bishnoi,
M. Zaki, and N. M. A. Krishnan, “Understanding the compositional control
on electrical, mechanical, optical, and physical properties of inorganic
glasses with interpretable machine learning,” \textit{Acta materialia},
vol. 242, p. 118439, 2023, doi: \href{https://doi.org/10.1016/j.actamat.2022.118439}{10.1016/j.actamat.2022.118439}.}

\hypertarget{citeproc_bib_item_10}{{[}10{]} C. Le Losq, A. P. Valentine,
B. O. Mysen, and D. R. Neuville, “Structure and properties of alkali
aluminosilicate glasses and melts: Insights from deep learning,” \textit{Geochimica
et cosmochimica acta}, vol. 314, pp. 27--54, 2021, doi: \href{https://doi.org/10.1016/j.gca.2021.08.023}{10.1016/j.gca.2021.08.023}.}

\hypertarget{citeproc_bib_item_11}{{[}11{]} R. Caruana, “Multitask
Learning,” \textit{Machine learning}, vol. 28, no. 1, pp. 41--75,
1997, doi: \href{https://doi.org/10.1023/A:1007379606734}{10.1023/A:1007379606734}.}

\hypertarget{citeproc_bib_item_12}{{[}12{]} C. Dreyfus and G. Dreyfus,
“A machine learning approach to the estimation of the liquidus temperature
of glass-forming oxide blends,” \textit{Journal of non-crystalline
solids}, vol. 318, no. 1-2, pp. 63--78, 2003, doi: \href{https://doi.org/10.1016/S0022-3093(02)01859-8}{10.1016/S0022-3093(02)01859-8}.}

\hypertarget{citeproc_bib_item_13}{{[}13{]} D. R. Cassar, A. C. P.
L. F. de Carvalho, and E. D. Zanotto, “Predicting glass transition
temperatures using neural networks,” \textit{Acta materialia}, vol.
159, pp. 249--256, 2018, doi: \href{https://doi.org/10.1016/j.actamat.2018.08.022}{10.1016/j.actamat.2018.08.022}.}

\hypertarget{citeproc_bib_item_14}{{[}14{]} K. Yang \textit{et al.},
“Predicting the Young’s Modulus of Silicate Glasses using High-Throughput
Molecular Dynamics Simulations and Machine Learning,” \textit{Scientific
reports}, vol. 9, no. 1, p. 8739, 2019, doi: \href{https://doi.org/10.1038/s41598-019-45344-3}{10.1038/s41598-019-45344-3}.}

\hypertarget{citeproc_bib_item_15}{{[}15{]} T. Han, N. Stone-Weiss,
J. Huang, A. Goel, and A. Kumar, “Machine learning as a tool to design
glasses with controlled dissolution for healthcare applications,”
\textit{Acta biomaterialia}, vol. 107, pp. 286--298, 2020, doi: \href{https://doi.org/10.1016/j.actbio.2020.02.037}{10.1016/j.actbio.2020.02.037}.}

\hypertarget{citeproc_bib_item_16}{{[}16{]} J. N. P. Lillington,
T. L. Goût, M. T. Harrison, and I. Farnan, “Predicting radioactive
waste glass dissolution with machine learning,” \textit{Journal of
non-crystalline solids}, vol. 533, p. 119852, 2020, doi: \href{https://doi.org/10.1016/j.jnoncrysol.2019.119852}{10.1016/j.jnoncrysol.2019.119852}.}

\hypertarget{citeproc_bib_item_17}{{[}17{]} R. Ravinder \textit{et
al.}, “Deep learning aided rational design of oxide glasses,” \textit{Materials
horizons}, vol. 7, no. 7, pp. 1819--1827, 2020, doi: \href{https://doi.org/10.1039/D0MH00162G}{10.1039/D0MH00162G}.}

\hypertarget{citeproc_bib_item_18}{{[}18{]} S. K. Ahmmad, N. Jabeen,
S. T. Uddin Ahmed, S. A. Ahmed, and S. Rahman, “Artificial intelligence
density model for oxide glasses,” \textit{Ceramics international},
vol. 47, no. 6, pp. 7946--7956, 2021, doi: \href{https://doi.org/10.1016/j.ceramint.2020.11.144}{10.1016/j.ceramint.2020.11.144}.}

\hypertarget{citeproc_bib_item_19}{{[}19{]} D. R. Cassar, “ViscNet:
Neural network for predicting the fragility index and the temperature-dependency
of viscosity,” \textit{Acta materialia}, vol. 206, p. 116602, 2021,
doi: \href{https://doi.org/10.1016/j.actamat.2020.116602}{10.1016/j.actamat.2020.116602}.}

\hypertarget{citeproc_bib_item_20}{{[}20{]} D. R. Cassar, G. G. Santos,
and E. D. Zanotto, “Designing optical glasses by machine learning
coupled with a genetic algorithm,” \textit{Ceramics international},
vol. 47, no. 8, pp. 10555--10564, 2021, doi: \href{https://doi.org/10.1016/j.ceramint.2020.12.167}{10.1016/j.ceramint.2020.12.167}.}

\hypertarget{citeproc_bib_item_21}{{[}21{]} Y. Tokuda, M. Fujisawa,
J. Ogawa, and Y. Ueda, “A machine learning approach to the prediction
of the dispersion property of oxide glass,” \textit{Aip advances},
vol. 11, no. 12, p. 125127, 2021, doi: \href{https://doi.org/10.1063/5.0075425}{10.1063/5.0075425}.}

\hypertarget{citeproc_bib_item_22}{{[}22{]} M. Zaki \textit{et al.},
“Interpreting the optical properties of oxide glasses with machine
learning and Shapely additive explanations,” \textit{Journal of the
american ceramic society}, vol. 105, no. 6, pp. 4046--4057, 2022,
doi: \href{https://doi.org/10.1111/jace.18345}{10.1111/jace.18345}.}

\hypertarget{citeproc_bib_item_23}{{[}23{]} “Epam/SciGlass.” EPAM
Systems, 2019. Accessed: Nov. 06, 2019. {[}Online{]}. Available: \url{https://github.com/epam/SciGlass}}

\hypertarget{citeproc_bib_item_24}{{[}24{]} D. E. Rumelhart, G. E.
Hinton, and R. J. Williams, “Learning representations by back-propagating
errors,” \textit{Nature}, vol. 323, no. 6088, pp. 533--536, 1986,
doi: \href{https://doi.org/10.1038/323533a0}{10.1038/323533a0}.}

\hypertarget{citeproc_bib_item_25}{{[}25{]} S. Kaufman, S. Rosset,
C. Perlich, and O. Stitelman, “Leakage in data mining: Formulation,
detection, and avoidance,” \textit{Acm transactions on knowledge discovery
from data}, vol. 6, no. 4, pp. 15:1--15:21, 2012, doi: \href{https://doi.org/10.1145/2382577.2382579}{10.1145/2382577.2382579}.}

\hypertarget{citeproc_bib_item_26}{{[}26{]} L. Ward, A. Agrawal,
A. Choudhary, and C. Wolverton, “A general-purpose machine learning
framework for predicting properties of inorganic materials,” \textit{Npj
computational materials}, vol. 2, no. 1, pp. 1--7, 2016, doi: \href{https://doi.org/10.1038/npjcompumats.2016.28}{10.1038/npjcompumats.2016.28}.}

\hypertarget{citeproc_bib_item_27}{{[}27{]} Y.-J. Hu \textit{et al.},
“Predicting densities and elastic moduli of SiO2-based glasses by
machine learning,” \textit{Npj computational materials}, vol. 6, no.
1, p. 25, 2020, doi: \href{https://doi.org/10.1038/s41524-020-0291-z}{10.1038/s41524-020-0291-z}.}

\hypertarget{citeproc_bib_item_28}{{[}28{]} K. Nakamura, N. Otani,
and T. Koike, “Multi-objective Bayesian optimization of optical glass
compositions,” \textit{Ceramics international}, 2021, doi: \href{https://doi.org/10.1016/j.ceramint.2021.02.155}{10.1016/j.ceramint.2021.02.155}.}

\hypertarget{citeproc_bib_item_29}{{[}29{]} K. Nakamura, N. Otani,
and T. Koike, “Search for oxide glass compositions using Bayesian
optimization with elemental-property-based descriptors,” \textit{Journal
of the ceramic society of japan}, vol. 128, no. 8, pp. 569--572,
2020, doi: \href{https://doi.org/10.2109/jcersj2.20118}{10.2109/jcersj2.20118}.}

\hypertarget{citeproc_bib_item_30}{{[}30{]} Ł. Mentel, “mendeleev
A Python resource for properties of chemical elements, ions and isotopes.”
2014. Available: \url{https://github.com/lmmentel/mendeleev}}

\hypertarget{citeproc_bib_item_31}{{[}31{]} L. Ward \textit{et al.},
“Matminer: An open source toolkit for materials data mining,” \textit{Computational
materials science}, vol. 152, pp. 60--69, 2018, doi: \href{https://doi.org/10.1016/j.commatsci.2018.05.018}{10.1016/j.commatsci.2018.05.018}.}

\hypertarget{citeproc_bib_item_32}{{[}32{]} D. W. Marquardt, “Generalized
Inverses, Ridge Regression, Biased Linear Estimation, and Nonlinear
Estimation,” \textit{Technometrics}, vol. 12, no. 3, pp. 591--612,
1970, doi: \href{https://doi.org/10.1080/00401706.1970.10488699}{10.1080/00401706.1970.10488699}.}

\hypertarget{citeproc_bib_item_33}{{[}33{]} H. Liu, “Feature Selection,”
in \textit{Encyclopedia of Machine Learning}, C. Sammut and G. I.
Webb, Eds., Boston, MA: Springer US, 2010, pp. 402--406.}

\hypertarget{citeproc_bib_item_34}{{[}34{]} “Scikit-learn 1.3 User
Guide: Preprocessing data,” \textit{Scikit-learn}. 2023. Accessed:
Jul. 06, 2023. {[}Online{]}. Available: \url{https://scikit-learn.org/1.3/modules/preprocessing.html}}

\hypertarget{citeproc_bib_item_35}{{[}35{]} I. Goodfellow, Y. Bengio,
A. Courville, and Y. Bengio, \textit{Deep learning}, vol. 1. MIT press
Cambridge, 2016. Available: \url{https://www.deeplearningbook.org/}}

\hypertarget{citeproc_bib_item_36}{{[}36{]} A. Paszke \textit{et
al.}, “PyTorch: An imperative style, high-performance deep learning
library,” in \textit{Advances in neural information processing systems
32}, H. Wallach, H. Larochelle, A. Beygelzimer, F. dAlché-Buc, E.
Fox, and R. Garnett, Eds., Curran Associates, Inc., 2019, pp. 8024--8035.
Available: \url{http://papers.neurips.cc/paper/9015-pytorch-an-imperative-style-high-performance-deep-learning-library.pdf}}

\hypertarget{citeproc_bib_item_37}{{[}37{]} W. Falcon and T. P. L.
team, “PyTorch Lightning.” Zenodo, 2023. doi: \href{https://doi.org/10.5281/zenodo.7688620}{10.5281/zenodo.7688620}.}

\hypertarget{citeproc_bib_item_38}{{[}38{]} L. Liebel and M. Körner,
“Auxiliary Tasks in Multi-task Learning.” arXiv, 2018. doi: \href{https://doi.org/10.48550/arXiv.1805.06334}{10.48550/arXiv.1805.06334}.}

\hypertarget{citeproc_bib_item_39}{{[}39{]} P. Moritz \textit{et
al.}, “Ray: A Distributed Framework for Emerging AI Applications.”
arXiv, 2018. doi: \href{https://doi.org/10.48550/arXiv.1712.05889}{10.48550/arXiv.1712.05889}.}

\hypertarget{citeproc_bib_item_40}{{[}40{]} L. Li \textit{et al.},
“A system for massively parallel hyperparameter tuning,” \textit{Proceedings
of machine learning and systems}, vol. 2, pp. 230--246, 2020.}

\hypertarget{citeproc_bib_item_41}{{[}41{]} J. Bergstra, D. Yamins,
and D. Cox, “Making a science of model search: Hyperparameter optimization
in hundreds of dimensions for vision architectures,” in \textit{International
Conference on Machine Learning}, 2013, pp. 115--123.}

\hypertarget{citeproc_bib_item_42}{{[}42{]} S. Raschka, “Model Evaluation,
Model Selection, and Algorithm Selection in Machine Learning.” arXiv,
2020. doi: \href{https://doi.org/10.48550/arXiv.1811.12808}{10.48550/arXiv.1811.12808}.}

\hypertarget{citeproc_bib_item_43}{{[}43{]} S. M. Lundberg and S.-I.
Lee, “A unified approach to interpreting model predictions,” \textit{Advances
in neural information processing systems}, vol. 30, pp. 4765--4774,
2017, Accessed: Feb. 10, 2021. {[}Online{]}. Available: \url{https://proceedings.neurips.cc/paper/2017/hash/8a20a8621978632d76c43dfd28b67767-Abstract.html}}

\hypertarget{citeproc_bib_item_44}{{[}44{]} S. M. Lundberg \textit{et
al.}, “From local explanations to global understanding with explainable
AI for trees,” \textit{Nature machine intelligence}, vol. 2, no. 1,
pp. 56--67, 2020, doi: \href{https://doi.org/10.1038/s42256-019-0138-9}{10.1038/s42256-019-0138-9}.}

\hypertarget{citeproc_bib_item_45}{{[}45{]} S. M. Lundberg \textit{et
al.}, “Explainable machine-learning predictions for the prevention
of hypoxaemia during surgery,” \textit{Nature biomedical engineering},
vol. 2, no. 10, pp. 749--760, 2018, doi: \href{https://doi.org/10.1038/s41551-018-0304-0}{10.1038/s41551-018-0304-0}.}

\hypertarget{citeproc_bib_item_46}{{[}46{]} J. C. Mauro, Y. Yue,
A. J. Ellison, P. K. Gupta, and D. C. Allan, “Viscosity of glass-forming
liquids.,” \textit{Proceedings of the national academy of sciences
of the united states of america}, vol. 106, no. 47, pp. 19780--19784,
2009, doi: \href{https://doi.org/10.1073/pnas.0911705106}{10.1073/pnas.0911705106}.}

\hypertarget{citeproc_bib_item_47}{{[}47{]} C. A. Angell, “Strong
and fragile liquids,” in \textit{Relaxation in complex systems}, K.
L. Ngai and G. B. Wright, Eds., Springfield: Naval Research Laboratory,
1985, pp. 3--12.}

\hypertarget{citeproc_bib_item_48}{{[}48{]} W. McKinney, “Data structures
for statistical computing in Python,” in \textit{Proceedings of the
9th Python in Science Conference}, Austin, Texas, 2010, pp. 51--56.
Accessed: Mar. 27, 2014. {[}Online{]}. Available: \url{http://jarrodmillman.com/scipy2010/pdfs/mckinney.pdf}}

\hypertarget{citeproc_bib_item_49}{{[}49{]} J. Jiusti, E. D. Zanotto,
D. R. Cassar, and M. R. B. Andreeta, “Viscosity and liquidus-based
predictor of glass-forming ability of oxide glasses,” \textit{Journal
of the american ceramic society}, vol. 103, no. 2, pp. 921--932,
2020, doi: \href{https://doi.org/10.1111/jace.16732}{10.1111/jace.16732}.}

\hypertarget{citeproc_bib_item_50}{{[}50{]} E. D. Zanotto, “A bright
future for glass-ceramics,” \textit{American ceramic society bulletin},
vol. 89, no. 8, pp. 19--27, 2010.}

\hypertarget{citeproc_bib_item_51}{{[}51{]} A. Sehlke and A. G. Whittington,
“The viscosity of planetary tholeiitic melts: A configurational entropy
model,” \textit{Geochimica et cosmochimica acta}, vol. 191, pp. 277--299,
2016, doi: \href{https://doi.org/10.1016/j.gca.2016.07.027}{10.1016/j.gca.2016.07.027}.}

\hypertarget{citeproc_bib_item_52}{{[}52{]} C. Le Losq and D. R.
Neuville, “Molecular structure, configurational entropy and viscosity
of silicate melts: Link through the Adam and Gibbs theory of viscous
flow,” \textit{Journal of non-crystalline solids}, vol. 463, pp. 175--188,
2017, doi: \href{https://doi.org/10.1016/j.jnoncrysol.2017.02.010}{10.1016/j.jnoncrysol.2017.02.010}.}

\hypertarget{citeproc_bib_item_53}{{[}53{]} K. Starodub, G. Wu, E.
Yazhenskikh, M. Müller, A. Khvan, and A. Kondratiev, “An Avramov-based
viscosity model for the SiO2-Al2O3-Na2O-K2O system in a wide temperature
range,” \textit{Ceramics international}, vol. 45, no. 9, pp. 12169--12181,
2019, doi: \href{https://doi.org/10.1016/j.ceramint.2019.03.121}{10.1016/j.ceramint.2019.03.121}.}

\hypertarget{citeproc_bib_item_54}{{[}54{]} S. Bozinovski and A.
Fulgosi, “The influence of pattern similarity and transfer learning
upon training of a base perceptron b2,” in \textit{Proceedings of
Symposium Informatica}, 1976, pp. 121--126.}

\hypertarget{citeproc_bib_item_55}{{[}55{]} S. Bozinovski, “Reminder
of the First Paper on Transfer Learning in Neural Networks, 1976,”
\textit{Informatica}, vol. 44, no. 3, 2020, doi: \href{https://doi.org/10.31449/inf.v44i3.2828}{10.31449/inf.v44i3.2828}.}

\hypertarget{citeproc_bib_item_56}{{[}56{]} A. K. Rappe, C. J. Casewit,
K. S. Colwell, W. A. Goddard, and W. M. Skiff, “UFF, a full periodic
table force field for molecular mechanics and molecular dynamics simulations,”
\textit{Journal of the american chemical society}, vol. 114, no. 25,
pp. 10024--10035, 1992, doi: \href{https://doi.org/10.1021/ja00051a040}{10.1021/ja00051a040}.}

\hypertarget{citeproc_bib_item_57}{{[}57{]} M. Rahm, R. Hoffmann,
and N. W. Ashcroft, “Atomic and Ionic Radii of Elements 1 96,” \textit{Chemistry
a european journal}, vol. 22, no. 41, pp. 14625--14632, 2016, doi:
\href{https://doi.org/10.1002/chem.201602949}{10.1002/chem.201602949}.}

\hypertarget{citeproc_bib_item_58}{{[}58{]} W. M. Haynes, \textit{CRC
Handbook of Chemistry and Physics}. CRC Press, 2014.}

\hypertarget{citeproc_bib_item_59}{{[}59{]} T. Andersen, “Atomic
negative ions: Structure, dynamics and collisions,” \textit{Physics
reports}, vol. 394, no. 4, pp. 157--313, 2004, doi: \href{https://doi.org/10.1016/j.physrep.2004.01.001}{10.1016/j.physrep.2004.01.001}.}

\hypertarget{citeproc_bib_item_60}{{[}60{]} R. T. Sanderson, “An
Explanation of Chemical Variations within Periodic Major Groups,”
\textit{Journal of the american chemical society}, vol. 74, no. 19,
pp. 4792--4794, 1952, doi: \href{https://doi.org/10.1021/ja01139a020}{10.1021/ja01139a020}.}

\hypertarget{citeproc_bib_item_61}{{[}61{]} R. T. Sanderson, “An
Interpretation of Bond Lengths and a Classification of Bonds,” \textit{Science},
vol. 114, no. 2973, pp. 670--672, 1951, doi: \href{https://doi.org/10.1126/science.114.2973.670}{10.1126/science.114.2973.670}.}

\hypertarget{citeproc_bib_item_62}{{[}62{]} C. Tantardini and A.
R. Oganov, “Thermochemical electronegativities of the elements,” \textit{Nature
communications}, vol. 12, no. 1, p. 2087, 2021, doi: \href{https://doi.org/10.1038/s41467-021-22429-0}{10.1038/s41467-021-22429-0}.}

\hypertarget{citeproc_bib_item_63}{{[}63{]} X. Chu and A. Dalgarno,
“Linear response time-dependent density functional theory for van
der Waals coefficients,” \textit{The journal of chemical physics},
vol. 121, no. 9, pp. 4083--4088, 2004, doi: \href{https://doi.org/10.1063/1.1779576}{10.1063/1.1779576}.}

\hypertarget{citeproc_bib_item_64}{{[}64{]} K. T. Tang, J. M. Norbeck,
and P. R. Certain, “Upper and lower bounds of two- and three-body
dipole, quadrupole, and octupole van der Waals coefficients for hydrogen,
noble gas, and alkali atom interactions,” \textit{The journal of chemical
physics}, vol. 64, no. 7, pp. 3063--3074, 1976, doi: \href{https://doi.org/10.1063/1.432569}{10.1063/1.432569}.}

\hypertarget{citeproc_bib_item_65}{{[}65{]} J. C. Slater, “Atomic
Radii in Crystals,” \textit{The journal of chemical physics}, vol.
41, no. 10, pp. 3199--3204, 1964, doi: \href{https://doi.org/10.1063/1.1725697}{10.1063/1.1725697}.}

\hypertarget{citeproc_bib_item_66}{{[}66{]} B. Cordero \textit{et
al.}, “Covalent radii revisited,” \textit{Dalton transactions}, no.
21, pp. 2832--2838, 2008, doi: \href{https://doi.org/10.1039/B801115J}{10.1039/B801115J}.}

\hypertarget{citeproc_bib_item_67}{{[}67{]} P. Pyykkö and M. Atsumi,
“Molecular Single-Bond Covalent Radii for Elements 1 118,” \textit{Chemistry
a european journal}, vol. 15, no. 1, pp. 186--197, 2009, doi: \href{https://doi.org/10.1002/chem.200800987}{10.1002/chem.200800987}.}

\hypertarget{citeproc_bib_item_68}{{[}68{]} P. Schwerdtfeger and
J. K. Nagle, “2018 Table of static dipole polarizabilities of the
neutral elements in the periodic table,” \textit{Molecular physics},
vol. 117, no. 9-12, pp. 1200--1225, 2019, doi: \href{https://doi.org/10.1080/00268976.2018.1535143}{10.1080/00268976.2018.1535143}.}

\hypertarget{citeproc_bib_item_69}{{[}69{]} A. L. Allred and E. G.
Rochow, “A scale of electronegativity based on electrostatic force,”
\textit{Journal of inorganic and nuclear chemistry}, vol. 5, no. 4,
pp. 264--268, 1958, doi: \href{https://doi.org/10.1016/0022-1902(58)80003-2}{10.1016/0022-1902(58)80003-2}.}

\hypertarget{citeproc_bib_item_70}{{[}70{]} T. L. Cottrell, L. E.
Sutton, and M. G. Evans, “Covalency, electrovalency and electronegativity,”
\textit{Proceedings of the royal society of london. series a. mathematical
and physical sciences}, vol. 207, no. 1088, pp. 49--63, 1997, doi:
\href{https://doi.org/10.1098/rspa.1951.0098}{10.1098/rspa.1951.0098}.}

\hypertarget{citeproc_bib_item_71}{{[}71{]} W. Gordy, “A New Method
of Determining Electronegativity from Other Atomic Properties,” \textit{Physical
review}, vol. 69, no. 11-12, pp. 604--607, 1946, doi: \href{https://doi.org/10.1103/PhysRev.69.604}{10.1103/PhysRev.69.604}.}

\hypertarget{citeproc_bib_item_72}{{[}72{]} D. C. Ghosh, “A new scale
of electronegativity based on absolute radii of atoms,” \textit{Journal
of theoretical and computational chemistry}, vol. 04, no. 01, pp.
21--33, 2005, doi: \href{https://doi.org/10.1142/S0219633605001556}{10.1142/S0219633605001556}.}

\hypertarget{citeproc_bib_item_73}{{[}73{]} S. S. Batsanov, “Dielectric
Methods of Studying the Chemical Bond and the Concept of Electronegativity,”
\textit{Russian chemical reviews}, vol. 51, no. 7, p. 684, 1982, doi:
\href{https://doi.org/10.1070/RC1982v051n07ABEH002900}{10.1070/RC1982v051n07ABEH002900}.}

\hypertarget{citeproc_bib_item_74}{{[}74{]} J. K. Nagle, “Atomic
polarizability and electronegativity,” \textit{Journal of the american
chemical society}, vol. 112, no. 12, pp. 4741--4747, 1990, doi: \href{https://doi.org/10.1021/ja00168a019}{10.1021/ja00168a019}.}

\hypertarget{citeproc_bib_item_75}{{[}75{]} H. Glawe, A. Sanna, E.
K. U. Gross, and M. A. L. Marques, “The optimal one dimensional periodic
table: A modified Pettifor chemical scale from data mining,” \textit{New
journal of physics}, vol. 18, no. 9, p. 093011, 2016, doi: \href{https://doi.org/10.1088/1367-2630/18/9/093011}{10.1088/1367-2630/18/9/093011}.}

\hypertarget{citeproc_bib_item_76}{{[}76{]} D. G. Pettifor, “A chemical
scale for crystal-structure maps,” \textit{Solid state communications},
vol. 51, no. 1, pp. 31--34, 1984, doi: \href{https://doi.org/10.1016/0038-1098(84)90765-8}{10.1016/0038-1098(84)90765-8}.}

\hypertarget{citeproc_bib_item_77}{{[}77{]} P. Villars, K. Cenzual,
J. Daams, Y. Chen, and S. Iwata, “Data-driven atomic environment prediction
for binaries using the Mendeleev number: Part 1. Composition AB,”
\textit{Journal of alloys and compounds}, vol. 367, no. 1, pp. 167--175,
2004, doi: \href{https://doi.org/10.1016/j.jallcom.2003.08.060}{10.1016/j.jallcom.2003.08.060}.}

\hypertarget{citeproc_bib_item_78}{{[}78{]} S. Alvarez, “A cartography
of the van der Waals territories,” \textit{Dalton transactions}, vol.
42, no. 24, pp. 8617--8636, 2013, doi: \href{https://doi.org/10.1039/C3DT50599E}{10.1039/C3DT50599E}.}

\hypertarget{citeproc_bib_item_79}{{[}79{]} N. L. Allinger, X. Zhou,
and J. Bergsma, “Molecular mechanics parameters,” \textit{Journal
of molecular structure: Theochem}, vol. 312, no. 1, pp. 69--83, 1994,
doi: \href{https://doi.org/10.1016/S0166-1280(09)80008-0}{10.1016/S0166-1280(09)80008-0}.}

\hypertarget{citeproc_bib_item_80}{{[}80{]} T. Chen and C. Guestrin,
“XGBoost: A scalable tree boosting system,” in \textit{Proceedings
of the 22nd ACM SIGKDD international conference on knowledge discovery
and data mining}, in KDD ’16. New York, NY, USA: ACM, 2016, pp. 785--794.
doi: \href{https://doi.org/10.1145/2939672.2939785}{10.1145/2939672.2939785}.}
\newpage\endgroup

\section*{Tables and figures}

\label{sec:orgf3ff32e} \newpage{}

\begin{landscape}

\begin{longtable}[c]{ll}
\caption{\label{tab:symbols}Meaning and units of some symbols used in this
work. Reference for the glass and melt data is the SciGlass database,
and references for the physicochemical data are given in the table.
For a complete list of all the properties predicted in this work,
see Table 3. For a complete list of all the attributes of the neural
network, see Table 5.}
\tabularnewline
\hline 
Symbol  & Meaning\tabularnewline
\endfirsthead
\hline 
\multicolumn{2}{l}{Continued from previous page}\tabularnewline
\hline 
Symbol  & Meaning \tabularnewline
\hline 
\endhead
\hline 
\multicolumn{2}{r}{Continued on next page}\tabularnewline
\hline 
\endfoot
\endlastfoot
\hline 
$T$  & Temperature (K)\tabularnewline
$\eta$  & Equilibrium shear viscosity (Pa.s)\tabularnewline
$T_{n}$ ($n\in\mathbb{Z}$)  & Temperature (K) where $\eta=10^{n}$ Pa.s\tabularnewline
$T_{g}$  & Glass transition temperature (K)\tabularnewline
$T_{\mathrm{melt}}$  & Melting temperature (K)\tabularnewline
$T_{\mathrm{liq}}$  & Liquidus temperature (K)\tabularnewline
$T_{\mathrm{Lit}}$  & Littleton softening point (K)\tabularnewline
$T_{\mathrm{ann}}$  & Annealing point (K)\tabularnewline
$T_{\mathrm{strain}}$  & Strain point (K)\tabularnewline
$T_{\mathrm{soft}}$  & Softening point (K)\tabularnewline
$T_{\mathrm{dil}}$  & Dilatometric softening temperature (K)\tabularnewline
$V_{D}$  & Abbe's number\tabularnewline
$n_{D}$  & Refractive index\tabularnewline
$n$ (low)  & Refractive index measured at a wavelength between 0.6 and 1 micron
at 293 K\tabularnewline
$n$ (high)  & Refractive index measured at a wavelength greater than 1 micron at
293 K\tabularnewline
$n_{F}-n_{C}$  & Mean dispersion\tabularnewline
$\varepsilon$  & Relative permittivity at ambient temperature and frequency of 1 MHz\tabularnewline
$T_{\rho=10^{6}\,\Omega.m}$  & Temperature (K) where the specific electrical resistivity is 10\textsuperscript{6}
$\Omega$.m\tabularnewline
$\rho$  & Specific electrical resistivity ($\Omega$.m)\tabularnewline
$E$  & Young's modulus (GPa)\tabularnewline
$G$  & Shear modulus (GPa)\tabularnewline
$H$  & Microhardness measured by Knoop or Vickers indentation (GPa)\tabularnewline
$\nu$  & Poisson's ratio\tabularnewline
$d$  & Density (g.cm\textsuperscript{\textminus 3})\tabularnewline
$\kappa$  & Thermal conductivity (W.m\textsuperscript{\textminus 1}.K\textsuperscript{\textminus 1})\tabularnewline
$\Delta T$  & Thermal shock resistance (K)\tabularnewline
$\alpha_{L}$  & Linear coefficient of thermal expansion (K\textsuperscript{\textminus 1})\tabularnewline
$C_{p}$  & Heat capacity at constant pressure (J.kg\textsuperscript{\textminus 1}.K\textsuperscript{\textminus 1})\tabularnewline
$T_{\mathrm{max}(U)}$  & Temperature (K) of maximum crystal growth velocity\tabularnewline
$U_{\mathrm{max}}$  & Maximum crystal growth velocity (m.s\textsuperscript{\textminus 1})\tabularnewline
$T_{c}$  & Temperature (K) of the crystallization peak (differential thermal
analysis)\tabularnewline
$T_{x}$  & Temperature (K) of the crystallization onset (differential thermal
analysis)\tabularnewline
$\gamma$  & Surface tension (J.m\textsuperscript{\textminus 2})\tabularnewline
\hline 
\hline 
$r_{W}$  & Van der Walls radius (pm) \citeprocitem{56}{{[}56{]}}\tabularnewline
$r_{R}$  & Atomic radius (pm) \citeprocitem{57}{{[}57{]}}\tabularnewline
$V_{at}$  & Atomic volume (cm\textsuperscript{3}.mol\textsuperscript{--1}) \citeprocitem{30}{{[}30{]}}\tabularnewline
$E_{ea}$  & Electron affinity (eV) \citeprocitem{58}{{[}58{]}}, \citeprocitem{59}{{[}59{]}}\tabularnewline
$E_{g}$  & DFT bandgap energy of $T=0\,\mathrm{K}$ ground state (eV) \citeprocitem{26}{{[}26{]}}\tabularnewline
$E_{at}$  & DFT energy per atom of $T=0\,\mathrm{K}$ ground state (eV.atom\textsuperscript{--1})
\citeprocitem{26}{{[}26{]}}\tabularnewline
$m_{m}$  & DFT magnetic moment of $T=0\,\mathrm{K}$ ground state \citeprocitem{26}{{[}26{]}}\tabularnewline
$\mathrm{FCC}_{lp}$  & Estimated FCC lattice parameter based on the DFT volume of the OQMD
ground state \citeprocitem{26}{{[}26{]}}\tabularnewline
$Z_{\mathrm{eff}}$  & Effective nuclear charge \citeprocitem{30}{{[}30{]}}\tabularnewline
$\chi_{S}$  & Electronegativity in the Sanderson scale \citeprocitem{60}{{[}60{]}},
\citeprocitem{61}{{[}61{]}}\tabularnewline
$\chi_{TO}$  & Electronegativity in the Tardini--Organov scale \citeprocitem{62}{{[}62{]}}\tabularnewline
$T_{b}$  & Boiling point (K) \citeprocitem{58}{{[}58{]}}\tabularnewline
$\Delta H_{m}$  & Melting Enthalpy (kJ.mol\textsuperscript{\textminus 1}) \citeprocitem{26}{{[}26{]}}\tabularnewline
$C_{6}$  & C\textsubscript{6} coefficient (a.u) \citeprocitem{63}{{[}63{]}},
\citeprocitem{64}{{[}64{]}}\tabularnewline
$N_{v}$  & Number of valence electrons \citeprocitem{30}{{[}30{]}}\tabularnewline
$N_{ox}$  & Number of oxidation states \citeprocitem{30}{{[}30{]}}\tabularnewline
$N_{u}$  & Number of unfilled valence orbitals \citeprocitem{26}{{[}26{]}}\tabularnewline
$N_{u,s}$  & Number of unfilled s valence orbitals \citeprocitem{26}{{[}26{]}}\tabularnewline
$N_{u,p}$  & Number of unfilled p valence orbitals \citeprocitem{26}{{[}26{]}}\tabularnewline
$N_{u,d}$  & Number of unfilled d valence orbitals \citeprocitem{26}{{[}26{]}}\tabularnewline
$N_{u,f}$  & Number of unfilled f valence orbitals \citeprocitem{26}{{[}26{]}}\tabularnewline
$N_{f,s}$  & Number of filled s valence orbitals \citeprocitem{26}{{[}26{]}}\tabularnewline
$N_{f,p}$  & Number of filled p valence orbitals \citeprocitem{26}{{[}26{]}}\tabularnewline
$N_{f,d}$  & Number of filled d valence orbitals \citeprocitem{26}{{[}26{]}}\tabularnewline
$N_{f,f}$  & Number of filled f valence orbitals \citeprocitem{26}{{[}26{]}}\tabularnewline
\hline 
\end{longtable}

\end{landscape} 
\begin{table}[htbp]
\caption{\label{tab:search_space}Search space used during hyperparameter tuning
and selected hyperparameter values. The learning rate values were
drawn in logarithmic space instead of linear space to sample the entire
interval uniformly.}
\centering %
\begin{tabular}{lll}
\hline 
Hyperparameter  & Search space  & Selected\tabularnewline
\hline 
\hline 
Number of layers  & \{1, 2, 3, 4, 5\}  & 4\tabularnewline
Number of neurons in layer 1  & \{10, 20, \ldots , 500\}  & 280\tabularnewline
Number of neurons in layer 2  & \{10, 20, \ldots , 500\}  & 500\tabularnewline
Number of neurons in layer 3  & \{10, 20, \ldots , 500\}  & 390\tabularnewline
Number of neurons in layer 4  & \{10, 20, \ldots , 500\}  & 480\tabularnewline
Number of neurons in layer 5  & \{10, 20, \ldots , 500\}  & --\tabularnewline
Activation function of layer 1  & See text  & Softplus\tabularnewline
Activation function of layer 2  & See text  & Mish\tabularnewline
Activation function of layer 3  & See text  & Leaky ReLU\tabularnewline
Activation function of layer 4  & See text  & PReLU\tabularnewline
Activation function of layer 5  & See text  & --\tabularnewline
Use batch normalization in layer 1  & \{true, false\}  & true\tabularnewline
Use batch normalization in layer 2  & \{true, false\}  & true\tabularnewline
Use batch normalization in layer 3  & \{true, false\}  & false\tabularnewline
Use batch normalization in layer 4  & \{true, false\}  & false\tabularnewline
Use batch normalization in layer 5  & \{true, false\}  & --\tabularnewline
Dropout percentage in layer 1  & {[}0, 0.5{]}  & 0.081\tabularnewline
Dropout percentage in layer 2  & {[}0, 0.5{]}  & 0.001\tabularnewline
Dropout percentage in layer 3  & {[}0, 0.5{]}  & 0.087\tabularnewline
Dropout percentage in layer 4  & {[}0, 0.5{]}  & 0.168\tabularnewline
Dropout percentage in layer 5  & {[}0, 0.5{]}  & --\tabularnewline
Learning rate  & {[}10\textsuperscript{-5}, 10\textsuperscript{-1}{]}  & 1.33 \texttimes{} 10\textsuperscript{-5}\tabularnewline
Batch size  & \{256, 512, 1024, 2048, 4096\}  & 256\tabularnewline
Early stopping patience  & \{5, 6, \ldots , 35\}  & 27\tabularnewline
\hline 
\end{tabular}
\end{table}

\newpage{}

\begin{landscape}

{\scriptsize{}}%
\begin{longtable}[c]{lrrrrrrrlll}
\caption{\label{tab:descript_metrics}Descriptive statistics and metrics (RMSE,
see text) for all the properties studied in this work. The lower the
RMSE, the better. The last three columns show the results of the t-test
(95\% confidence) used to compare the performance of the models. In
these last three columns, $=$ means that no statistical difference
was observed and the name of the model with the better performance
is shown if a statistical difference was observed.}
\tabularnewline
\hline 
{\scriptsize{}Symbol } & {\scriptsize{}Count } & {\scriptsize{}Min } & {\scriptsize{}Mean } & {\scriptsize{}Max } & {\scriptsize{}MT-MLP RMSE } & {\scriptsize{}MT-MH RMSE } & {\scriptsize{}ST-NN RMSE } & {\scriptsize{}MT-MLP vs. MT-MH } & {\scriptsize{}MT-MLP vs. ST-NN } & {\scriptsize{}MT-MH vs. ST-NN}\tabularnewline
\endfirsthead
\hline 
\multicolumn{11}{l}{{\scriptsize{}Continued from previous page}}\tabularnewline
\hline 
{\scriptsize{}Symbol } & {\scriptsize{}Count } & {\scriptsize{}Min } & {\scriptsize{}Mean } & {\scriptsize{}Max } & {\scriptsize{}MT-MLP RMSE } & {\scriptsize{}MT-MH RMSE } & {\scriptsize{}ST-NN RMSE } & {\scriptsize{}MT-MLP vs. MT-MH } & {\scriptsize{}MT-MLP vs. ST-NN } & {\scriptsize{}MT-MH vs. ST-NN }\tabularnewline
\hline 
\endhead
\hline 
\multicolumn{11}{r}{{\scriptsize{}Continued on next page}}\tabularnewline
\hline 
\endfoot
\endlastfoot
\hline 
{\scriptsize{}$T_{0}$ } & {\scriptsize{}1598 } & {\scriptsize{}412 } & {\scriptsize{}1520 } & {\scriptsize{}2620 } & {\scriptsize{}100(20) } & {\scriptsize{}90(20) } & {\scriptsize{}70(20) } & {\scriptsize{}$=$ } & {\scriptsize{}ST-NN } & {\scriptsize{}$=$}\tabularnewline
{\scriptsize{}$T_{1}$ } & {\scriptsize{}7043 } & {\scriptsize{}373 } & {\scriptsize{}1620 } & {\scriptsize{}2500 } & {\scriptsize{}70(10) } & {\scriptsize{}65(8) } & {\scriptsize{}50(90) } & {\scriptsize{}$=$ } & {\scriptsize{}$=$ } & {\scriptsize{}$=$}\tabularnewline
{\scriptsize{}$T_{2}$ } & {\scriptsize{}14250 } & {\scriptsize{}391 } & {\scriptsize{}1540 } & {\scriptsize{}2500 } & {\scriptsize{}48(6) } & {\scriptsize{}43(5) } & {\scriptsize{}54(9) } & {\scriptsize{}$=$ } & {\scriptsize{}$=$ } & {\scriptsize{}MT-MH}\tabularnewline
{\scriptsize{}$T_{3}$ } & {\scriptsize{}13216 } & {\scriptsize{}358 } & {\scriptsize{}1380 } & {\scriptsize{}2240 } & {\scriptsize{}51(4) } & {\scriptsize{}48(5) } & {\scriptsize{}60(7) } & {\scriptsize{}$=$ } & {\scriptsize{}MT-MLP } & {\scriptsize{}MT-MH}\tabularnewline
{\scriptsize{}$T_{4}$ } & {\scriptsize{}7375 } & {\scriptsize{}344 } & {\scriptsize{}1240 } & {\scriptsize{}1800 } & {\scriptsize{}44(7) } & {\scriptsize{}40(10) } & {\scriptsize{}54(6) } & {\scriptsize{}$=$ } & {\scriptsize{}MT-MLP } & {\scriptsize{}MT-MH}\tabularnewline
{\scriptsize{}$T_{5}$ } & {\scriptsize{}5452 } & {\scriptsize{}334 } & {\scriptsize{}1060 } & {\scriptsize{}2210 } & {\scriptsize{}39(9) } & {\scriptsize{}37(7) } & {\scriptsize{}42(7) } & {\scriptsize{}$=$ } & {\scriptsize{}$=$ } & {\scriptsize{}$=$}\tabularnewline
{\scriptsize{}$T_{6}$ } & {\scriptsize{}4790 } & {\scriptsize{}326 } & {\scriptsize{}973 } & {\scriptsize{}2040 } & {\scriptsize{}40(7) } & {\scriptsize{}39(5) } & {\scriptsize{}40(10) } & {\scriptsize{}$=$ } & {\scriptsize{}$=$ } & {\scriptsize{}$=$}\tabularnewline
{\scriptsize{}$T_{7}$ } & {\scriptsize{}4977 } & {\scriptsize{}318 } & {\scriptsize{}907 } & {\scriptsize{}1900 } & {\scriptsize{}37(5) } & {\scriptsize{}34(3) } & {\scriptsize{}39(9) } & {\scriptsize{}$=$ } & {\scriptsize{}$=$ } & {\scriptsize{}$=$}\tabularnewline
{\scriptsize{}$T_{8}$ } & {\scriptsize{}5158 } & {\scriptsize{}310 } & {\scriptsize{}870 } & {\scriptsize{}1770 } & {\scriptsize{}35(3) } & {\scriptsize{}34(3) } & {\scriptsize{}49(8) } & {\scriptsize{}$=$ } & {\scriptsize{}MT-MLP } & {\scriptsize{}MT-MH}\tabularnewline
{\scriptsize{}$T_{9}$ } & {\scriptsize{}5561 } & {\scriptsize{}303 } & {\scriptsize{}847 } & {\scriptsize{}1690 } & {\scriptsize{}32(3) } & {\scriptsize{}31(2) } & {\scriptsize{}40(5) } & {\scriptsize{}$=$ } & {\scriptsize{}MT-MLP } & {\scriptsize{}MT-MH}\tabularnewline
{\scriptsize{}$T_{10}$ } & {\scriptsize{}5899 } & {\scriptsize{}296 } & {\scriptsize{}827 } & {\scriptsize{}1640 } & {\scriptsize{}31(3) } & {\scriptsize{}32(3) } & {\scriptsize{}50(7) } & {\scriptsize{}$=$ } & {\scriptsize{}MT-MLP } & {\scriptsize{}MT-MH}\tabularnewline
{\scriptsize{}$T_{11}$ } & {\scriptsize{}5589 } & {\scriptsize{}290 } & {\scriptsize{}806 } & {\scriptsize{}1650 } & {\scriptsize{}33(3) } & {\scriptsize{}31(3) } & {\scriptsize{}46(6) } & {\scriptsize{}$=$ } & {\scriptsize{}MT-MLP } & {\scriptsize{}MT-MH}\tabularnewline
{\scriptsize{}$T_{12}$ } & {\scriptsize{}4357 } & {\scriptsize{}281 } & {\scriptsize{}779 } & {\scriptsize{}1520 } & {\scriptsize{}32(4) } & {\scriptsize{}32(4) } & {\scriptsize{}43(7) } & {\scriptsize{}$=$ } & {\scriptsize{}MT-MLP } & {\scriptsize{}MT-MH}\tabularnewline
{\scriptsize{}$\log_{10}(\eta(773\,\mathrm{K}))$ } & {\scriptsize{}1838 } & {\scriptsize{}-0.977 } & {\scriptsize{}8.46 } & {\scriptsize{}12.5 } & {\scriptsize{}1.1(2) } & {\scriptsize{}1.0(3) } & {\scriptsize{}1.1(1) } & {\scriptsize{}$=$ } & {\scriptsize{}$=$ } & {\scriptsize{}$=$}\tabularnewline
{\scriptsize{}$\log_{10}(\eta(873\,\mathrm{K}))$ } & {\scriptsize{}3217 } & {\scriptsize{}-0.676 } & {\scriptsize{}7.88 } & {\scriptsize{}12.5 } & {\scriptsize{}0.8(1) } & {\scriptsize{}0.8(2) } & {\scriptsize{}0.80(9) } & {\scriptsize{}$=$ } & {\scriptsize{}$=$ } & {\scriptsize{}$=$}\tabularnewline
{\scriptsize{}$\log_{10}(\eta(973\,\mathrm{K}))$ } & {\scriptsize{}3703 } & {\scriptsize{}-0.945 } & {\scriptsize{}6.42 } & {\scriptsize{}12.5 } & {\scriptsize{}0.73(6) } & {\scriptsize{}0.71(7) } & {\scriptsize{}0.7(1) } & {\scriptsize{}$=$ } & {\scriptsize{}$=$ } & {\scriptsize{}$=$}\tabularnewline
{\scriptsize{}$\log_{10}(\eta(1073\,\mathrm{K}))$ } & {\scriptsize{}4121 } & {\scriptsize{}-0.919 } & {\scriptsize{}5.27 } & {\scriptsize{}12.5 } & {\scriptsize{}0.59(7) } & {\scriptsize{}0.55(6) } & {\scriptsize{}0.56(7) } & {\scriptsize{}$=$ } & {\scriptsize{}$=$ } & {\scriptsize{}$=$}\tabularnewline
{\scriptsize{}$\log_{10}(\eta(1173\,\mathrm{K}))$ } & {\scriptsize{}4612 } & {\scriptsize{}-0.98 } & {\scriptsize{}4.2 } & {\scriptsize{}12.4 } & {\scriptsize{}0.48(6) } & {\scriptsize{}0.46(4) } & {\scriptsize{}0.44(7) } & {\scriptsize{}$=$ } & {\scriptsize{}$=$ } & {\scriptsize{}$=$}\tabularnewline
{\scriptsize{}$\log_{10}(\eta(1273\,\mathrm{K}))$ } & {\scriptsize{}5551 } & {\scriptsize{}-0.901 } & {\scriptsize{}3.4 } & {\scriptsize{}12.4 } & {\scriptsize{}0.44(4) } & {\scriptsize{}0.43(5) } & {\scriptsize{}0.35(4) } & {\scriptsize{}$=$ } & {\scriptsize{}ST-NN } & {\scriptsize{}ST-NN}\tabularnewline
{\scriptsize{}$\log_{10}(\eta(1373\,\mathrm{K}))$ } & {\scriptsize{}7130 } & {\scriptsize{}-0.966 } & {\scriptsize{}2.84 } & {\scriptsize{}12.4 } & {\scriptsize{}0.33(3) } & {\scriptsize{}0.32(4) } & {\scriptsize{}0.31(6) } & {\scriptsize{}$=$ } & {\scriptsize{}$=$ } & {\scriptsize{}$=$}\tabularnewline
{\scriptsize{}$\log_{10}(\eta(1473\,\mathrm{K}))$ } & {\scriptsize{}9426 } & {\scriptsize{}-0.882 } & {\scriptsize{}2.52 } & {\scriptsize{}12.5 } & {\scriptsize{}0.28(2) } & {\scriptsize{}0.28(3) } & {\scriptsize{}0.21(4) } & {\scriptsize{}$=$ } & {\scriptsize{}ST-NN } & {\scriptsize{}ST-NN}\tabularnewline
{\scriptsize{}$\log_{10}(\eta(1573\,\mathrm{K}))$ } & {\scriptsize{}10576 } & {\scriptsize{}-1 } & {\scriptsize{}2.09 } & {\scriptsize{}11.8 } & {\scriptsize{}0.24(1) } & {\scriptsize{}0.23(2) } & {\scriptsize{}0.27(8) } & {\scriptsize{}$=$ } & {\scriptsize{}$=$ } & {\scriptsize{}$=$}\tabularnewline
{\scriptsize{}$\log_{10}(\eta(1673\,\mathrm{K}))$ } & {\scriptsize{}10285 } & {\scriptsize{}-0.955 } & {\scriptsize{}1.71 } & {\scriptsize{}10.8 } & {\scriptsize{}0.24(1) } & {\scriptsize{}0.23(2) } & {\scriptsize{}0.17(5) } & {\scriptsize{}$=$ } & {\scriptsize{}ST-NN } & {\scriptsize{}ST-NN}\tabularnewline
{\scriptsize{}$\log_{10}(\eta(1773\,\mathrm{K}))$ } & {\scriptsize{}7577 } & {\scriptsize{}-0.996 } & {\scriptsize{}1.48 } & {\scriptsize{}7.92 } & {\scriptsize{}0.21(2) } & {\scriptsize{}0.21(2) } & {\scriptsize{}0.24(6) } & {\scriptsize{}$=$ } & {\scriptsize{}$=$ } & {\scriptsize{}$=$}\tabularnewline
{\scriptsize{}$\log_{10}(\eta(1873\,\mathrm{K}))$ } & {\scriptsize{}4117 } & {\scriptsize{}-0.983 } & {\scriptsize{}1.37 } & {\scriptsize{}7.21 } & {\scriptsize{}0.23(3) } & {\scriptsize{}0.21(3) } & {\scriptsize{}0.20(2) } & {\scriptsize{}$=$ } & {\scriptsize{}ST-NN } & {\scriptsize{}$=$}\tabularnewline
{\scriptsize{}$\log_{10}(\eta(2073\,\mathrm{K}))$ } & {\scriptsize{}172 } & {\scriptsize{}-0.976 } & {\scriptsize{}0.844 } & {\scriptsize{}5.8 } & {\scriptsize{}0.2(2) } & {\scriptsize{}0.2(2) } & {\scriptsize{}2(1) } & {\scriptsize{}$=$ } & {\scriptsize{}MT-MLP } & {\scriptsize{}MT-MH}\tabularnewline
{\scriptsize{}$\log_{10}(\eta(2273\,\mathrm{K}))$ } & {\scriptsize{}44 } & {\scriptsize{}-0.984 } & {\scriptsize{}0.452 } & {\scriptsize{}4.56 } & {\scriptsize{}0.2(5) } & {\scriptsize{}0.2(5) } & {\scriptsize{}3(1) } & {\scriptsize{}$=$ } & {\scriptsize{}MT-MLP } & {\scriptsize{}MT-MH}\tabularnewline
{\scriptsize{}$\log_{10}(\eta(2473\,\mathrm{K}))$ } & {\scriptsize{}17 } & {\scriptsize{}-0.998 } & {\scriptsize{}0.494 } & {\scriptsize{}3.67 } & {\scriptsize{}0.1(6) } & {\scriptsize{}0(1) } & {\scriptsize{}3(2) } & {\scriptsize{}$=$ } & {\scriptsize{}MT-MLP } & {\scriptsize{}MT-MH}\tabularnewline
{\scriptsize{}$T_{g}$ } & {\scriptsize{}73504 } & {\scriptsize{}213 } & {\scriptsize{}713 } & {\scriptsize{}1460 } & {\scriptsize{}48(1) } & {\scriptsize{}45(1) } & {\scriptsize{}26(2) } & {\scriptsize{}MT-MH } & {\scriptsize{}ST-NN } & {\scriptsize{}ST-NN}\tabularnewline
{\scriptsize{}$T_{\mathrm{melt}}$ } & {\scriptsize{}15552 } & {\scriptsize{}160 } & {\scriptsize{}1040 } & {\scriptsize{}3230 } & {\scriptsize{}115(8) } & {\scriptsize{}99(5) } & {\scriptsize{}100(20) } & {\scriptsize{}MT-MH } & {\scriptsize{}ST-NN } & {\scriptsize{}$=$}\tabularnewline
{\scriptsize{}$T_{\mathrm{liq}}$ } & {\scriptsize{}40260 } & {\scriptsize{}303 } & {\scriptsize{}1340 } & {\scriptsize{}3090 } & {\scriptsize{}99(4) } & {\scriptsize{}86(2) } & {\scriptsize{}50(10) } & {\scriptsize{}MT-MH } & {\scriptsize{}ST-NN } & {\scriptsize{}ST-NN}\tabularnewline
{\scriptsize{}$T_{\mathrm{Lit}}$ } & {\scriptsize{}4226 } & {\scriptsize{}533 } & {\scriptsize{}979 } & {\scriptsize{}1890 } & {\scriptsize{}31(4) } & {\scriptsize{}30(5) } & {\scriptsize{}40(10) } & {\scriptsize{}$=$ } & {\scriptsize{}MT-MLP } & {\scriptsize{}MT-MH}\tabularnewline
{\scriptsize{}$T_{\mathrm{ann}}$ } & {\scriptsize{}8292 } & {\scriptsize{}334 } & {\scriptsize{}918 } & {\scriptsize{}1400 } & {\scriptsize{}21(1) } & {\scriptsize{}20(2) } & {\scriptsize{}33(6) } & {\scriptsize{}$=$ } & {\scriptsize{}MT-MLP } & {\scriptsize{}MT-MH}\tabularnewline
{\scriptsize{}$T_{\mathrm{strain}}$ } & {\scriptsize{}10377 } & {\scriptsize{}332 } & {\scriptsize{}884 } & {\scriptsize{}1390 } & {\scriptsize{}24(2) } & {\scriptsize{}23(2) } & {\scriptsize{}36(6) } & {\scriptsize{}$=$ } & {\scriptsize{}MT-MLP } & {\scriptsize{}MT-MH}\tabularnewline
{\scriptsize{}$T_{\mathrm{soft}}$ } & {\scriptsize{}12815 } & {\scriptsize{}311 } & {\scriptsize{}926 } & {\scriptsize{}1520 } & {\scriptsize{}45(4) } & {\scriptsize{}42(4) } & {\scriptsize{}32(5) } & {\scriptsize{}$=$ } & {\scriptsize{}ST-NN } & {\scriptsize{}ST-NN}\tabularnewline
{\scriptsize{}$T_{\mathrm{dil}}$ } & {\scriptsize{}18312 } & {\scriptsize{}284 } & {\scriptsize{}819 } & {\scriptsize{}1420 } & {\scriptsize{}44(2) } & {\scriptsize{}40(1) } & {\scriptsize{}36(4) } & {\scriptsize{}MT-MH } & {\scriptsize{}ST-NN } & {\scriptsize{}ST-NN}\tabularnewline
{\scriptsize{}$V_{D}$ } & {\scriptsize{}26019 } & {\scriptsize{}7.94 } & {\scriptsize{}46.3 } & {\scriptsize{}110 } & {\scriptsize{}3.3(3) } & {\scriptsize{}2.8(2) } & {\scriptsize{}1.9(4) } & {\scriptsize{}MT-MH } & {\scriptsize{}ST-NN } & {\scriptsize{}ST-NN}\tabularnewline
{\scriptsize{}$n_{D}$ } & {\scriptsize{}54442 } & {\scriptsize{}1.12 } & {\scriptsize{}1.68 } & {\scriptsize{}3.87 } & {\scriptsize{}0.059(5) } & {\scriptsize{}0.052(4) } & {\scriptsize{}0.029(4) } & {\scriptsize{}MT-MH } & {\scriptsize{}ST-NN } & {\scriptsize{}ST-NN}\tabularnewline
{\scriptsize{}$n$ (low) } & {\scriptsize{}359 } & {\scriptsize{}1.43 } & {\scriptsize{}2.35 } & {\scriptsize{}3.11 } & {\scriptsize{}0.13(4) } & {\scriptsize{}0.12(4) } & {\scriptsize{}0.1(3) } & {\scriptsize{}$=$ } & {\scriptsize{}$=$ } & {\scriptsize{}$=$}\tabularnewline
{\scriptsize{}$n$ (high) } & {\scriptsize{}410 } & {\scriptsize{}1.9 } & {\scriptsize{}2.56 } & {\scriptsize{}3.5 } & {\scriptsize{}0.09(1) } & {\scriptsize{}0.08(2) } & {\scriptsize{}0.11(4) } & {\scriptsize{}$=$ } & {\scriptsize{}$=$ } & {\scriptsize{}$=$}\tabularnewline
{\scriptsize{}$\log_{10}(n_{F}-n_{C})$ } & {\scriptsize{}25873 } & {\scriptsize{}-2.8 } & {\scriptsize{}-1.79 } & {\scriptsize{}-0.749 } & {\scriptsize{}0.044(5) } & {\scriptsize{}0.040(5) } & {\scriptsize{}0.03(1) } & {\scriptsize{}$=$ } & {\scriptsize{}ST-NN } & {\scriptsize{}ST-NN}\tabularnewline
{\scriptsize{}$\varepsilon$ } & {\scriptsize{}3113 } & {\scriptsize{}1.74 } & {\scriptsize{}10.3 } & {\scriptsize{}50 } & {\scriptsize{}2.2(3) } & {\scriptsize{}2.1(4) } & {\scriptsize{}2.2(3) } & {\scriptsize{}$=$ } & {\scriptsize{}$=$ } & {\scriptsize{}$=$}\tabularnewline
{\scriptsize{}$\log_{10}(\tan(\delta))$ } & {\scriptsize{}2335 } & {\scriptsize{}-4 } & {\scriptsize{}-2.6 } & {\scriptsize{}-0.815 } & {\scriptsize{}0.18(2) } & {\scriptsize{}0.20(3) } & {\scriptsize{}0.22(4) } & {\scriptsize{}$=$ } & {\scriptsize{}MT-MLP } & {\scriptsize{}$=$}\tabularnewline
{\scriptsize{}$T_{\rho=10^{6}\,\Omega.m}$ } & {\scriptsize{}12592 } & {\scriptsize{}0.15 } & {\scriptsize{}473 } & {\scriptsize{}1580 } & {\scriptsize{}65(4) } & {\scriptsize{}63(4) } & {\scriptsize{}50(8) } & {\scriptsize{}$=$ } & {\scriptsize{}ST-NN } & {\scriptsize{}ST-NN}\tabularnewline
{\scriptsize{}$\log_{10}(\rho(273\,\mathrm{K}))$ } & {\scriptsize{}13130 } & {\scriptsize{}-28.2 } & {\scriptsize{}4.98 } & {\scriptsize{}31.2 } & {\scriptsize{}2.5(2) } & {\scriptsize{}2.3(2) } & {\scriptsize{}2.6(4) } & {\scriptsize{}MT-MH } & {\scriptsize{}$=$ } & {\scriptsize{}$=$}\tabularnewline
{\scriptsize{}$\log_{10}(\rho(373\,\mathrm{K}))$ } & {\scriptsize{}13681 } & {\scriptsize{}-29.2 } & {\scriptsize{}3.8 } & {\scriptsize{}26.3 } & {\scriptsize{}2.2(2) } & {\scriptsize{}2.1(2) } & {\scriptsize{}1.8(4) } & {\scriptsize{}$=$ } & {\scriptsize{}ST-NN } & {\scriptsize{}$=$}\tabularnewline
{\scriptsize{}$\log_{10}(\rho(423\,\mathrm{K}))$ } & {\scriptsize{}14499 } & {\scriptsize{}-26.6 } & {\scriptsize{}3.54 } & {\scriptsize{}24.6 } & {\scriptsize{}2.1(1) } & {\scriptsize{}2.0(2) } & {\scriptsize{}1.7(4) } & {\scriptsize{}$=$ } & {\scriptsize{}ST-NN } & {\scriptsize{}$=$}\tabularnewline
{\scriptsize{}$\log_{10}(\rho(573\,\mathrm{K}))$ } & {\scriptsize{}11351 } & {\scriptsize{}-21.7 } & {\scriptsize{}2.66 } & {\scriptsize{}17.2 } & {\scriptsize{}1.89(7) } & {\scriptsize{}1.82(9) } & {\scriptsize{}2.0(2) } & {\scriptsize{}$=$ } & {\scriptsize{}$=$ } & {\scriptsize{}$=$}\tabularnewline
{\scriptsize{}$\log_{10}(\rho(1073\,\mathrm{K}))$ } & {\scriptsize{}1880 } & {\scriptsize{}-8.3 } & {\scriptsize{}-2.78 } & {\scriptsize{}3.73 } & {\scriptsize{}0.32(6) } & {\scriptsize{}0.33(7) } & {\scriptsize{}0.5(1) } & {\scriptsize{}$=$ } & {\scriptsize{}MT-MLP } & {\scriptsize{}MT-MH}\tabularnewline
{\scriptsize{}$\log_{10}(\rho(1273\,\mathrm{K}))$ } & {\scriptsize{}2334 } & {\scriptsize{}-8 } & {\scriptsize{}-3.16 } & {\scriptsize{}4.09 } & {\scriptsize{}0.65(8) } & {\scriptsize{}0.7(1) } & {\scriptsize{}0.43(8) } & {\scriptsize{}$=$ } & {\scriptsize{}ST-NN } & {\scriptsize{}ST-NN}\tabularnewline
{\scriptsize{}$\log_{10}(\rho(1473\,\mathrm{K}))$ } & {\scriptsize{}2074 } & {\scriptsize{}-8.05 } & {\scriptsize{}-3.24 } & {\scriptsize{}4.16 } & {\scriptsize{}1.10(4) } & {\scriptsize{}1.09(9) } & {\scriptsize{}0.5(1) } & {\scriptsize{}$=$ } & {\scriptsize{}ST-NN } & {\scriptsize{}ST-NN}\tabularnewline
{\scriptsize{}$\log_{10}(\rho(1673\,\mathrm{K}))$ } & {\scriptsize{}2086 } & {\scriptsize{}-7.74 } & {\scriptsize{}-3.44 } & {\scriptsize{}3.73 } & {\scriptsize{}0.76(7) } & {\scriptsize{}0.72(7) } & {\scriptsize{}0.47(9) } & {\scriptsize{}$=$ } & {\scriptsize{}ST-NN } & {\scriptsize{}ST-NN}\tabularnewline
{\scriptsize{}$E$ } & {\scriptsize{}14476 } & {\scriptsize{}1.62 } & {\scriptsize{}74.5 } & {\scriptsize{}175 } & {\scriptsize{}7.5(5) } & {\scriptsize{}6.9(5) } & {\scriptsize{}6.0(9) } & {\scriptsize{}MT-MH } & {\scriptsize{}ST-NN } & {\scriptsize{}ST-NN}\tabularnewline
{\scriptsize{}$G$ } & {\scriptsize{}6733 } & {\scriptsize{}1.37 } & {\scriptsize{}25.6 } & {\scriptsize{}72 } & {\scriptsize{}3.4(4) } & {\scriptsize{}3.2(4) } & {\scriptsize{}3.6(6) } & {\scriptsize{}$=$ } & {\scriptsize{}$=$ } & {\scriptsize{}$=$}\tabularnewline
{\scriptsize{}$H$ } & {\scriptsize{}13842 } & {\scriptsize{}0.0024 } & {\scriptsize{}4.2 } & {\scriptsize{}14.7 } & {\scriptsize{}0.85(6) } & {\scriptsize{}0.80(6) } & {\scriptsize{}0.53(5) } & {\scriptsize{}$=$ } & {\scriptsize{}ST-NN } & {\scriptsize{}ST-NN}\tabularnewline
{\scriptsize{}$\nu$ } & {\scriptsize{}6491 } & {\scriptsize{}0.052 } & {\scriptsize{}0.256 } & {\scriptsize{}0.821 } & {\scriptsize{}0.037(3) } & {\scriptsize{}0.037(3) } & {\scriptsize{}0.028(3) } & {\scriptsize{}$=$ } & {\scriptsize{}ST-NN } & {\scriptsize{}ST-NN}\tabularnewline
{\scriptsize{}$d(293\,\mathrm{K})$ } & {\scriptsize{}80509 } & {\scriptsize{}1.14 } & {\scriptsize{}3.49 } & {\scriptsize{}9.94 } & {\scriptsize{}0.322(9) } & {\scriptsize{}0.286(8) } & {\scriptsize{}0.17(1) } & {\scriptsize{}MT-MH } & {\scriptsize{}ST-NN } & {\scriptsize{}ST-NN}\tabularnewline
{\scriptsize{}$d(1073\,\mathrm{K})$ } & {\scriptsize{}808 } & {\scriptsize{}1.53 } & {\scriptsize{}3.62 } & {\scriptsize{}9.79 } & {\scriptsize{}0.3(1) } & {\scriptsize{}0.39(8) } & {\scriptsize{}0.40(6) } & {\scriptsize{}MT-MLP } & {\scriptsize{}MT-MLP } & {\scriptsize{}$=$}\tabularnewline
{\scriptsize{}$d(1273\,\mathrm{K})$ } & {\scriptsize{}995 } & {\scriptsize{}1.41 } & {\scriptsize{}3.2 } & {\scriptsize{}8.17 } & {\scriptsize{}0.22(3) } & {\scriptsize{}0.21(3) } & {\scriptsize{}0.23(3) } & {\scriptsize{}$=$ } & {\scriptsize{}$=$ } & {\scriptsize{}$=$}\tabularnewline
{\scriptsize{}$d(1473\,\mathrm{K})$ } & {\scriptsize{}843 } & {\scriptsize{}1.5 } & {\scriptsize{}3.14 } & {\scriptsize{}7.62 } & {\scriptsize{}0.12(3) } & {\scriptsize{}0.13(3) } & {\scriptsize{}0.22(4) } & {\scriptsize{}$=$ } & {\scriptsize{}MT-MLP } & {\scriptsize{}MT-MH}\tabularnewline
{\scriptsize{}$d(1673\,\mathrm{K})$ } & {\scriptsize{}840 } & {\scriptsize{}1.35 } & {\scriptsize{}2.9 } & {\scriptsize{}4.85 } & {\scriptsize{}0.13(3) } & {\scriptsize{}0.14(3) } & {\scriptsize{}0.13(3) } & {\scriptsize{}$=$ } & {\scriptsize{}$=$ } & {\scriptsize{}$=$}\tabularnewline
{\scriptsize{}$\kappa$ } & {\scriptsize{}1246 } & {\scriptsize{}0.002 } & {\scriptsize{}0.832 } & {\scriptsize{}5.94 } & {\scriptsize{}0.5(1) } & {\scriptsize{}0.5(1) } & {\scriptsize{}0.5(1) } & {\scriptsize{}$=$ } & {\scriptsize{}$=$ } & {\scriptsize{}$=$}\tabularnewline
{\scriptsize{}$\Delta T$ } & {\scriptsize{}1058 } & {\scriptsize{}25 } & {\scriptsize{}175 } & {\scriptsize{}1000 } & {\scriptsize{}50(10) } & {\scriptsize{}38(9) } & {\scriptsize{}50(20) } & {\scriptsize{}MT-MH } & {\scriptsize{}$=$ } & {\scriptsize{}MT-MH}\tabularnewline
{\scriptsize{}$\log_{10}(\alpha_{L}(T<T_{g}))$ } & {\scriptsize{}59351 } & {\scriptsize{}-8 } & {\scriptsize{}-5.1 } & {\scriptsize{}-3.75 } & {\scriptsize{}0.073(3) } & {\scriptsize{}0.073(3) } & {\scriptsize{}0.08(4) } & {\scriptsize{}$=$ } & {\scriptsize{}$=$ } & {\scriptsize{}$=$}\tabularnewline
{\scriptsize{}$\log_{10}(\alpha_{L}(328\,\mathrm{K}))$ } & {\scriptsize{}2220 } & {\scriptsize{}-6.48 } & {\scriptsize{}-5.07 } & {\scriptsize{}-4 } & {\scriptsize{}0.062(8) } & {\scriptsize{}0.06(1) } & {\scriptsize{}0.08(2) } & {\scriptsize{}$=$ } & {\scriptsize{}MT-MLP } & {\scriptsize{}MT-MH}\tabularnewline
{\scriptsize{}$\log_{10}(\alpha_{L}(373\,\mathrm{K}))$ } & {\scriptsize{}2300 } & {\scriptsize{}-6.41 } & {\scriptsize{}-4.99 } & {\scriptsize{}-3.95 } & {\scriptsize{}0.08(2) } & {\scriptsize{}0.08(2) } & {\scriptsize{}0.09(2) } & {\scriptsize{}$=$ } & {\scriptsize{}$=$ } & {\scriptsize{}$=$}\tabularnewline
{\scriptsize{}$\log_{10}(\alpha_{L}(433\,\mathrm{K}))$ } & {\scriptsize{}17600 } & {\scriptsize{}-7.7 } & {\scriptsize{}-5.14 } & {\scriptsize{}-3.89 } & {\scriptsize{}0.060(5) } & {\scriptsize{}0.063(5) } & {\scriptsize{}0.08(2) } & {\scriptsize{}$=$ } & {\scriptsize{}MT-MLP } & {\scriptsize{}MT-MH}\tabularnewline
{\scriptsize{}$\log_{10}(\alpha_{L}(483\,\mathrm{K}))$ } & {\scriptsize{}14618 } & {\scriptsize{}-6.81 } & {\scriptsize{}-5.13 } & {\scriptsize{}-3.85 } & {\scriptsize{}0.047(3) } & {\scriptsize{}0.043(3) } & {\scriptsize{}0.05(2) } & {\scriptsize{}MT-MH } & {\scriptsize{}$=$ } & {\scriptsize{}$=$}\tabularnewline
{\scriptsize{}$\log_{10}(\alpha_{L}(623\,\mathrm{K}))$ } & {\scriptsize{}1164 } & {\scriptsize{}-7.32 } & {\scriptsize{}-5.13 } & {\scriptsize{}-3.75 } & {\scriptsize{}0.09(2) } & {\scriptsize{}0.09(2) } & {\scriptsize{}0.14(3) } & {\scriptsize{}$=$ } & {\scriptsize{}MT-MLP } & {\scriptsize{}MT-MH}\tabularnewline
{\scriptsize{}$C_{p}(293\,\mathrm{K})$ } & {\scriptsize{}710 } & {\scriptsize{}0.392 } & {\scriptsize{}671 } & {\scriptsize{}1560 } & {\scriptsize{}140(30) } & {\scriptsize{}140(40) } & {\scriptsize{}120(30) } & {\scriptsize{}$=$ } & {\scriptsize{}$=$ } & {\scriptsize{}$=$}\tabularnewline
{\scriptsize{}$C_{p}(473\,\mathrm{K})$ } & {\scriptsize{}652 } & {\scriptsize{}0.495 } & {\scriptsize{}793 } & {\scriptsize{}1900 } & {\scriptsize{}150(50) } & {\scriptsize{}160(50) } & {\scriptsize{}100(300) } & {\scriptsize{}$=$ } & {\scriptsize{}$=$ } & {\scriptsize{}$=$}\tabularnewline
{\scriptsize{}$C_{p}(673\,\mathrm{K})$ } & {\scriptsize{}499 } & {\scriptsize{}2.63 } & {\scriptsize{}1000 } & {\scriptsize{}2460 } & {\scriptsize{}160(20) } & {\scriptsize{}170(40) } & {\scriptsize{}200(200) } & {\scriptsize{}$=$ } & {\scriptsize{}$=$ } & {\scriptsize{}$=$}\tabularnewline
{\scriptsize{}$C_{p}(1073\,\mathrm{K})$ } & {\scriptsize{}297 } & {\scriptsize{}524 } & {\scriptsize{}1350 } & {\scriptsize{}2220 } & {\scriptsize{}180(20) } & {\scriptsize{}180(20) } & {\scriptsize{}800(500) } & {\scriptsize{}$=$ } & {\scriptsize{}MT-MLP } & {\scriptsize{}MT-MH}\tabularnewline
{\scriptsize{}$C_{p}(1273\,\mathrm{K})$ } & {\scriptsize{}181 } & {\scriptsize{}748 } & {\scriptsize{}1430 } & {\scriptsize{}2790 } & {\scriptsize{}300(200) } & {\scriptsize{}300(200) } & {\scriptsize{}400(300) } & {\scriptsize{}$=$ } & {\scriptsize{}$=$ } & {\scriptsize{}$=$}\tabularnewline
{\scriptsize{}$C_{p}(1473\,\mathrm{K})$ } & {\scriptsize{}167 } & {\scriptsize{}765 } & {\scriptsize{}1450 } & {\scriptsize{}2620 } & {\scriptsize{}390(50) } & {\scriptsize{}400(50) } & {\scriptsize{}1100(300) } & {\scriptsize{}$=$ } & {\scriptsize{}MT-MLP } & {\scriptsize{}MT-MH}\tabularnewline
{\scriptsize{}$C_{p}(1673\,\mathrm{K})$ } & {\scriptsize{}116 } & {\scriptsize{}781 } & {\scriptsize{}1390 } & {\scriptsize{}2050 } & {\scriptsize{}70(50) } & {\scriptsize{}60(50) } & {\scriptsize{}900(200) } & {\scriptsize{}$=$ } & {\scriptsize{}MT-MLP } & {\scriptsize{}MT-MH}\tabularnewline
{\scriptsize{}$T_{\mathrm{max}(U)}$ } & {\scriptsize{}788 } & {\scriptsize{}592 } & {\scriptsize{}1210 } & {\scriptsize{}1820 } & {\scriptsize{}70(10) } & {\scriptsize{}70(10) } & {\scriptsize{}70(20) } & {\scriptsize{}$=$ } & {\scriptsize{}$=$ } & {\scriptsize{}$=$}\tabularnewline
{\scriptsize{}$\log_{10}(U_{\mathrm{max}})$ } & {\scriptsize{}775 } & {\scriptsize{}-9.85 } & {\scriptsize{}-7.34 } & {\scriptsize{}-2.53 } & {\scriptsize{}0.6(1) } & {\scriptsize{}0.6(1) } & {\scriptsize{}0.7(1) } & {\scriptsize{}$=$ } & {\scriptsize{}$=$ } & {\scriptsize{}$=$}\tabularnewline
{\scriptsize{}$T_{c}$ } & {\scriptsize{}20049 } & {\scriptsize{}263 } & {\scriptsize{}838 } & {\scriptsize{}1630 } & {\scriptsize{}74(3) } & {\scriptsize{}69(3) } & {\scriptsize{}52(2) } & {\scriptsize{}MT-MH } & {\scriptsize{}ST-NN } & {\scriptsize{}ST-NN}\tabularnewline
{\scriptsize{}$T_{x}$ } & {\scriptsize{}11688 } & {\scriptsize{}297 } & {\scriptsize{}789 } & {\scriptsize{}1810 } & {\scriptsize{}58(3) } & {\scriptsize{}54(3) } & {\scriptsize{}40(10) } & {\scriptsize{}MT-MH } & {\scriptsize{}ST-NN } & {\scriptsize{}ST-NN}\tabularnewline
{\scriptsize{}$\gamma(T>T_{g})$ } & {\scriptsize{}3584 } & {\scriptsize{}0.0439 } & {\scriptsize{}0.319 } & {\scriptsize{}0.79 } & {\scriptsize{}0.040(3) } & {\scriptsize{}0.038(4) } & {\scriptsize{}0.031(4) } & {\scriptsize{}$=$ } & {\scriptsize{}ST-NN } & {\scriptsize{}ST-NN}\tabularnewline
{\scriptsize{}$\gamma(1173\,\mathrm{K})$ } & {\scriptsize{}707 } & {\scriptsize{}0.0613 } & {\scriptsize{}0.182 } & {\scriptsize{}0.425 } & {\scriptsize{}0.014(4) } & {\scriptsize{}0.013(6) } & {\scriptsize{}0.021(5) } & {\scriptsize{}$=$ } & {\scriptsize{}MT-MLP } & {\scriptsize{}MT-MH}\tabularnewline
{\scriptsize{}$\gamma(1473\,\mathrm{K})$ } & {\scriptsize{}877 } & {\scriptsize{}0.0483 } & {\scriptsize{}0.226 } & {\scriptsize{}0.456 } & {\scriptsize{}0.030(2) } & {\scriptsize{}0.028(3) } & {\scriptsize{}0.020(3) } & {\scriptsize{}$=$ } & {\scriptsize{}ST-NN } & {\scriptsize{}ST-NN}\tabularnewline
{\scriptsize{}$\gamma(1573\,\mathrm{K})$ } & {\scriptsize{}1187 } & {\scriptsize{}0.0856 } & {\scriptsize{}0.277 } & {\scriptsize{}0.619 } & {\scriptsize{}0.029(2) } & {\scriptsize{}0.027(3) } & {\scriptsize{}0.027(4) } & {\scriptsize{}$=$ } & {\scriptsize{}$=$ } & {\scriptsize{}$=$}\tabularnewline
{\scriptsize{}$\gamma(1673\,\mathrm{K})$ } & {\scriptsize{}760 } & {\scriptsize{}0.0851 } & {\scriptsize{}0.364 } & {\scriptsize{}0.632 } & {\scriptsize{}0.043(5) } & {\scriptsize{}0.044(4) } & {\scriptsize{}0.031(4) } & {\scriptsize{}$=$ } & {\scriptsize{}ST-NN } & {\scriptsize{}ST-NN}\tabularnewline
\hline 
\end{longtable}{\scriptsize\par}

\end{landscape} \newpage{}

\begin{longtable}[c]{lrrrrrrr}
\caption{\label{tab:elements}Descriptive statistics of the chemical elements
present in the entire dataset (before the holdout split). All values
(except the second column) are rounded to the third decimal place.
SD is the standard deviation, and Q\protect\textsubscript{1} and
Q\protect\textsubscript{3} are the first and third quartiles, respectively.}
\tabularnewline
\hline 
Element  & Count  & Mean  & SD  & Q\textsubscript{1}  & Median  & Q\textsubscript{3}  & Maximum\tabularnewline
\endfirsthead
\hline 
\multicolumn{8}{l}{Continued from previous page}\tabularnewline
\hline 
Element  & Count  & Mean  & SD  & Q\textsubscript{1}  & Median  & Q\textsubscript{3}  & Maximum \tabularnewline
\hline 
\endhead
\hline 
\multicolumn{8}{r}{Continued on next page}\tabularnewline
\hline 
\endfoot
\endlastfoot
\hline 
Re  & 1  & 0.001  & --  & 0.001  & 0.001  & 0.001  & 0.001\tabularnewline
Pd  & 4  & 0.002  & 0.001  & 0.001  & 0.002  & 0.002  & 0.003\tabularnewline
Ru  & 15  & 0.003  & 0.002  & 0.002  & 0.003  & 0.004  & 0.007\tabularnewline
Au  & 19  & 0.015  & 0.028  & 0.001  & 0.003  & 0.006  & 0.1\tabularnewline
C  & 249  & 0.03  & 0.043  & 0.003  & 0.015  & 0.041  & 0.367\tabularnewline
Lu  & 257  & 0.054  & 0.062  & 0.005  & 0.029  & 0.089  & 0.4\tabularnewline
Sc  & 333  & 0.054  & 0.064  & 0.013  & 0.032  & 0.07  & 0.4\tabularnewline
Tm  & 488  & 0.009  & 0.017  & 0.001  & 0.003  & 0.007  & 0.167\tabularnewline
Hg  & 501  & 0.098  & 0.085  & 0.037  & 0.082  & 0.132  & 0.5\tabularnewline
Ho  & 511  & 0.042  & 0.078  & 0.003  & 0.006  & 0.044  & 0.4\tabularnewline
Tb  & 609  & 0.049  & 0.06  & 0.005  & 0.03  & 0.076  & 0.4\tabularnewline
Eu  & 830  & 0.024  & 0.05  & 0.003  & 0.005  & 0.016  & 0.4\tabularnewline
Hf  & 915  & 0.077  & 0.067  & 0.014  & 0.059  & 0.138  & 0.333\tabularnewline
Dy  & 924  & 0.053  & 0.08  & 0.003  & 0.008  & 0.08  & 0.4\tabularnewline
Sm  & 1136  & 0.032  & 0.058  & 0.003  & 0.007  & 0.032  & 0.4\tabularnewline
Pr  & 1228  & 0.033  & 0.063  & 0.003  & 0.006  & 0.029  & 0.4\tabularnewline
Ni  & 1264  & 0.024  & 0.063  & 0.002  & 0.004  & 0.018  & 0.5\tabularnewline
Co  & 1377  & 0.038  & 0.064  & 0.003  & 0.01  & 0.044  & 0.5\tabularnewline
N  & 1570  & 0.068  & 0.065  & 0.02  & 0.052  & 0.087  & 0.438\tabularnewline
H  & 1598  & 0.063  & 0.078  & 0.01  & 0.037  & 0.08  & 0.559\tabularnewline
Rb  & 1931  & 0.135  & 0.112  & 0.05  & 0.107  & 0.191  & 1\tabularnewline
Cr  & 2120  & 0.037  & 0.059  & 0.002  & 0.009  & 0.039  & 0.4\tabularnewline
Br  & 2267  & 0.228  & 0.245  & 0.02  & 0.1  & 0.5  & 0.8\tabularnewline
Be  & 2347  & 0.123  & 0.097  & 0.045  & 0.091  & 0.19  & 0.5\tabularnewline
Yb  & 2392  & 0.022  & 0.039  & 0.004  & 0.008  & 0.025  & 0.5\tabularnewline
In  & 2655  & 0.075  & 0.089  & 0.02  & 0.052  & 0.1  & 1\tabularnewline
Er  & 2780  & 0.022  & 0.052  & 0.002  & 0.004  & 0.011  & 0.4\tabularnewline
Cs  & 3302  & 0.106  & 0.108  & 0.025  & 0.073  & 0.155  & 1\tabularnewline
Tl  & 3304  & 0.168  & 0.16  & 0.05  & 0.12  & 0.235  & 1\tabularnewline
Nd  & 3389  & 0.028  & 0.053  & 0.003  & 0.006  & 0.026  & 0.4\tabularnewline
Ce  & 3641  & 0.019  & 0.042  & 0.001  & 0.004  & 0.02  & 0.5\tabularnewline
Mo  & 4068  & 0.084  & 0.07  & 0.015  & 0.069  & 0.149  & 0.25\tabularnewline
I  & 4396  & 0.148  & 0.139  & 0.05  & 0.105  & 0.2  & 1\tabularnewline
Sn  & 4809  & 0.067  & 0.092  & 0.002  & 0.021  & 0.116  & 1\tabularnewline
Cd  & 5012  & 0.07  & 0.077  & 0.017  & 0.044  & 0.096  & 0.993\tabularnewline
Mn  & 5294  & 0.049  & 0.065  & 0.004  & 0.021  & 0.068  & 0.5\tabularnewline
Cu  & 5596  & 0.072  & 0.108  & 0.006  & 0.023  & 0.095  & 0.928\tabularnewline
Ta  & 5903  & 0.022  & 0.03  & 0.007  & 0.015  & 0.025  & 0.286\tabularnewline
Ga  & 6106  & 0.104  & 0.098  & 0.04  & 0.08  & 0.133  & 1\tabularnewline
Ag  & 6126  & 0.182  & 0.144  & 0.062  & 0.162  & 0.263  & 1\tabularnewline
Cl  & 6203  & 0.119  & 0.172  & 0.006  & 0.044  & 0.15  & 0.745\tabularnewline
Gd  & 6306  & 0.035  & 0.043  & 0.012  & 0.023  & 0.042  & 0.4\tabularnewline
Y  & 8075  & 0.032  & 0.041  & 0.008  & 0.018  & 0.04  & 0.4\tabularnewline
W  & 8175  & 0.034  & 0.045  & 0.006  & 0.014  & 0.042  & 0.25\tabularnewline
V  & 8710  & 0.13  & 0.079  & 0.059  & 0.138  & 0.193  & 0.287\tabularnewline
Sb  & 8998  & 0.079  & 0.118  & 0.001  & 0.011  & 0.12  & 1\tabularnewline
Se  & 9705  & 0.522  & 0.22  & 0.4  & 0.565  & 0.65  & 1\tabularnewline
S  & 10870  & 0.328  & 0.269  & 0.014  & 0.384  & 0.581  & 1\tabularnewline
As  & 11904  & 0.176  & 0.169  & 0.002  & 0.15  & 0.318  & 1\tabularnewline
Bi  & 12245  & 0.117  & 0.097  & 0.033  & 0.095  & 0.182  & 1\tabularnewline
Nb  & 13397  & 0.051  & 0.054  & 0.012  & 0.031  & 0.076  & 0.286\tabularnewline
Fe  & 14397  & 0.05  & 0.074  & 0.003  & 0.02  & 0.066  & 0.5\tabularnewline
Te  & 15351  & 0.24  & 0.199  & 0.099  & 0.204  & 0.286  & 1\tabularnewline
Ge  & 16506  & 0.176  & 0.112  & 0.09  & 0.175  & 0.246  & 1\tabularnewline
La  & 19465  & 0.049  & 0.045  & 0.012  & 0.041  & 0.074  & 0.4\tabularnewline
Sr  & 20495  & 0.029  & 0.037  & 0.006  & 0.016  & 0.036  & 0.5\tabularnewline
Zr  & 24871  & 0.026  & 0.041  & 0.005  & 0.011  & 0.02  & 0.333\tabularnewline
Pb  & 25993  & 0.096  & 0.087  & 0.032  & 0.074  & 0.139  & 1\tabularnewline
F  & 26165  & 0.304  & 0.279  & 0.043  & 0.197  & 0.61  & 0.833\tabularnewline
Ti  & 27353  & 0.038  & 0.052  & 0.006  & 0.017  & 0.049  & 0.357\tabularnewline
Zn  & 37706  & 0.048  & 0.047  & 0.014  & 0.033  & 0.068  & 0.5\tabularnewline
P  & 39374  & 0.138  & 0.08  & 0.075  & 0.15  & 0.2  & 1\tabularnewline
Li  & 39671  & 0.082  & 0.081  & 0.025  & 0.055  & 0.111  & 0.667\tabularnewline
Mg  & 43682  & 0.034  & 0.036  & 0.012  & 0.021  & 0.044  & 0.5\tabularnewline
Ba  & 44149  & 0.043  & 0.044  & 0.011  & 0.028  & 0.063  & 1\tabularnewline
K  & 47969  & 0.055  & 0.068  & 0.011  & 0.032  & 0.068  & 0.667\tabularnewline
Ca  & 62503  & 0.055  & 0.057  & 0.017  & 0.034  & 0.078  & 0.5\tabularnewline
Na  & 77163  & 0.082  & 0.068  & 0.033  & 0.068  & 0.109  & 0.667\tabularnewline
B  & 80769  & 0.138  & 0.102  & 0.05  & 0.118  & 0.216  & 0.401\tabularnewline
Al  & 85456  & 0.053  & 0.046  & 0.018  & 0.045  & 0.075  & 0.527\tabularnewline
Si  & 118216  & 0.164  & 0.076  & 0.116  & 0.186  & 0.221  & 1\tabularnewline
O  & 192270  & 0.586  & 0.085  & 0.573  & 0.6  & 0.623  & 0.75\tabularnewline
\hline 
\end{longtable}\newpage{}

\begin{landscape}

\begin{longtable}[c]{llll}
\caption{\label{tab:selected_features}Selected physicochemical features.}
\tabularnewline
\hline 
Feature  & Origin  & Aggregator  & Symbol\tabularnewline
\endfirsthead
\hline 
\multicolumn{4}{l}{Continued from previous page}\tabularnewline
\hline 
Feature  & Origin  & Aggregator  & Symbol \tabularnewline
\hline 
\endhead
\hline 
\multicolumn{4}{r}{Continued on next page}\tabularnewline
\hline 
\endfoot
\endlastfoot
\hline 
Atomic fraction of Ag  & Glass composition  & --  & Ag\tabularnewline
Atomic fraction of Al  & Glass composition  & --  & Al\tabularnewline
Atomic fraction of As  & Glass composition  & --  & As\tabularnewline
Atomic fraction of B  & Glass composition  & --  & B\tabularnewline
Atomic fraction of Ba  & Glass composition  & --  & Ba\tabularnewline
Atomic fraction of Be  & Glass composition  & --  & Be\tabularnewline
Atomic fraction of Bi  & Glass composition  & --  & Bi\tabularnewline
Atomic fraction of Br  & Glass composition  & --  & Br\tabularnewline
Atomic fraction of C  & Glass composition  & --  & C\tabularnewline
Atomic fraction of Ca  & Glass composition  & --  & Ca\tabularnewline
Atomic fraction of Cd  & Glass composition  & --  & Cd\tabularnewline
Atomic fraction of Ce  & Glass composition  & --  & Ce\tabularnewline
Atomic fraction of Cl  & Glass composition  & --  & Cl\tabularnewline
Atomic fraction of Co  & Glass composition  & --  & Co\tabularnewline
Atomic fraction of Cr  & Glass composition  & --  & Cr\tabularnewline
Atomic fraction of Cs  & Glass composition  & --  & Cs\tabularnewline
Atomic fraction of Cu  & Glass composition  & --  & Cu\tabularnewline
Atomic fraction of Dy  & Glass composition  & --  & Dy\tabularnewline
Atomic fraction of Er  & Glass composition  & --  & Er\tabularnewline
Atomic fraction of Eu  & Glass composition  & --  & Eu\tabularnewline
Atomic fraction of Fe  & Glass composition  & --  & Fe\tabularnewline
Atomic fraction of Ga  & Glass composition  & --  & Ga\tabularnewline
Atomic fraction of Gd  & Glass composition  & --  & Gd\tabularnewline
Atomic fraction of Ge  & Glass composition  & --  & Ge\tabularnewline
Atomic fraction of H  & Glass composition  & --  & H\tabularnewline
Atomic fraction of Hf  & Glass composition  & --  & Hf\tabularnewline
Atomic fraction of Hg  & Glass composition  & --  & Hg\tabularnewline
Atomic fraction of Ho  & Glass composition  & --  & Ho\tabularnewline
Atomic fraction of I  & Glass composition  & --  & I\tabularnewline
Atomic fraction of In  & Glass composition  & --  & In\tabularnewline
Atomic fraction of K  & Glass composition  & --  & K\tabularnewline
Atomic fraction of La  & Glass composition  & --  & La\tabularnewline
Atomic fraction of Li  & Glass composition  & --  & Li\tabularnewline
Atomic fraction of Lu  & Glass composition  & --  & Lu\tabularnewline
Atomic fraction of Mg  & Glass composition  & --  & Mg\tabularnewline
Atomic fraction of Mn  & Glass composition  & --  & Mn\tabularnewline
Atomic fraction of Mo  & Glass composition  & --  & Mo\tabularnewline
Atomic fraction of N  & Glass composition  & --  & N\tabularnewline
Atomic fraction of Na  & Glass composition  & --  & Na\tabularnewline
Atomic fraction of Nb  & Glass composition  & --  & Nb\tabularnewline
Atomic fraction of Nd  & Glass composition  & --  & Nd\tabularnewline
Atomic fraction of Ni  & Glass composition  & --  & Ni\tabularnewline
Atomic fraction of P  & Glass composition  & --  & P\tabularnewline
Atomic fraction of Pb  & Glass composition  & --  & Pb\tabularnewline
Atomic fraction of Pr  & Glass composition  & --  & Pr\tabularnewline
Atomic fraction of Rb  & Glass composition  & --  & Rb\tabularnewline
Atomic fraction of S  & Glass composition  & --  & S\tabularnewline
Atomic fraction of Sb  & Glass composition  & --  & Sb\tabularnewline
Atomic fraction of Sc  & Glass composition  & --  & Sc\tabularnewline
Atomic fraction of Se  & Glass composition  & --  & Se\tabularnewline
Atomic fraction of Sm  & Glass composition  & --  & Sm\tabularnewline
Atomic fraction of Sn  & Glass composition  & --  & Sn\tabularnewline
Atomic fraction of Sr  & Glass composition  & --  & Sr\tabularnewline
Atomic fraction of Ta  & Glass composition  & --  & Ta\tabularnewline
Atomic fraction of Tb  & Glass composition  & --  & Tb\tabularnewline
Atomic fraction of Te  & Glass composition  & --  & Te\tabularnewline
Atomic fraction of Ti  & Glass composition  & --  & Ti\tabularnewline
Atomic fraction of Tl  & Glass composition  & --  & Tl\tabularnewline
Atomic fraction of V  & Glass composition  & --  & V\tabularnewline
Atomic fraction of W  & Glass composition  & --  & W\tabularnewline
Atomic fraction of Y  & Glass composition  & --  & Y\tabularnewline
Atomic fraction of Yb  & Glass composition  & --  & Yb\tabularnewline
Atomic fraction of Zn  & Glass composition  & --  & Zn\tabularnewline
Atomic fraction of Zr  & Glass composition  & --  & Zr\tabularnewline
\hline 
\hline 
$C_{6}$  & Eq. (\ref{eq:feat_w})  & Minimum  & $\min(C_{6})$\tabularnewline
Electron affinity  & Eq. (\ref{eq:feat_w})  & Minimum  & $\min(E_{ea})$\tabularnewline
Number of unfilled valence orbitals  & Eq. (\ref{eq:feat_w})  & Standard dev.  & $\mathrm{std}(N_{u})$\tabularnewline
Number of filled d valence orbitals  & Eq. (\ref{eq:feat_w})  & Minimum  & $\min(N_{f,d})$\tabularnewline
Number of filled f valence orbitals  & Eq. (\ref{eq:feat_w})  & Minimum  & $\min(N_{f,f})$\tabularnewline
Number of unfilled p valence orbitals  & Eq. (\ref{eq:feat_w})  & Minimum  & $\min(N_{u,p})$\tabularnewline
Number of unfilled s valence orbitals  & Eq. (\ref{eq:feat_w})  & Minimum  & $\min(N_{u,s})$\tabularnewline
Number of valence electrons  & Eq. (\ref{eq:feat_w})  & Maximum  & $\max(N_{v})$\tabularnewline
Atomic volume  & Eq. (\ref{eq:feat_w})  & Minimum  & $\min(V_{at})$\tabularnewline
DFT bandgap energy of $T=0\,\mathrm{K}$ ground state  & Eq. (\ref{eq:feat_w})  & Minimum  & $\min(E_{g})$\tabularnewline
DFT energy per atom of $T=0\,\mathrm{K}$ ground state  & Eq. (\ref{eq:feat_w})  & Maximum  & $\max(E_{at})$\tabularnewline
Fusion Enthalpy  & Eq. (\ref{eq:feat_w})  & Minimum  & $\min(\Delta H_{m})$\tabularnewline
\hline 
\hline 
Number of oxidation states  & Eq. (\ref{eq:feat_a})  & Maximum  & $\max(\lceil N_{ox}\rceil)$\tabularnewline
Number of unfilled valence orbitals  & Eq. (\ref{eq:feat_a})  & Minimum  & $\min(\lceil N_{u}\rceil)$\tabularnewline
Number of unfilled d valence orbitals  & Eq. (\ref{eq:feat_a})  & Maximum  & $\max(\lceil N_{u,d}\rceil)$\tabularnewline
Number of filled d valence orbitals  & Eq. (\ref{eq:feat_a})  & Maximum  & $\max(\lceil N_{f,d}\rceil)$\tabularnewline
Number of unfilled f valence orbitals  & Eq. (\ref{eq:feat_a})  & Sum  & $\mathrm{sum}(\lceil N_{u,f}\rceil)$\tabularnewline
Number of filled f valence orbitals  & Eq. (\ref{eq:feat_a})  & Sum  & $\mathrm{sum}(\lceil N_{f}\rceil)$\tabularnewline
Number of unfilled p valence orbitals  & Eq. (\ref{eq:feat_a})  & Sum  & $\mathrm{sum}(\lceil N_{u,p}\rceil)$\tabularnewline
Number of unfilled p valence orbitals  & Eq. (\ref{eq:feat_a})  & Maximum  & $\max(\lceil N_{u,p}\rceil)$\tabularnewline
Number of filled p valence orbitals  & Eq. (\ref{eq:feat_a})  & Maximum  & $\max(\lceil N_{f,p}\rceil)$\tabularnewline
Number of unfilled s valence orbitals  & Eq. (\ref{eq:feat_a})  & Sum  & $\mathrm{sum}(\lceil N_{u,s}\rceil)$\tabularnewline
Number of filled s valence orbitals  & Eq. (\ref{eq:feat_a})  & Minimum  & $\min(\lceil N_{f,s}\rceil)$\tabularnewline
Number of valence electrons  & Eq. (\ref{eq:feat_a})  & Maximum  & $\max(\lceil N_{v}\rceil)$\tabularnewline
Van der Walls radius  & Eq. (\ref{eq:feat_a})  & Maximum  & $\max(\lceil r_{W}\rceil)$\tabularnewline
Atomic radius (Rahm)  & Eq. (\ref{eq:feat_a})  & Maximum  & $\max(\lceil r_{R}\rceil)$\tabularnewline
DFT bandgap energy of $T=0\,\mathrm{K}$ ground state  & Eq. (\ref{eq:feat_a})  & Minimum  & $\min(\lceil E_{g}\rceil)$\tabularnewline
FCC lattice parameter  & Eq. (\ref{eq:feat_a})  & Standard dev.  & $\mathrm{std}(\lceil\mathrm{FCC}_{lp}\rceil)$\tabularnewline
DFT magnetic moment of $T=0\,\mathrm{K}$ ground state  & Eq. (\ref{eq:feat_a})  & Sum  & $\mathrm{sum}(\lceil m_{m}\rceil)$\tabularnewline
Electronegativity in the Sanderson scale  & Eq. (\ref{eq:feat_a})  & Maximum  & $\max(\lceil\chi_{S}\rceil)$\tabularnewline
Electronegativity in the Tardini--Organov scale  & Eq. (\ref{eq:feat_a})  & Minimum  & $\min(\lceil\chi_{TO}\rceil)$\tabularnewline
Effective nuclear charge  & Eq. (\ref{eq:feat_a})  & Standard dev.  & $\mathrm{std}(\lceil Z_{\mathrm{eff}}\rceil)$\tabularnewline
Boiling point  & Eq. (\ref{eq:feat_a})  & Standard dev.  & $\mathrm{std}(\lceil T_{b}\rceil)$\tabularnewline
Fusion Enthalpy  & Eq. (\ref{eq:feat_a})  & Maximum  & $\max(\lceil\Delta H_{m}\rceil)$\tabularnewline
\hline 
\end{longtable}

\end{landscape} 

\newpage{}

\begin{landscape}

\begin{table}[htbp]
\caption{\label{tab:shap_analysis}The most frequent SHAP features (overall
and for property groups), taking into account the top 10 features
for each studied property.}
\centering %
\begin{tabular}{lllllllll}
\hline 
Overall  & Viscosity and relaxation  & Optical  & Electrical and dielectric  & Mechanical  & Density  & Thermal  & Crystallization  & Surface tension\tabularnewline
\hline 
\hline 
Na  & B  & Pb  & Na  & $N_{u,s}$  & $N_{f,d}$  & Na  & $N_{f,p}$  & Ca\tabularnewline
B  & Na  & $N_{u}$  & $N_{f,s}$  & Na  & $N_{u}$  & $\mathrm{FCC}_{lp}$  & $N_{u,p}$  & B\tabularnewline
Pb  & Al  & Bi  & Li  & $\Delta H_{m}$  & $r_{R}$  & B  & $\Delta H_{m}$  & $r_{R}$\tabularnewline
Li  & K  & $N_{f,d}$  & B  & $\mathrm{FCC}_{lp}$  & B  & Ba  & B  & $\mathrm{FCC}_{lp}$\tabularnewline
Al  & Li  & $N_{f,p}$  & $\Delta H_{m}$  & P  & Pb  & Ca  & Na  & Na\tabularnewline
K  & Pb  & Ti  & Se  & Mg  & Na  & Pb  & Se  & Fe\tabularnewline
$\Delta H_{m}$  & $N_{f,s}$  & Nb  & V  & Ca  & $N_{f,f}$  & K  & S  & Ba\tabularnewline
Ca  & Ca  & $N_{f,f}$  & Fe  & $r_{R}$  & Ge  & $N_{f,d}$  & Te  & $N_{f,d}$\tabularnewline
$N_{u}$  & $\Delta H_{m}$  & La  & K  & $N_{f,d}$  & Bi  & P  & $T_{b}$  & $N_{u}$\tabularnewline
$N_{f,d}$  & P  & Ge  & Ag  & Al  & Li  & Li  & V  & Al\tabularnewline
\hline 
\end{tabular}
\end{table}

\end{landscape}

\appendix
\setcounter{figure}{0} 
\global\long\def\thetable{S.\arabic{table}}%
 
\global\long\def\thefigure{S.\arabic{figure}}%

\section*{Supplementary material}

\label{sec:org9c4018d} 

\subsection*{Data processing}

\label{sec:orge53ba82}

Table \ref{tab:data_minmax} shows the minimum and maximum values
accepted for some of the properties studied in this work, as described
in the Materials and Methods section of the manuscript. These values
were chosen after looking at the histogram of the properties to avoid
values that are unreasonably extreme (probably due to measurement
or typing errors).

\begin{longtable}[c]{lrr}
\caption{\label{tab:data_minmax}Minimum and maximum accepted values for some
of the properties studied in this work. Properties not included in
this list were not subjected to this restriction.}
\tabularnewline
\hline 
Property  & Minimum  & Maximum\tabularnewline
\endfirsthead
\hline 
\multicolumn{3}{l}{Continued from previous page}\tabularnewline
\hline 
Property  & Minimum  & Maximum \tabularnewline
\hline 
\endhead
\hline 
\multicolumn{3}{r}{Continued on next page}\tabularnewline
\hline 
\endfoot
\endlastfoot
\hline 
$T_{3}$  & --  & 2350\tabularnewline
$T_{4}$  & --  & 2000\tabularnewline
$\log_{10}(\eta(1773\,\mathrm{K}))$  & --  & 10\tabularnewline
$\log_{10}(\eta(1873\,\mathrm{K}))$  & --  & 10\tabularnewline
$\log_{10}(\eta(2073\,\mathrm{K}))$  & --  & 8\tabularnewline
$\log_{10}(\eta(2273\,\mathrm{K}))$  & --  & 8\tabularnewline
$T_{\mathrm{soft}}$  & --  & 1600\tabularnewline
$V_{D}$  & --  & 115\tabularnewline
$n_{D}$  & --  & 4\tabularnewline
$n$ (high)  & 1.7  & 3.5\tabularnewline
$\varepsilon$  & --  & 50\tabularnewline
$\log_{10}(\tan(\delta))$  & -4  & -0.796\tabularnewline
$T_{\rho=10^{6}\,\Omega.m}$  & --  & 2000\tabularnewline
$\log_{10}(\rho(273\,\mathrm{K}))$  & --  & 40\tabularnewline
$\log_{10}(\rho(373\,\mathrm{K}))$  & --  & 28\tabularnewline
$\log_{10}(\rho(1073\,\mathrm{K}))$  & --  & 4\tabularnewline
$\log_{10}(\rho(1273\,\mathrm{K}))$  & --  & 5\tabularnewline
$E$  & --  & 175\tabularnewline
$H$  & --  & 15\tabularnewline
$\nu$  & --  & 1\tabularnewline
$d(293\,\mathrm{K})$  & 1  & 10\tabularnewline
$\kappa$  & --  & 6\tabularnewline
$\log_{10}(\alpha_{L}(328\,\mathrm{K}))$  & -6.5  & --\tabularnewline
$\log_{10}(\alpha_{L}(373\,\mathrm{K}))$  & -6.5  & --\tabularnewline
$\log_{10}(\alpha_{L}(433\,\mathrm{K}))$  & -8  & --\tabularnewline
$\log_{10}(\alpha_{L}(483\,\mathrm{K}))$  & -7  & --\tabularnewline
$C_{p}(293\,\mathrm{K})$  & --  & 2000\tabularnewline
$C_{p}(473\,\mathrm{K})$  & --  & 2000\tabularnewline
$C_{p}(673\,\mathrm{K})$  & --  & 3000\tabularnewline
$C_{p}(1073\,\mathrm{K})$  & 500  & 2500\tabularnewline
$C_{p}(1273\,\mathrm{K})$  & 500  & 3000\tabularnewline
$C_{p}(1473\,\mathrm{K})$  & 500  & 3000\tabularnewline
$C_{p}(1673\,\mathrm{K})$  & 500  & 2250\tabularnewline
$\log_{10}(U_{\mathrm{max}})$  & -10  & --\tabularnewline
$\gamma(T>T_{g})$  & --  & 0.8\tabularnewline
$\gamma(1473\,\mathrm{K})$  & --  & 0.5\tabularnewline
$\gamma(1573\,\mathrm{K})$  & --  & 0.7\tabularnewline
$\gamma(1673\,\mathrm{K})$  & --  & 0.7\tabularnewline
\hline 
\end{longtable}

\subsection*{Physicochemical features considered, but not selected}

\label{sec:org8e211e6}

The physicochemical features considered in this work that were not
selected are:
\begin{itemize}
\item Atomic number \citeprocitem{30}{{[}30{]}} 
\item Atomic radius \citeprocitem{65}{{[}65{]}} 
\item Atomic weight \citeprocitem{30}{{[}30{]}} 
\item Estimated BCC lattice parameter based on the DFT volume of the OQMD
ground state for each element \citeprocitem{26}{{[}26{]}} 
\item Covalent radius \citeprocitem{66}{{[}66{]}} 
\item Single bond covalent radius \citeprocitem{67}{{[}67{]}} 
\item Density at 295 K \citeprocitem{58}{{[}58{]}} 
\item Dipole polarizability \citeprocitem{68}{{[}68{]}} 
\item Electronegativity in the Allred and Rochow scale \citeprocitem{69}{{[}69{]}} 
\item Electronegativity in the Cottrell and Sutton scale \citeprocitem{70}{{[}70{]}} 
\item Electronegativity in the Gordy scale \citeprocitem{71}{{[}71{]}} 
\item Electronegativity in the Gosh scale \citeprocitem{72}{{[}72{]}} 
\item Electronegativity in the Martynov and Batsanov scale \citeprocitem{73}{{[}73{]}} 
\item Electronegativity in the Nagle scale \citeprocitem{74}{{[}74{]}} 
\item Energy to remove the first electron \citeprocitem{26}{{[}26{]}} 
\item Glawe's number \citeprocitem{75}{{[}75{]}} 
\item Heat of formation \citeprocitem{58}{{[}58{]}} 
\item Mass number of the most abundant isotope \citeprocitem{30}{{[}30{]}} 
\item Maximum ionization energy \citeprocitem{30}{{[}30{]}} 
\item Melting point \citeprocitem{58}{{[}58{]}} 
\item Mendeleev's number \citeprocitem{76}{{[}76{]}}, \citeprocitem{77}{{[}77{]}} 
\item Number of electrons \citeprocitem{30}{{[}30{]}} 
\item Number of neutrons \citeprocitem{30}{{[}30{]}} 
\item Number of protons \citeprocitem{30}{{[}30{]}} 
\item Number of valence electrons \citeprocitem{26}{{[}26{]}} 
\item Pettifor's number \citeprocitem{76}{{[}76{]}} 
\item Van der Walls radius \citeprocitem{58}{{[}58{]}} 
\item Van der Walls radius \citeprocitem{78}{{[}78{]}} 
\item Van der Walls radius \citeprocitem{79}{{[}79{]}} 
\item DFT volume per atom of $T=0\,\mathrm{K}$ ground state \citeprocitem{26}{{[}26{]}} 
\end{itemize}
\FloatBarrier

\subsection*{Testing H\protect\textsubscript{1} with different algorithms}

\label{sec:org1094111}

In the main document, H\textsubscript{1} was tested using single-tast
NNs with the same architecture as the MT-MH model. Two other tests
were done comparing the multitask NNs with other single-task algorithms.

One of the algorithms used to induce this comparative model was a
random forest trained using the \texttt{scikit-learn} Python module
with the default set of hyperparameters. The models induced by this
algorithm had great performance in previous publications \citeprocitem{4}{{[}4{]}},
\citeprocitem{8}{{[}8{]}}. See the results in Table \ref{tab:descript_metrics_rf}.

Hypothesis H\textsubscript{1} was also tested using single-task eXtreme
Gradient Boosting (XGBoost) models with the default hyperparameters
of the Python module \texttt{xgboost} \citeprocitem{80}{{[}80{]}}.
The results of the test with XGBoost as the baseline showed that the
NN models were better or statistically similar at predicting the targets
compared to the models induced by the XGBoost algorithm.

\begin{longtable}[c]{lrll}
\caption{\label{tab:descript_metrics_rf}Metrics (RMSE) for the random forest
models. The last two columns show the results of the t-test (95\%
confidence) used to compare the performance of the models. In these
two columns, $=$ means that no statistical difference was observed
and the name of the model with the better performance is shown if
a statistical difference was observed. RF means random forest.}
\tabularnewline
\hline 
Symbol  & RF RMSE  & MT-MLP vs. RF  & MT-MH vs. RF\tabularnewline
\endfirsthead
\hline 
\multicolumn{4}{l}{Continued from previous page}\tabularnewline
\hline 
Symbol  & RF RMSE  & MT-MLP vs. RF  & MT-MH vs. RF \tabularnewline
\hline 
\endhead
\hline 
\multicolumn{4}{r}{Continued on next page}\tabularnewline
\hline 
\endfoot
\endlastfoot
\hline 
$T_{0}$  & 120(20)  & MT-MLP  & MT-MH\tabularnewline
$T_{1}$  & 68(6)  & $=$  & $=$\tabularnewline
$T_{2}$  & 55(4)  & MT-MLP  & MT-MH\tabularnewline
$T_{3}$  & 56(4)  & MT-MLP  & MT-MH\tabularnewline
$T_{4}$  & 58(7)  & MT-MLP  & MT-MH\tabularnewline
$T_{5}$  & 50(10)  & MT-MLP  & MT-MH\tabularnewline
$T_{6}$  & 50(10)  & MT-MLP  & MT-MH\tabularnewline
$T_{7}$  & 45(6)  & MT-MLP  & MT-MH\tabularnewline
$T_{8}$  & 50(6)  & MT-MLP  & MT-MH\tabularnewline
$T_{9}$  & 57(4)  & MT-MLP  & MT-MH\tabularnewline
$T_{10}$  & 58(4)  & MT-MLP  & MT-MH\tabularnewline
$T_{11}$  & 58(5)  & MT-MLP  & MT-MH\tabularnewline
$T_{12}$  & 52(3)  & MT-MLP  & MT-MH\tabularnewline
$\log_{10}(\eta(773\,\mathrm{K}))$  & 1.3(2)  & MT-MLP  & MT-MH\tabularnewline
$\log_{10}(\eta(873\,\mathrm{K}))$  & 1.0(1)  & MT-MLP  & MT-MH\tabularnewline
$\log_{10}(\eta(973\,\mathrm{K}))$  & 0.85(7)  & MT-MLP  & MT-MH\tabularnewline
$\log_{10}(\eta(1073\,\mathrm{K}))$  & 0.70(9)  & MT-MLP  & MT-MH\tabularnewline
$\log_{10}(\eta(1173\,\mathrm{K}))$  & 0.51(7)  & $=$  & $=$\tabularnewline
$\log_{10}(\eta(1273\,\mathrm{K}))$  & 0.47(8)  & $=$  & $=$\tabularnewline
$\log_{10}(\eta(1373\,\mathrm{K}))$  & 0.39(6)  & MT-MLP  & MT-MH\tabularnewline
$\log_{10}(\eta(1473\,\mathrm{K}))$  & 0.36(4)  & MT-MLP  & MT-MH\tabularnewline
$\log_{10}(\eta(1573\,\mathrm{K}))$  & 0.27(4)  & MT-MLP  & MT-MH\tabularnewline
$\log_{10}(\eta(1673\,\mathrm{K}))$  & 0.25(6)  & $=$  & $=$\tabularnewline
$\log_{10}(\eta(1773\,\mathrm{K}))$  & 0.21(6)  & $=$  & $=$\tabularnewline
$\log_{10}(\eta(1873\,\mathrm{K}))$  & 0.22(4)  & $=$  & $=$\tabularnewline
$\log_{10}(\eta(2073\,\mathrm{K}))$  & 0.3(3)  & $=$  & $=$\tabularnewline
$\log_{10}(\eta(2273\,\mathrm{K}))$  & 0.4(6)  & $=$  & $=$\tabularnewline
$\log_{10}(\eta(2473\,\mathrm{K}))$  & 0.5(8)  & $=$  & $=$\tabularnewline
$T_{g}$  & 34(1)  & RF  & RF\tabularnewline
$T_{\mathrm{melt}}$  & 73(8)  & RF  & RF\tabularnewline
$T_{\mathrm{liq}}$  & 53(3)  & RF  & RF\tabularnewline
$T_{\mathrm{Lit}}$  & 34(7)  & $=$  & $=$\tabularnewline
$T_{\mathrm{ann}}$  & 21(4)  & $=$  & $=$\tabularnewline
$T_{\mathrm{strain}}$  & 20(5)  & RF  & $=$\tabularnewline
$T_{\mathrm{soft}}$  & 40(4)  & RF  & $=$\tabularnewline
$T_{\mathrm{dil}}$  & 39(1)  & RF  & RF\tabularnewline
$V_{D}$  & 2.3(2)  & RF  & RF\tabularnewline
$n_{D}$  & 0.039(4)  & RF  & RF\tabularnewline
$n$ (low)  & 0.11(4)  & $=$  & $=$\tabularnewline
$n$ (high)  & 0.10(2)  & $=$  & MT-MH\tabularnewline
$\log_{10}(n_{F}-n_{C})$  & 0.035(5)  & RF  & RF\tabularnewline
$\varepsilon$  & 2.4(4)  & $=$  & $=$\tabularnewline
$\log_{10}(\tan(\delta))$  & 0.18(3)  & $=$  & $=$\tabularnewline
$T_{\rho=10^{6}\,\Omega.m}$  & 60(5)  & RF  & $=$\tabularnewline
$\log_{10}(\rho(273\,\mathrm{K}))$  & 1.9(2)  & RF  & RF\tabularnewline
$\log_{10}(\rho(373\,\mathrm{K}))$  & 1.7(2)  & RF  & RF\tabularnewline
$\log_{10}(\rho(423\,\mathrm{K}))$  & 1.8(2)  & RF  & RF\tabularnewline
$\log_{10}(\rho(573\,\mathrm{K}))$  & 1.5(2)  & RF  & RF\tabularnewline
$\log_{10}(\rho(1073\,\mathrm{K}))$  & 0.42(9)  & MT-MLP  & MT-MH\tabularnewline
$\log_{10}(\rho(1273\,\mathrm{K}))$  & 0.7(1)  & $=$  & $=$\tabularnewline
$\log_{10}(\rho(1473\,\mathrm{K}))$  & 1.15(7)  & $=$  & $=$\tabularnewline
$\log_{10}(\rho(1673\,\mathrm{K}))$  & 0.77(8)  & $=$  & $=$\tabularnewline
$E$  & 6.6(5)  & RF  & $=$\tabularnewline
$G$  & 3.2(4)  & $=$  & $=$\tabularnewline
$H$  & 0.65(5)  & RF  & RF\tabularnewline
$\nu$  & 0.035(4)  & $=$  & $=$\tabularnewline
$d(293\,\mathrm{K})$  & 0.22(1)  & RF  & RF\tabularnewline
$d(1073\,\mathrm{K})$  & 0.30(9)  & $=$  & RF\tabularnewline
$d(1273\,\mathrm{K})$  & 0.28(5)  & MT-MLP  & MT-MH\tabularnewline
$d(1473\,\mathrm{K})$  & 0.16(6)  & $=$  & $=$\tabularnewline
$d(1673\,\mathrm{K})$  & 0.17(3)  & MT-MLP  & MT-MH\tabularnewline
$\kappa$  & 0.4(2)  & $=$  & $=$\tabularnewline
$\Delta T$  & 50(20)  & $=$  & $=$\tabularnewline
$\log_{10}(\alpha_{L}(T<T_{g}))$  & 0.058(4)  & RF  & RF\tabularnewline
$\log_{10}(\alpha_{L}(328\,\mathrm{K}))$  & 0.08(1)  & MT-MLP  & MT-MH\tabularnewline
$\log_{10}(\alpha_{L}(373\,\mathrm{K}))$  & 0.10(2)  & MT-MLP  & MT-MH\tabularnewline
$\log_{10}(\alpha_{L}(433\,\mathrm{K}))$  & 0.047(4)  & RF  & RF\tabularnewline
$\log_{10}(\alpha_{L}(483\,\mathrm{K}))$  & 0.049(6)  & $=$  & MT-MH\tabularnewline
$\log_{10}(\alpha_{L}(623\,\mathrm{K}))$  & 0.13(4)  & MT-MLP  & MT-MH\tabularnewline
$C_{p}(293\,\mathrm{K})$  & 160(30)  & $=$  & $=$\tabularnewline
$C_{p}(473\,\mathrm{K})$  & 170(60)  & $=$  & $=$\tabularnewline
$C_{p}(673\,\mathrm{K})$  & 160(50)  & $=$  & $=$\tabularnewline
$C_{p}(1073\,\mathrm{K})$  & 220(30)  & MT-MLP  & MT-MH\tabularnewline
$C_{p}(1273\,\mathrm{K})$  & 300(100)  & $=$  & $=$\tabularnewline
$C_{p}(1473\,\mathrm{K})$  & 410(60)  & $=$  & $=$\tabularnewline
$C_{p}(1673\,\mathrm{K})$  & 60(60)  & $=$  & $=$\tabularnewline
$T_{\mathrm{max}(U)}$  & 70(20)  & $=$  & $=$\tabularnewline
$\log_{10}(U_{\mathrm{max}})$  & 0.6(1)  & $=$  & $=$\tabularnewline
$T_{c}$  & 51(3)  & RF  & RF\tabularnewline
$T_{x}$  & 44(3)  & RF  & RF\tabularnewline
$\gamma(T>T_{g})$  & 0.041(5)  & $=$  & $=$\tabularnewline
$\gamma(1173\,\mathrm{K})$  & 0.017(4)  & $=$  & $=$\tabularnewline
$\gamma(1473\,\mathrm{K})$  & 0.033(3)  & MT-MLP  & MT-MH\tabularnewline
$\gamma(1573\,\mathrm{K})$  & 0.031(3)  & $=$  & MT-MH\tabularnewline
$\gamma(1673\,\mathrm{K})$  & 0.044(6)  & $=$  & $=$\tabularnewline
\hline 
\end{longtable}

\subsection*{Training a multitask model without the Liebel and Körner loss weights}

\label{sec:org3eee21f}

A test was conducted to check the influence of the Liebel and Körner
(LK) loss weights on the performance of the model. For this test,
a NN with the same architecture of the MT-MH was used. The same procedure
was used to compare the RMSE metrics as discussed in the paper (10-fold
cross-validation with a t-test).

In this test, we observed that the MT-MH model (trained with LK loss
weights) performed better than a model with the MT-MH architecture
trained without LK loss weights for four properties ($\log_{10}(\eta(1573\,\mathrm{K}))$,
$T_{\mathrm{melt}}$, $C_{p}(1073\,\mathrm{K})$, and $\log_{10}(U_{\mathrm{max}})$).
There is no clear pattern here to explain why these properties would
perform better in this case.

The MT-MH model (trained with LK loss weights) performed worse than
a model with the MT-MH architecture trained without LK loss weights
for five properties ($T_{\mathrm{liq}}$, $\log_{10}(\rho(573\,\mathrm{K}))$,
$\log_{10}(\rho(1273\,\mathrm{K}))$, $\nu$, and $T_{c}$). Again,
no clear pattern emerges from this set of properties.

The remaining 76 properties showed no statistical difference in this
test.

The hybrid GlassNet (MT-MH plus single-task NNs) performs better or
with no statistical difference compared to the model trained without
the LK loss weights for 84 of the 85 targets examined in this paper.
The only property for which the model trained without the LK loss
weights outperformed the hybrid GlassNet was $\log_{10}(\rho(573\,\mathrm{K}))$.

\subsection*{How the different liquid chemistries were defined}

\label{sec:orgaf8d811}

The predictive power of the temperature-dependence of viscosity for
unseen data was tested and reported in the manuscript. Table \ref{tab:liq_chem}
shows the definition, number of examples and RMSE for different tests.
Of the 31,976 data points considered, GlassNet could not perform the
MYEGA regression for 98 of them (the regression did not converge).

\newpage{}

\begin{landscape}

\begin{table}[htbp]
\caption{\label{tab:liq_chem}Definition of liquid types, number of examples,
and RMSE for different tests of the viscosity prediction of GlassNet.
RMSE calculated using $\log_{10}(\eta)$ with $\eta$ in Pa.s. The
symbol $\land$ is the logical \textquotedblleft and\textquotedblright{}
operator and the symbol $\lor$ is the logical \textquotedblleft or\textquotedblright{}
operator. Chemical information was considered in mol\%.}
\centering %
\begin{tabular}{llrr}
\hline 
Liquid type  & Definition  & Num. examples  & RMSE\tabularnewline
\hline 
\hline 
Overall  & All examples  & 31878  & 0.90\tabularnewline
Silicate  & SiO\textsubscript{2} $\ge$ 60\%  & 18009  & 0.56\tabularnewline
Aluminosilicate  & SiO\textsubscript{2} $\ge$ 50\% $\land$ Al\textsubscript{2}O\textsubscript{3}
$\ge$ 10\%  & 6661  & 0.45\tabularnewline
Borate  & B\textsubscript{2}O\textsubscript{3} $\ge$ 60\%  & 948  & 1.4\tabularnewline
Borosilicate  & SiO\textsubscript{2} $\ge$ 50\% $\land$ B\textsubscript{2}O\textsubscript{3}
$\ge$ 10\%  & 2610  & 0.75\tabularnewline
Germanate  & GeO\textsubscript{2} $\ge$ 60\%  & 306  & 1.3\tabularnewline
Germanosilicate  & SiO\textsubscript{2} $\ge$ 50\% $\land$ GeO\textsubscript{2} $\ge$
10\%  & 32  & 1.2\tabularnewline
Phosphate  & P\textsubscript{2}O\textsubscript{5} $\ge$ 60\%  & 117  & 2.3\tabularnewline
Phosphosilicate  & SiO\textsubscript{2} $\ge$ 50\% $\land$ P\textsubscript{2}O\textsubscript{5}
$\ge$ 10\%  & 33  & 1.5\tabularnewline
Chalcogenide  & (Se > 0\% $\lor$ Te > 0\% $\lor$ S > 0\%) $\land$ (F = 0\% $\land$
Cl = 0\% $\land$ Br = 0\% $\land$ I = 0\% $\land$ O = 0\%)  & 80  & 1.4\tabularnewline
Halide  & (F > 0\% $\lor$ Cl > 0\% $\lor$ Br > 0\% $\lor$ I > 0\%) $\land$
(Se = 0\% $\land$ Te = 0\% $\land$ S = 0\% $\land$ O = 0\%)  & 153  & 1.6\tabularnewline
Chalcohalide  & (Se > 0\% $\lor$ Te > 0\% $\lor$ S > 0\%) $\land$ (F > 0\% $\lor$
Cl > 0\% $\lor$ Br > 0\% $\lor$ I > 0\%) $\land$ O = 0\%  & 32  & 1.7\tabularnewline
\hline 
\end{tabular}
\end{table}

\end{landscape}

\subsection*{SHAP values violin plots --- viscosity and relaxation}

\label{sec:org19665de}

\begin{figure}[H]
\centering \includegraphics[width=0.4\textwidth,keepaspectratio]{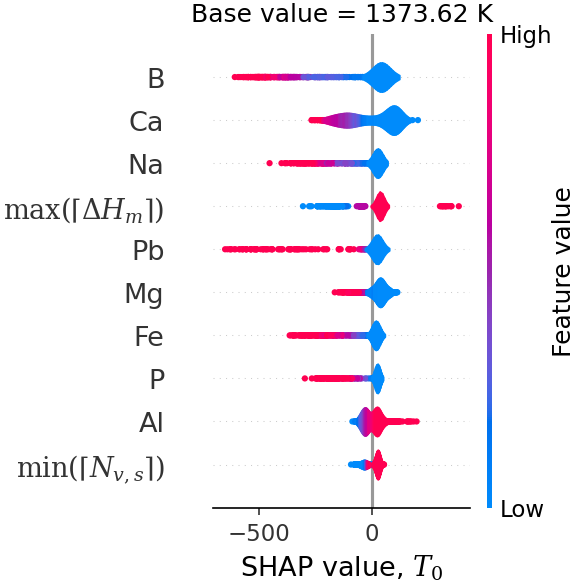}
\caption{\label{fig:shap_T0}Violin plot of SHAP values for $T_{0}$.}
\end{figure}

\begin{figure}[H]
\centering \includegraphics[width=0.4\textwidth,keepaspectratio]{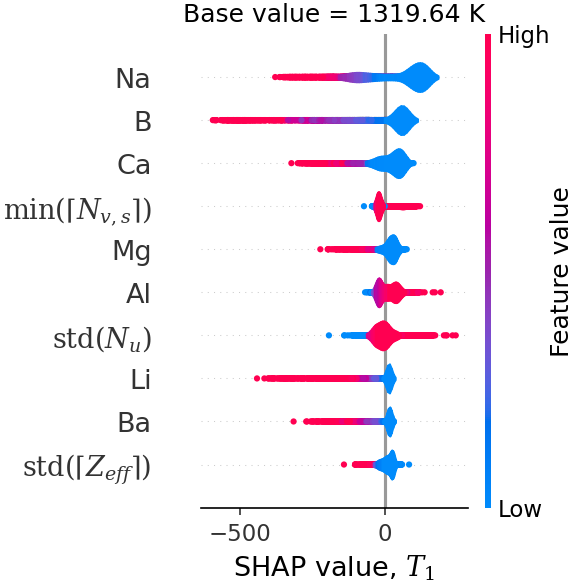}
\caption{\label{fig:shap_T1}Violin plot of SHAP values for $T_{1}$.}
\end{figure}

\begin{figure}[H]
\centering \includegraphics[width=0.4\textwidth,keepaspectratio]{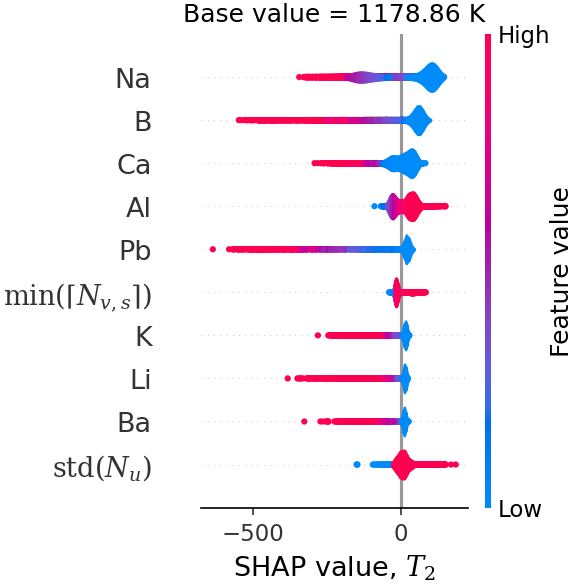}
\caption{\label{fig:shap_T2}Violin plot of SHAP values for $T_{2}$.}
\end{figure}

\begin{figure}[H]
\centering \includegraphics[width=0.4\textwidth,keepaspectratio]{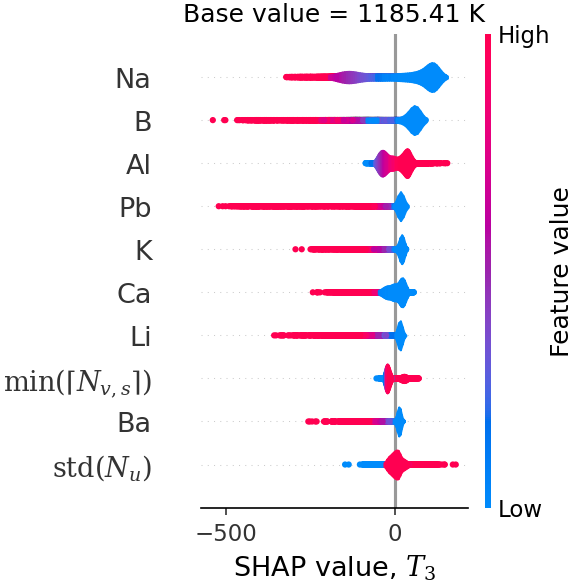}
\caption{\label{fig:shap_T3}Violin plot of SHAP values for $T_{3}$.}
\end{figure}

\begin{figure}[H]
\centering \includegraphics[width=0.4\textwidth,keepaspectratio]{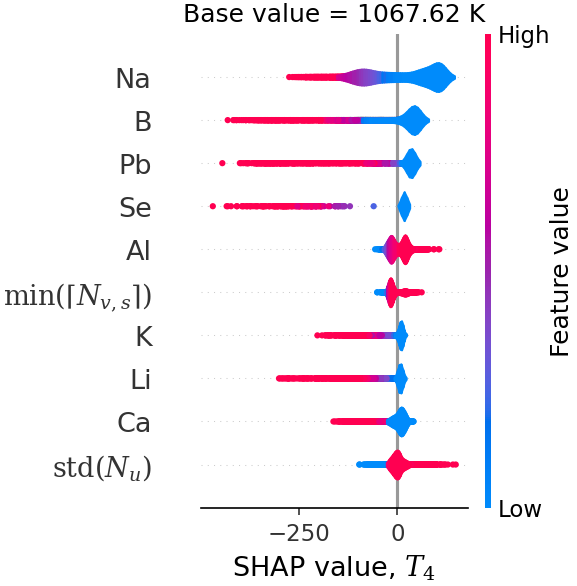}
\caption{\label{fig:shap_T4}Violin plot of SHAP values for $T_{4}$.}
\end{figure}

\begin{figure}[H]
\centering \includegraphics[width=0.4\textwidth,keepaspectratio]{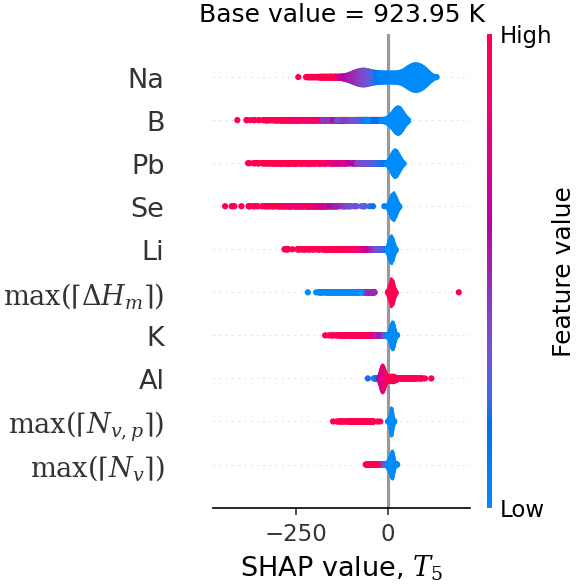}
\caption{\label{fig:shap_T5}Violin plot of SHAP values for $T_{5}$.}
\end{figure}

\begin{figure}[H]
\centering \includegraphics[width=0.4\textwidth,keepaspectratio]{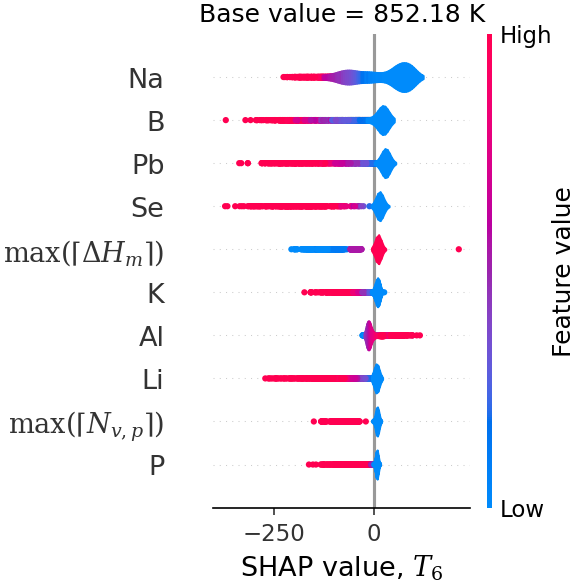}
\caption{\label{fig:shap_T6}Violin plot of SHAP values for $T_{6}$.}
\end{figure}

\begin{figure}[H]
\centering \includegraphics[width=0.4\textwidth,keepaspectratio]{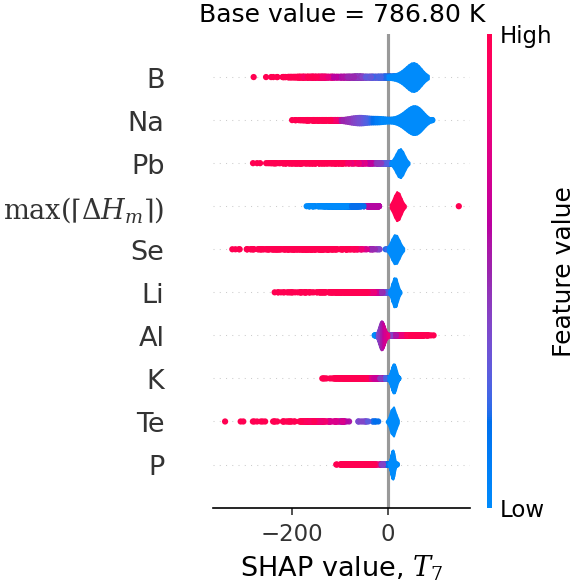}
\caption{\label{fig:shap_T7}Violin plot of SHAP values for $T_{7}$.}
\end{figure}

\begin{figure}[H]
\centering \includegraphics[width=0.4\textwidth,keepaspectratio]{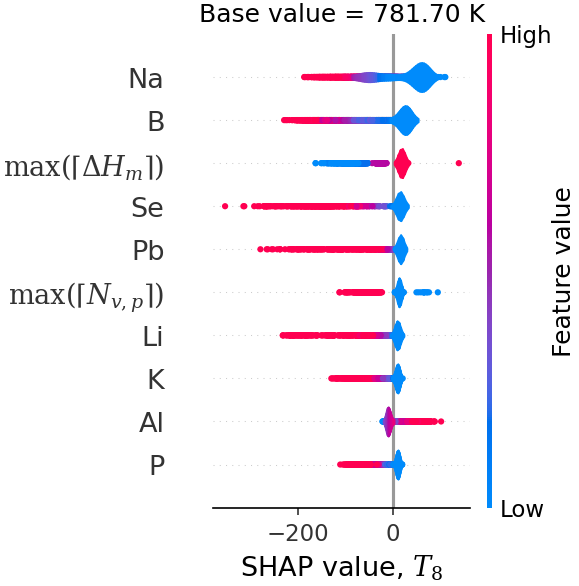}
\caption{\label{fig:shap_T8}Violin plot of SHAP values for $T_{8}$.}
\end{figure}

\begin{figure}[H]
\centering \includegraphics[width=0.4\textwidth,keepaspectratio]{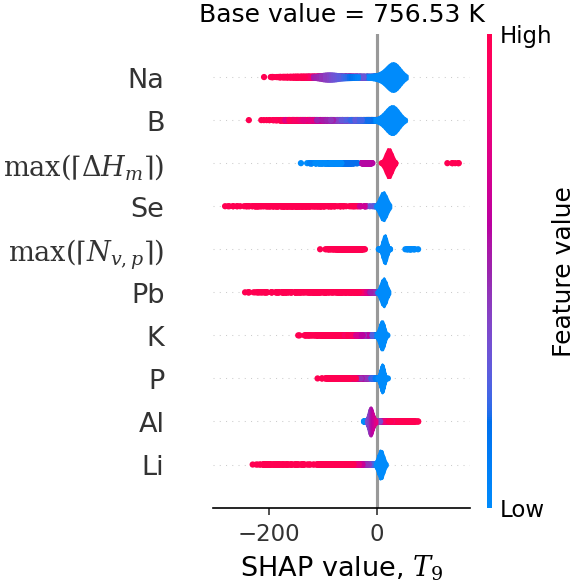}
\caption{\label{fig:shap_T9}Violin plot of SHAP values for $T_{9}$.}
\end{figure}

\begin{figure}[H]
\centering \includegraphics[width=0.4\textwidth,keepaspectratio]{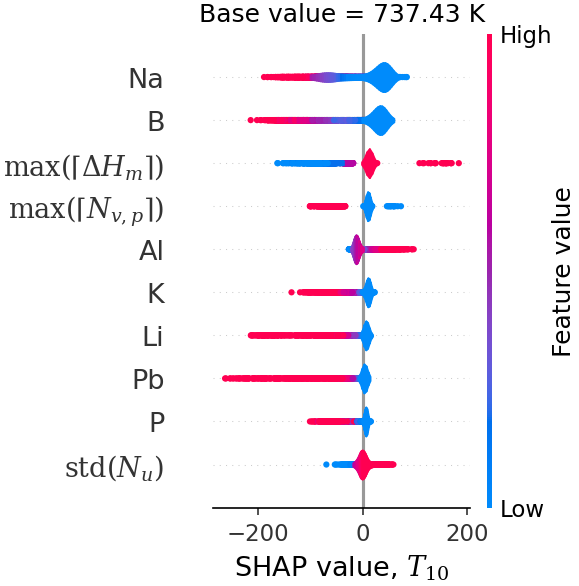}
\caption{\label{fig:shap_T10}Violin plot of SHAP values for $T_{10}$.}
\end{figure}

\begin{figure}[H]
\centering \includegraphics[width=0.4\textwidth,keepaspectratio]{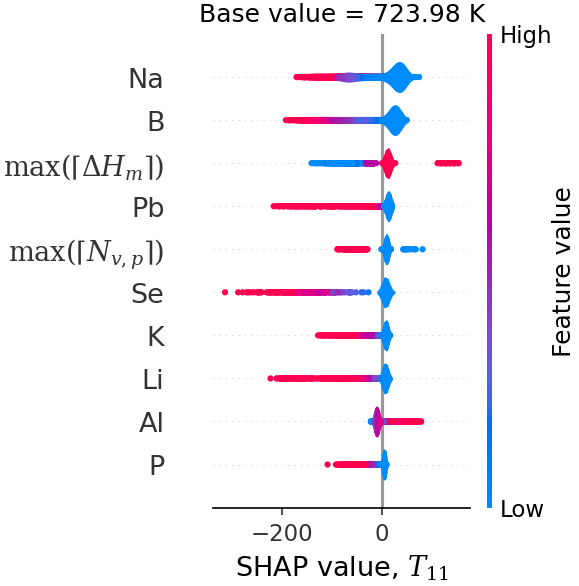}
\caption{\label{fig:shap_T11}Violin plot of SHAP values for $T_{11}$.}
\end{figure}

\begin{figure}[H]
\centering \includegraphics[width=0.4\textwidth,keepaspectratio]{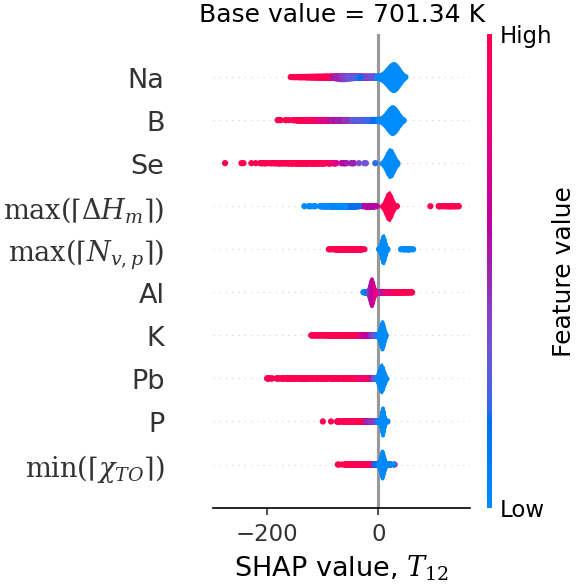}
\caption{\label{fig:shap_T12}Violin plot of SHAP values for $T_{12}$.}
\end{figure}

\begin{figure}[H]
\centering \includegraphics[width=0.4\textwidth,keepaspectratio]{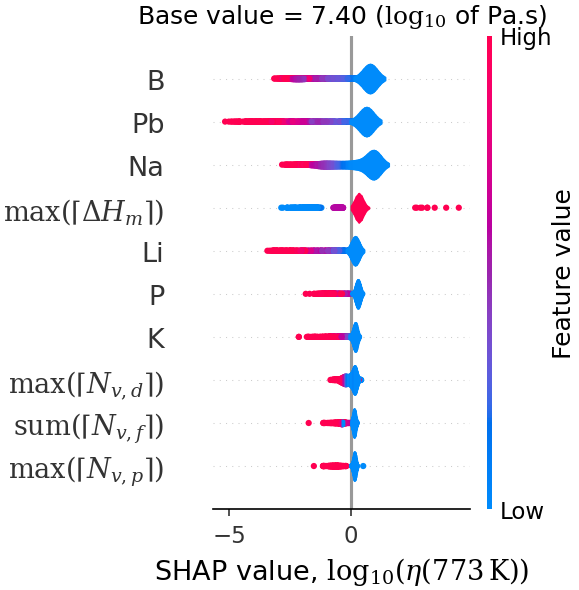}
\caption{\label{fig:shap_Viscosity773K}Violin plot of SHAP values for $\log_{10}(\eta(773\,\mathrm{K}))$.}
\end{figure}

\begin{figure}[H]
\centering \includegraphics[width=0.4\textwidth,keepaspectratio]{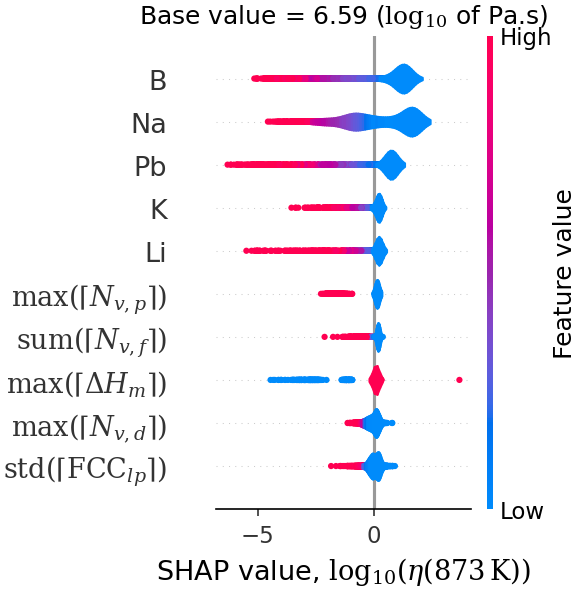}
\caption{\label{fig:shap_Viscosity873K}Violin plot of SHAP values for $\log_{10}(\eta(873\,\mathrm{K}))$.}
\end{figure}

\begin{figure}[H]
\centering \includegraphics[width=0.4\textwidth,keepaspectratio]{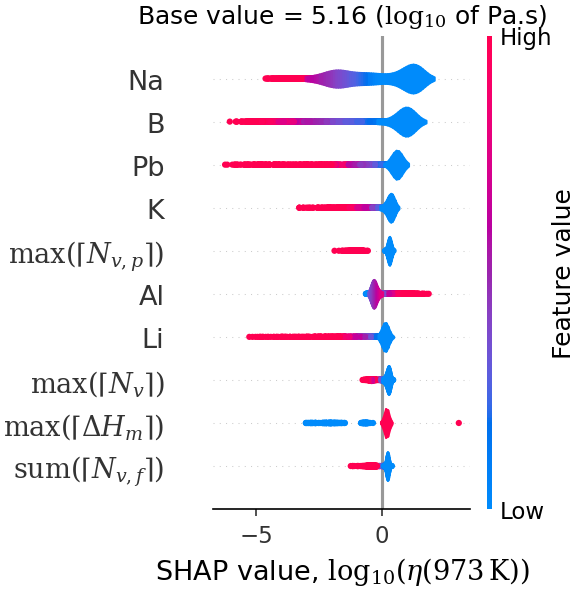}
\caption{\label{fig:shap_Viscosity973K}Violin plot of SHAP values for $\log_{10}(\eta(973\,\mathrm{K}))$.}
\end{figure}

\begin{figure}[H]
\centering \includegraphics[width=0.4\textwidth,keepaspectratio]{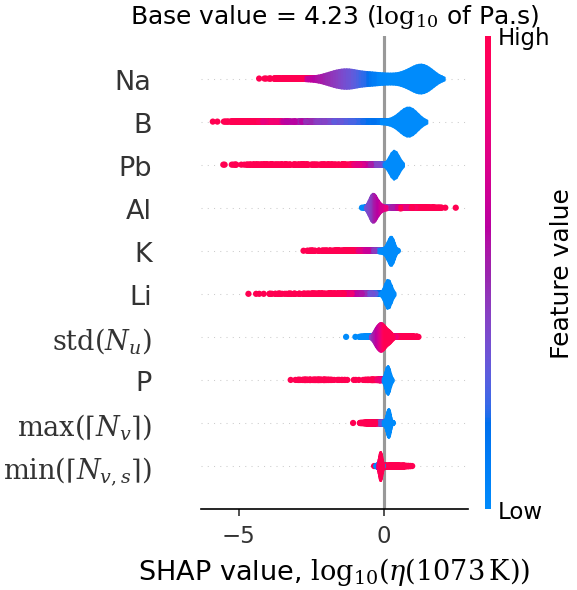}
\caption{\label{fig:shap_Viscosity1073K}Violin plot of SHAP values for $\log_{10}(\eta(1073\,\mathrm{K}))$.}
\end{figure}

\begin{figure}[H]
\centering \includegraphics[width=0.4\textwidth,keepaspectratio]{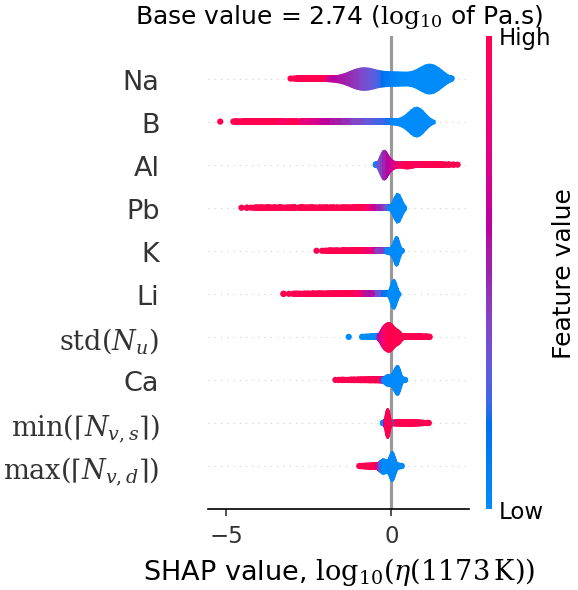}
\caption{\label{fig:shap_Viscosity1173K}Violin plot of SHAP values for $\log_{10}(\eta(1173\,\mathrm{K}))$.}
\end{figure}

\begin{figure}[H]
\centering \includegraphics[width=0.4\textwidth,keepaspectratio]{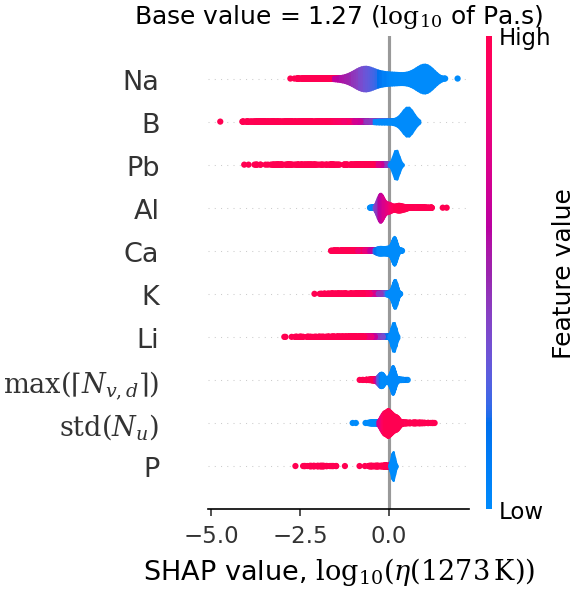}
\caption{\label{fig:shap_Viscosity1273K}Violin plot of SHAP values for $\log_{10}(\eta(1273\,\mathrm{K}))$.}
\end{figure}

\begin{figure}[H]
\centering \includegraphics[width=0.4\textwidth,keepaspectratio]{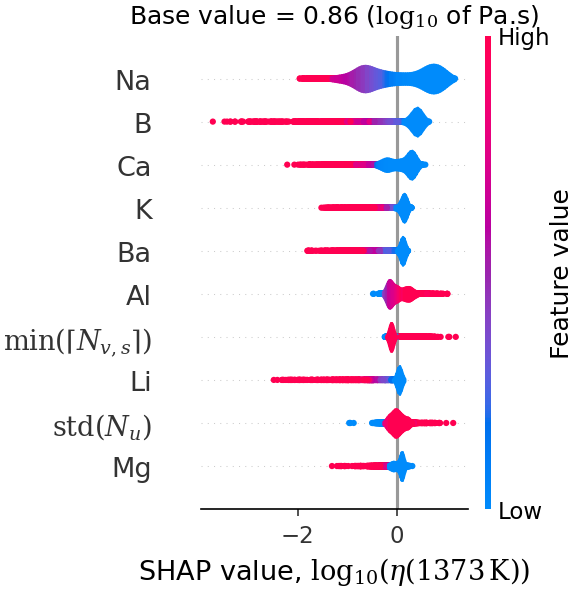}
\caption{\label{fig:shap_Viscosity1373K}Violin plot of SHAP values for $\log_{10}(\eta(1373\,\mathrm{K}))$.}
\end{figure}

\begin{figure}[H]
\centering \includegraphics[width=0.4\textwidth,keepaspectratio]{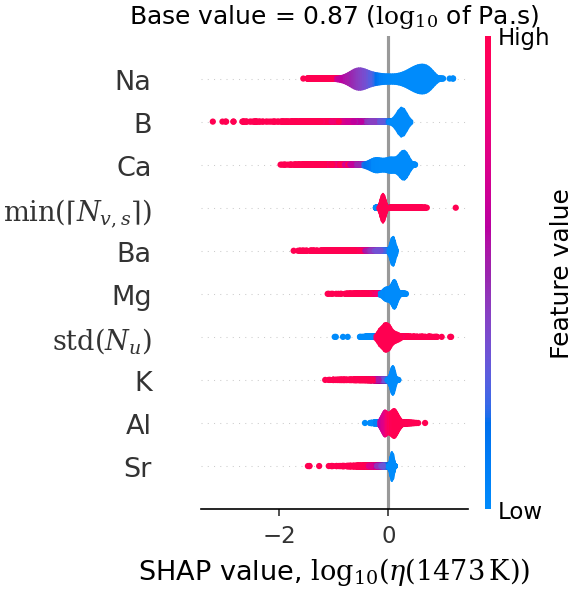}
\caption{\label{fig:shap_Viscosity1473K}Violin plot of SHAP values for $\log_{10}(\eta(1473\,\mathrm{K}))$.}
\end{figure}

\begin{figure}[H]
\centering \includegraphics[width=0.4\textwidth,keepaspectratio]{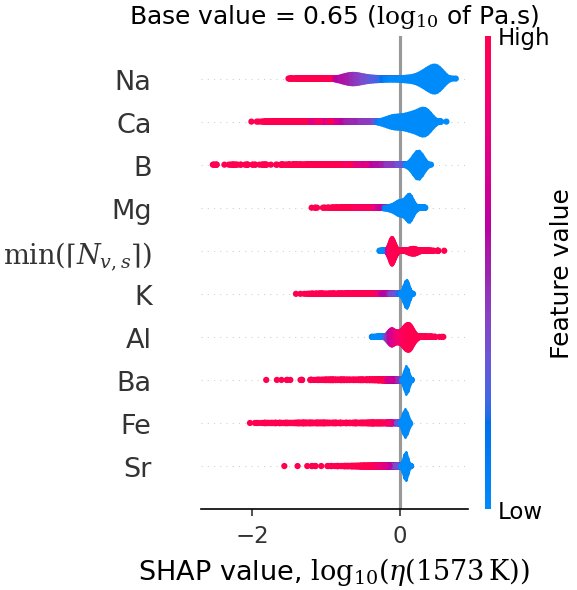}
\caption{\label{fig:shap_Viscosity1573K}Violin plot of SHAP values for $\log_{10}(\eta(1573\,\mathrm{K}))$.}
\end{figure}

\begin{figure}[H]
\centering \includegraphics[width=0.4\textwidth,keepaspectratio]{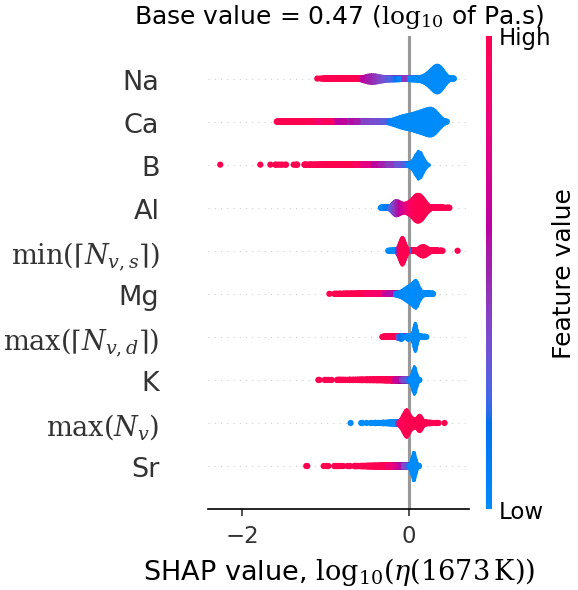}
\caption{\label{fig:shap_Viscosity1673K}Violin plot of SHAP values for $\log_{10}(\eta(1673\,\mathrm{K}))$.}
\end{figure}

\begin{figure}[H]
\centering \includegraphics[width=0.4\textwidth,keepaspectratio]{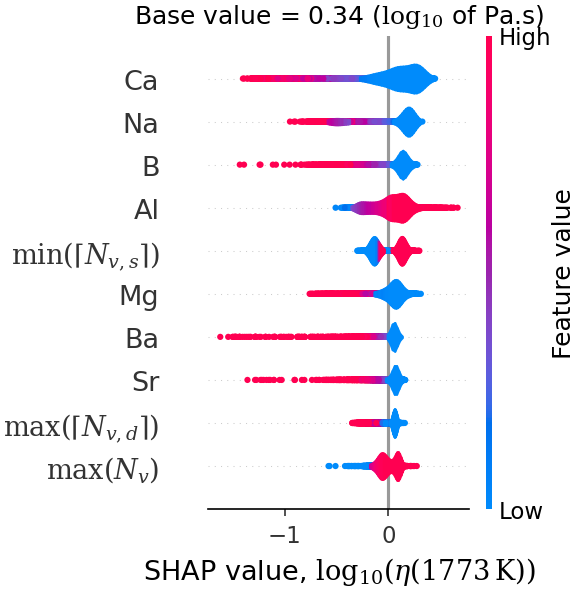}
\caption{\label{fig:shap_Viscosity1773K}Violin plot of SHAP values for $\log_{10}(\eta(1773\,\mathrm{K}))$.}
\end{figure}

\begin{figure}[H]
\centering \includegraphics[width=0.4\textwidth,keepaspectratio]{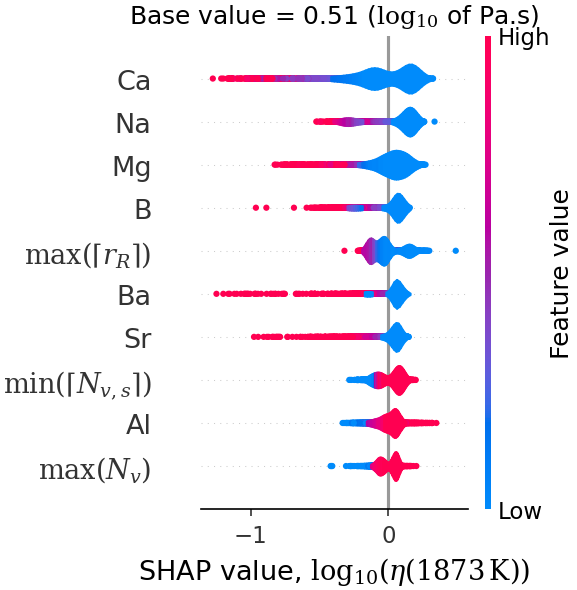}
\caption{\label{fig:shap_Viscosity1873K}Violin plot of SHAP values for $\log_{10}(\eta(1873\,\mathrm{K}))$.}
\end{figure}

\begin{figure}[H]
\centering \includegraphics[width=0.4\textwidth,keepaspectratio]{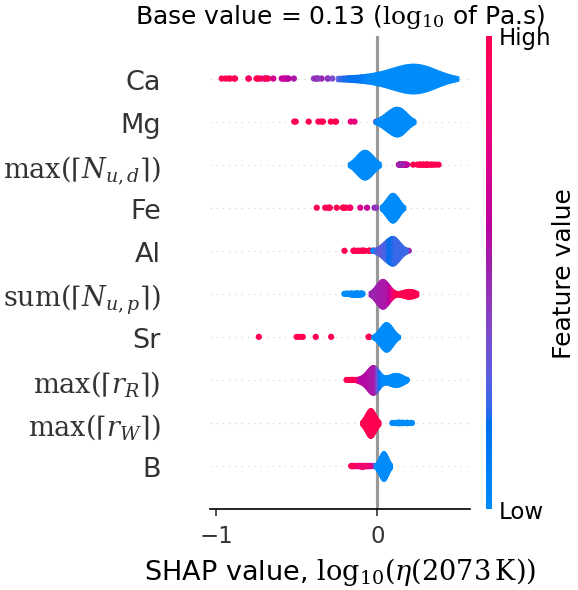}
\caption{\label{fig:shap_Viscosity2073K}Violin plot of SHAP values for $\log_{10}(\eta(2073\,\mathrm{K}))$.}
\end{figure}

\begin{figure}[H]
\centering \includegraphics[width=0.4\textwidth,keepaspectratio]{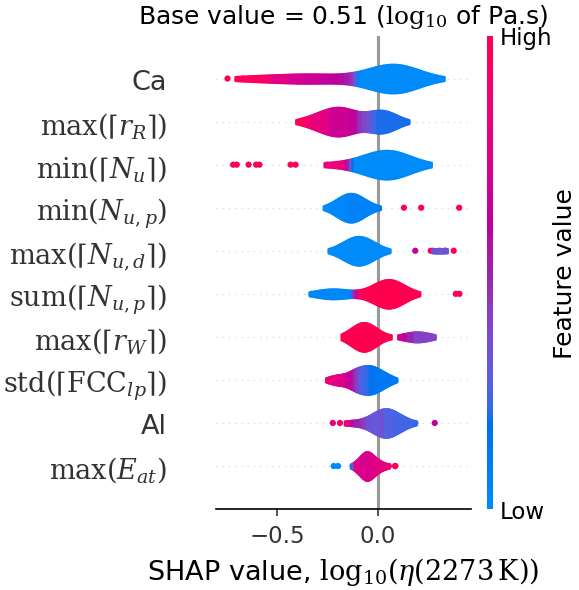}
\caption{\label{fig:shap_Viscosity2273K}Violin plot of SHAP values for $\log_{10}(\eta(2273\,\mathrm{K}))$.}
\end{figure}

\begin{figure}[H]
\centering \includegraphics[width=0.4\textwidth,keepaspectratio]{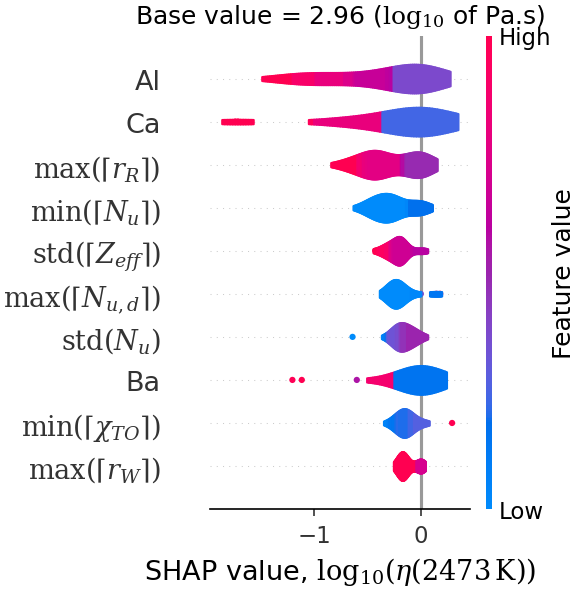}
\caption{\label{fig:shap_Viscosity2473K}Violin plot of SHAP values for $\log_{10}(\eta(2473\,\mathrm{K}))$.
This is the only violin plot in this set computed using the MLP model
instead of the multi-headed model.}
\end{figure}

\begin{figure}[H]
\centering \includegraphics[width=0.4\textwidth,keepaspectratio]{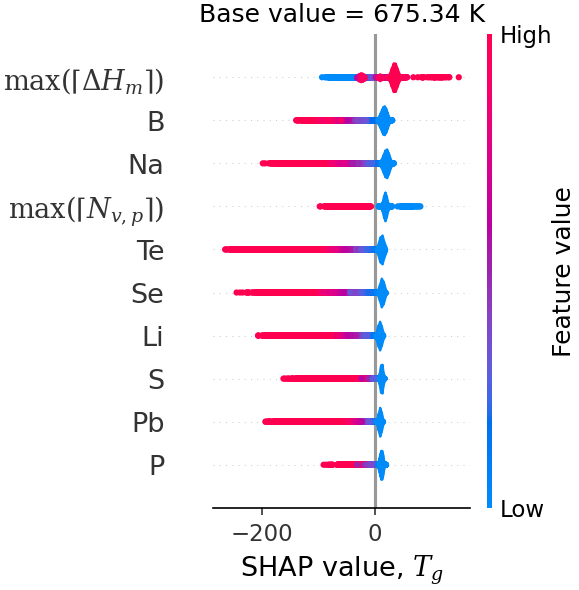}
\caption{\label{fig:shap_Tg}Violin plot of SHAP values for $T_{g}$.}
\end{figure}

\begin{figure}[H]
\centering \includegraphics[width=0.4\textwidth,keepaspectratio]{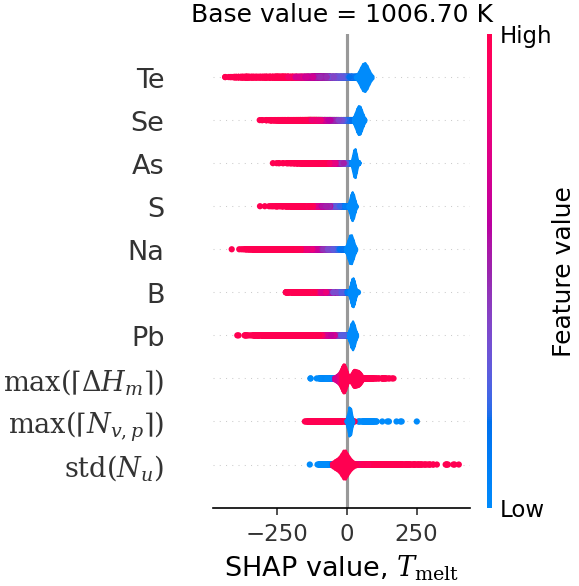}
\caption{\label{fig:shap_Tmelt}Violin plot of SHAP values for $T_{\mathrm{melt}}$.}
\end{figure}

\begin{figure}[H]
\centering \includegraphics[width=0.4\textwidth,keepaspectratio]{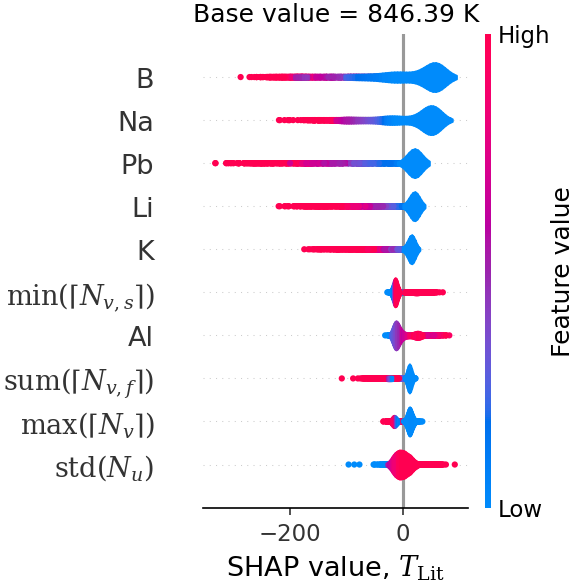}
\caption{\label{fig:shap_TLittletons}Violin plot of SHAP values for $T_{\mathrm{Lit}}$.}
\end{figure}

\begin{figure}[H]
\centering \includegraphics[width=0.4\textwidth,keepaspectratio]{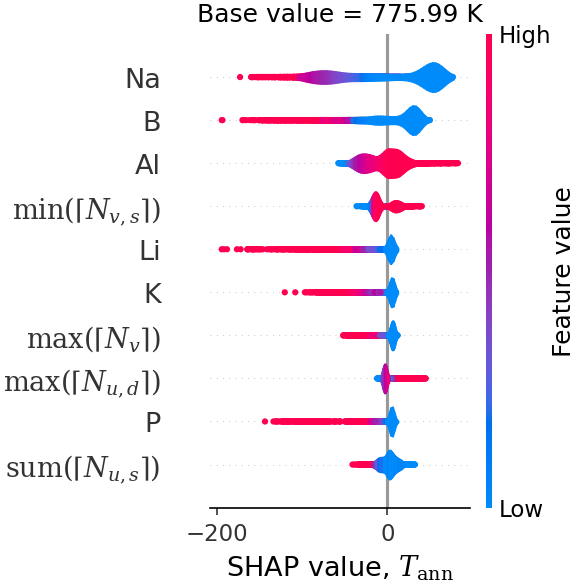}
\caption{\label{fig:shap_TAnnealing}Violin plot of SHAP values for $T_{\mathrm{ann}}$.}
\end{figure}

\begin{figure}[H]
\centering \includegraphics[width=0.4\textwidth,keepaspectratio]{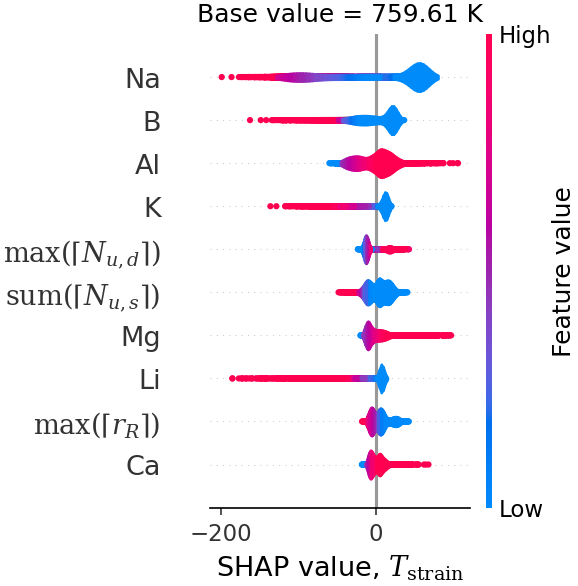}
\caption{\label{fig:shap_Tstrain}Violin plot of SHAP values for $T_{\mathrm{strain}}$.}
\end{figure}

\begin{figure}[H]
\centering \includegraphics[width=0.4\textwidth,keepaspectratio]{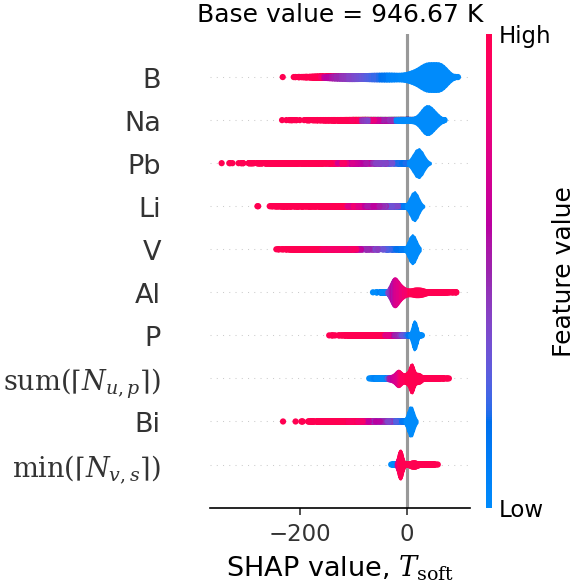}
\caption{\label{fig:shap_Tsoft}Violin plot of SHAP values for $T_{\mathrm{soft}}$.}
\end{figure}

\begin{figure}[H]
\centering \includegraphics[width=0.4\textwidth,keepaspectratio]{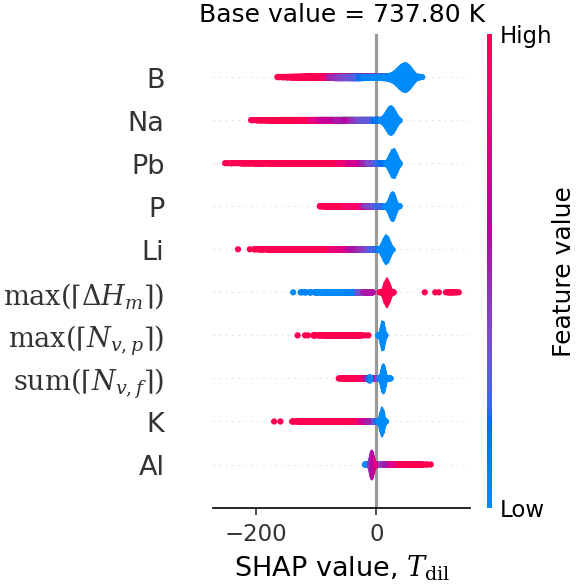}
\caption{\label{fig:shap_TdilatometricSoftening}Violin plot of SHAP values
for $T_{\mathrm{dil}}$.}
\end{figure}

Figures \ref{fig:shap_T0} to \ref{fig:shap_TdilatometricSoftening}
show the violin plots of the SHAP values for the viscosity and relaxation
related properties. The SHAP values were calculated using the multi-headed
GlassNet model, with the exception of $\log_{10}(\eta(2473\,\mathrm{K}))$,
which was calculated using the MLP model because it was not possible
to run the SHAP explainer with the multi-headed model.

\FloatBarrier

\subsection*{SHAP values violin plots --- optical properties}

\label{sec:org155ece0}

\begin{figure}[H]
\centering \includegraphics[width=0.4\textwidth,keepaspectratio]{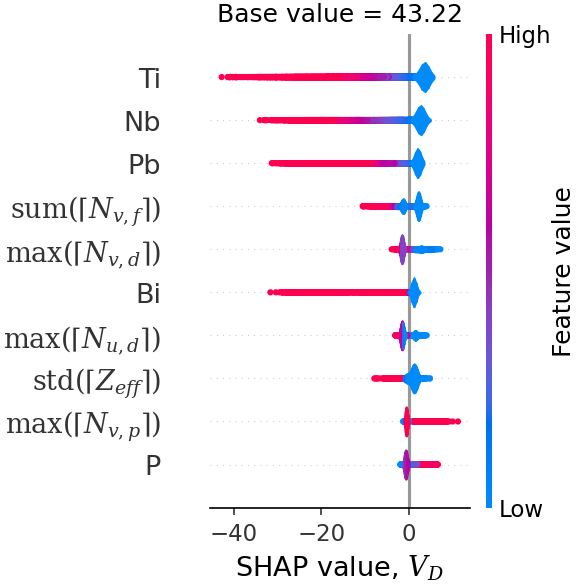}
\caption{\label{fig:shap_AbbeNum}Violin plot of SHAP values for $V_{D}$.}
\end{figure}

\begin{figure}[H]
\centering \includegraphics[width=0.4\textwidth,keepaspectratio]{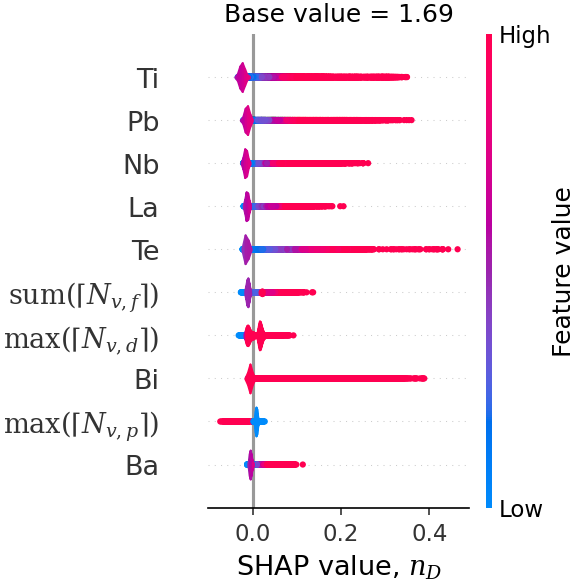}
\caption{\label{fig:shap_RefractiveIndex}Violin plot of SHAP values for $n_{D}$.}
\end{figure}

\begin{figure}[H]
\centering \includegraphics[width=0.4\textwidth,keepaspectratio]{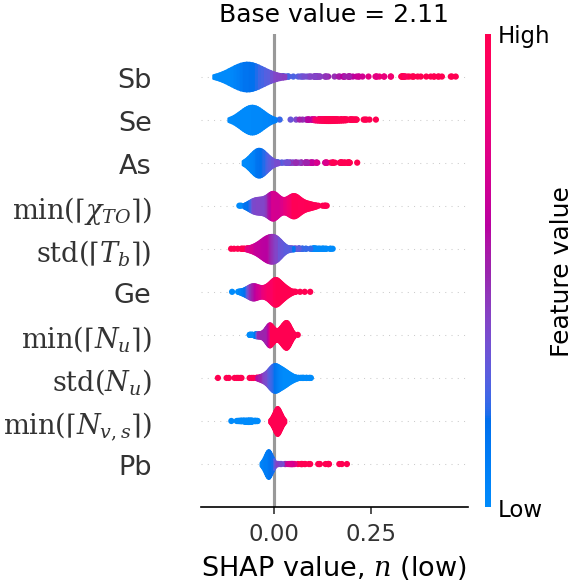}
\caption{\label{fig:shap_RefractiveIndexLow}Violin plot of SHAP values for
$n$ (low).}
\end{figure}

\begin{figure}[H]
\centering \includegraphics[width=0.4\textwidth,keepaspectratio]{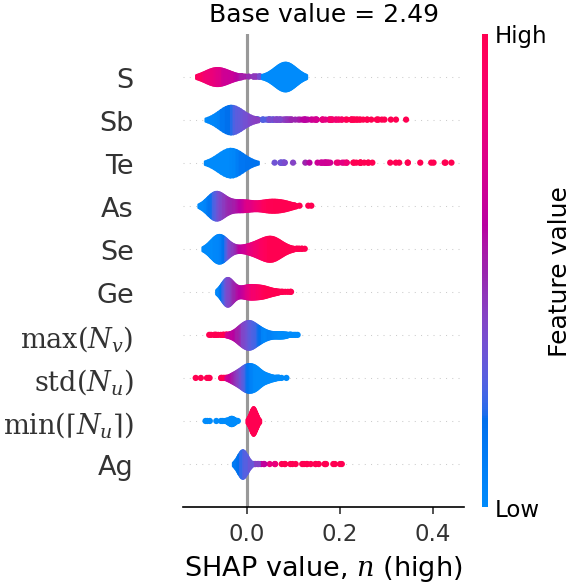}
\caption{\label{fig:shap_RefractiveIndexHigh}Violin plot of SHAP values for
$n$ (high).}
\end{figure}

\begin{figure}[H]
\centering \includegraphics[width=0.4\textwidth,keepaspectratio]{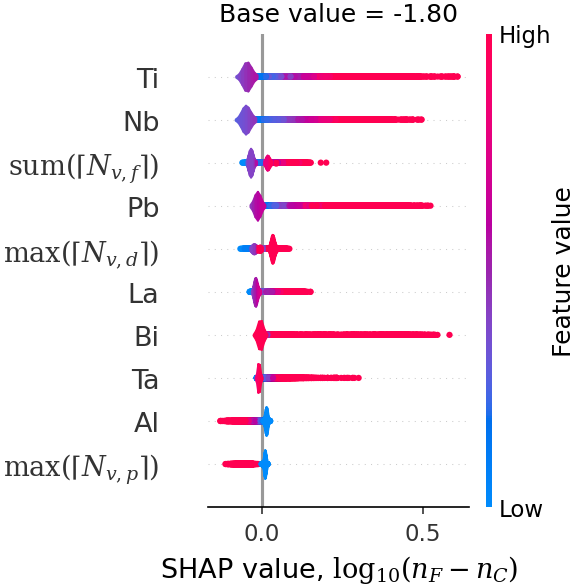}
\caption{\label{fig:shap_MeanDispersion}Violin plot of SHAP values for $\log_{10}(n_{F}-n_{C})$.}
\end{figure}

Figures \ref{fig:shap_AbbeNum} to \ref{fig:shap_MeanDispersion}
show the violin plots of the SHAP values for the optical properties.
The SHAP values were calculated using the multi-headed GlassNet model.

\FloatBarrier

\subsection*{SHAP values violin plots --- electrical and dielectric properties}

\label{sec:org94fcfab}

\begin{figure}[H]
\centering \includegraphics[width=0.4\textwidth,keepaspectratio]{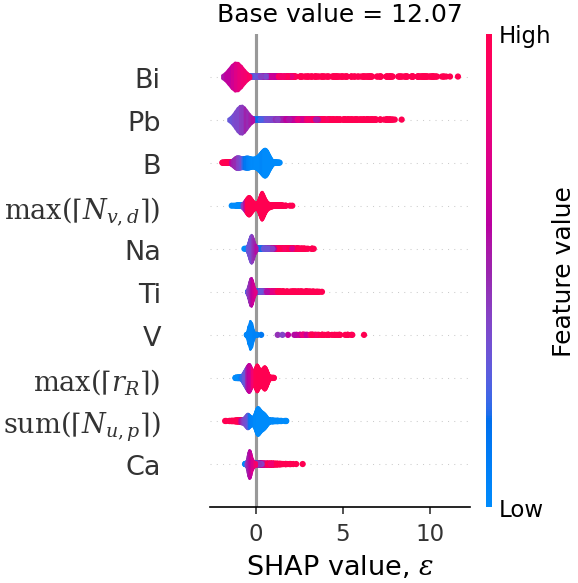}
\caption{\label{fig:shap_Permittivity}Violin plot of SHAP values for $\varepsilon$.}
\end{figure}

\begin{figure}[H]
\centering \includegraphics[width=0.4\textwidth,keepaspectratio]{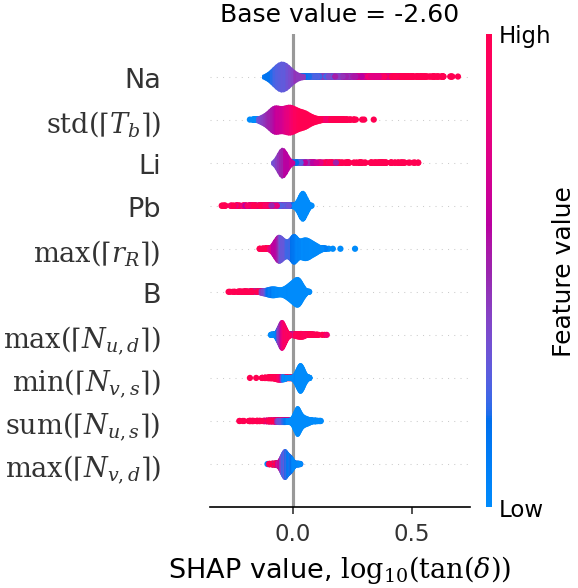}
\caption{\label{fig:shap_TangentOfLossAngle}Violin plot of SHAP values for
$\log_{10}(\tan(\delta))$.}
\end{figure}

\begin{figure}[H]
\centering \includegraphics[width=0.4\textwidth,keepaspectratio]{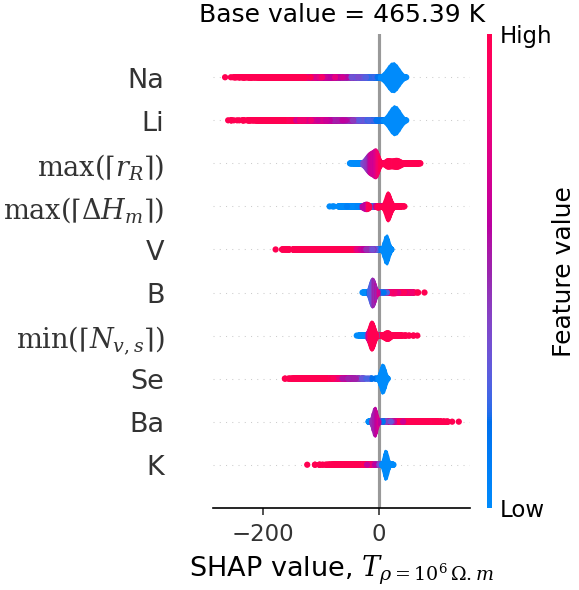}
\caption{\label{fig:shap_TresistivityIs1MOhm.m}Violin plot of SHAP values
for $T_{\rho=10^{6}\,\Omega.m}$.}
\end{figure}

\begin{figure}[H]
\centering \includegraphics[width=0.4\textwidth,keepaspectratio]{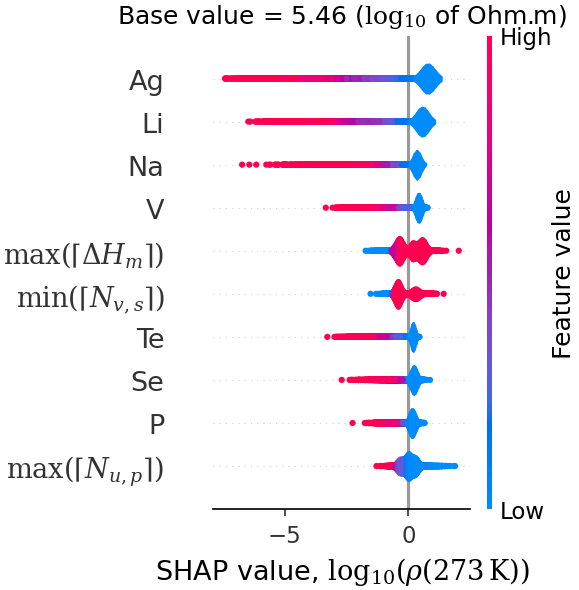}
\caption{\label{fig:shap_Resistivity273K}Violin plot of SHAP values for $\log_{10}(\rho(273\,\mathrm{K}))$.}
\end{figure}

\begin{figure}[H]
\centering \includegraphics[width=0.4\textwidth,keepaspectratio]{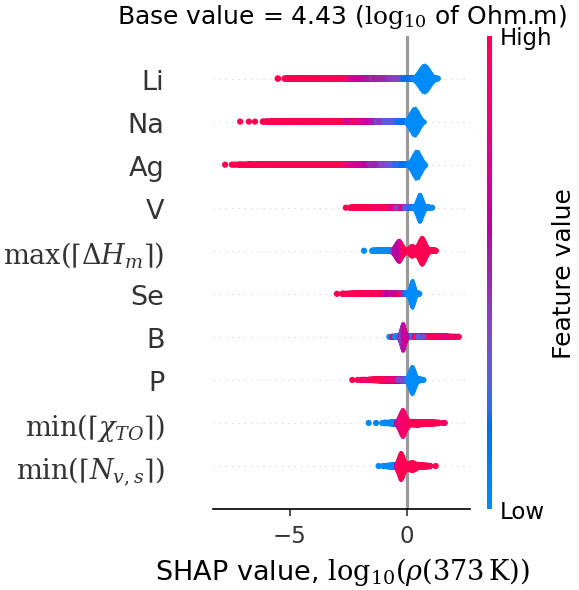}
\caption{\label{fig:shap_Resistivity373K}Violin plot of SHAP values for $\log_{10}(\rho(373\,\mathrm{K}))$.}
\end{figure}

\begin{figure}[H]
\centering \includegraphics[width=0.4\textwidth,keepaspectratio]{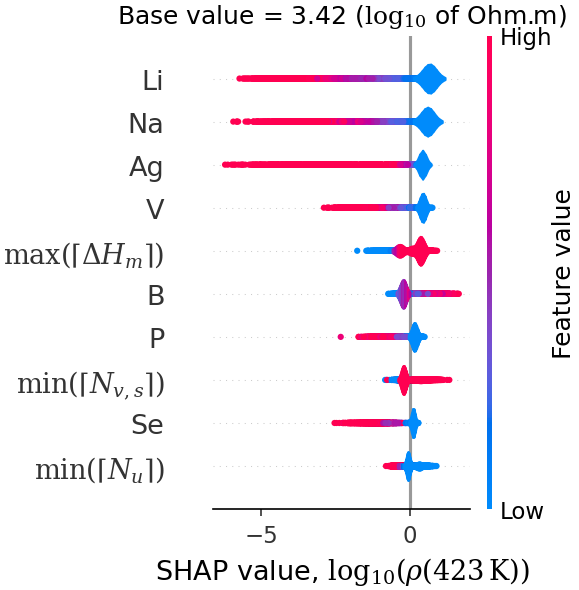}
\caption{\label{fig:shap_Resistivity423K}Violin plot of SHAP values for $\log_{10}(\rho(423\,\mathrm{K}))$.}
\end{figure}

\begin{figure}[H]
\centering \includegraphics[width=0.4\textwidth,keepaspectratio]{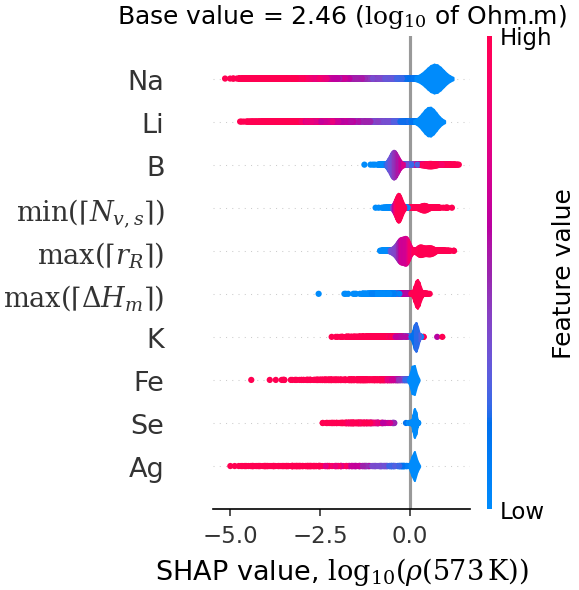}
\caption{\label{fig:shap_Resistivity573K}Violin plot of SHAP values for $\log_{10}(\rho(573\,\mathrm{K}))$.}
\end{figure}

\begin{figure}[H]
\centering \includegraphics[width=0.4\textwidth,keepaspectratio]{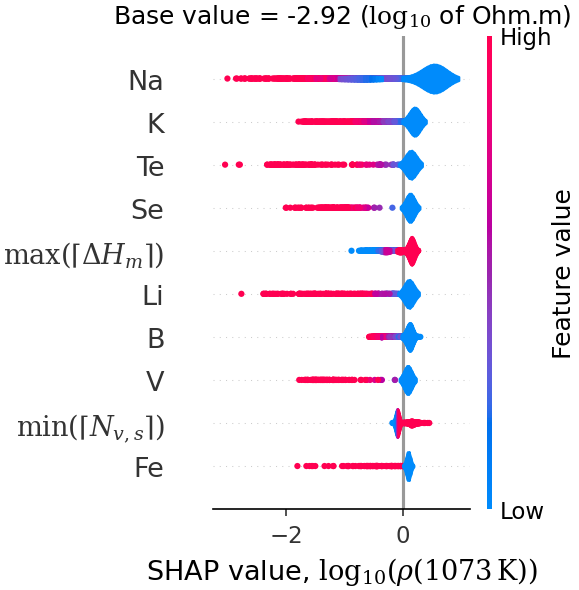}
\caption{\label{fig:shap_Resistivity1073K}Violin plot of SHAP values for $\log_{10}(\rho(1073\,\mathrm{K}))$.}
\end{figure}

\begin{figure}[H]
\centering \includegraphics[width=0.4\textwidth,keepaspectratio]{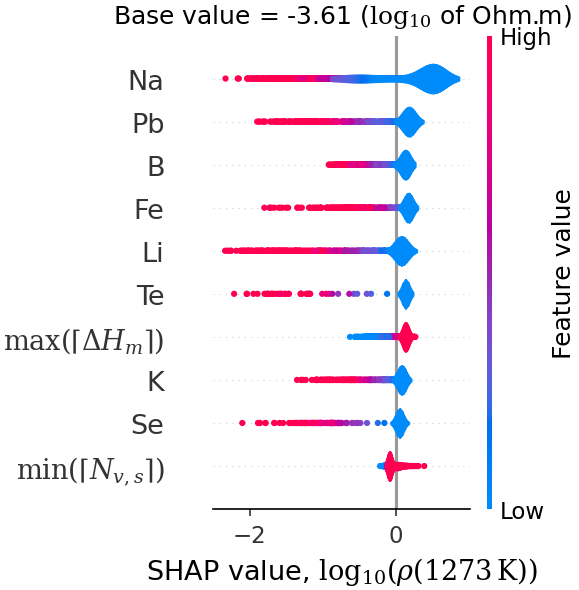}
\caption{\label{fig:shap_Resistivity1273K}Violin plot of SHAP values for $\log_{10}(\rho(1273\,\mathrm{K}))$.}
\end{figure}

\begin{figure}[H]
\centering \includegraphics[width=0.4\textwidth,keepaspectratio]{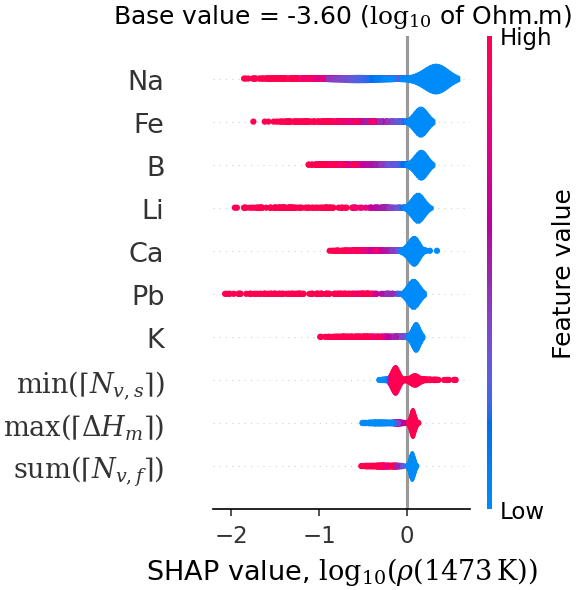}
\caption{\label{fig:shap_Resistivity1473K}Violin plot of SHAP values for $\log_{10}(\rho(1473\,\mathrm{K}))$.}
\end{figure}

\begin{figure}[H]
\centering \includegraphics[width=0.4\textwidth,keepaspectratio]{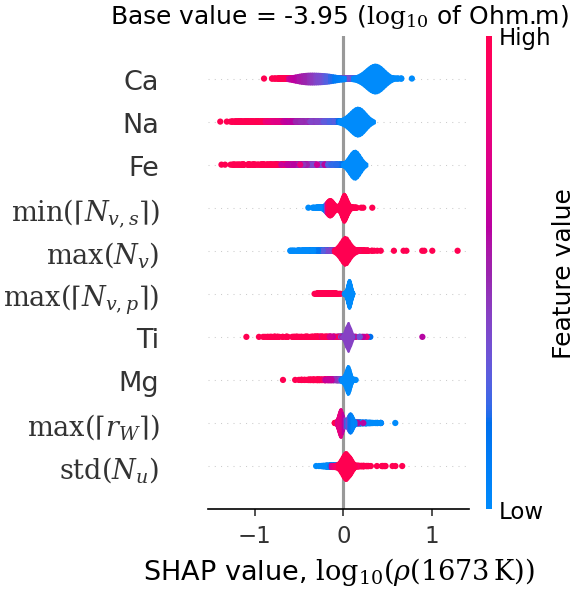}
\caption{\label{fig:shap_Resistivity1673K}Violin plot of SHAP values for $\log_{10}(\rho(1673\,\mathrm{K}))$.}
\end{figure}

Figures \ref{fig:shap_Permittivity} to \ref{fig:shap_Resistivity1673K}
show the violin plots of the SHAP values for the electrical and dielectric
properties. The SHAP values were calculated using the multi-headed
GlassNet model.

\FloatBarrier

\subsection*{SHAP values violin plots --- mechanical properties}

\label{sec:org0ff405e}

\begin{figure}[H]
\centering \includegraphics[width=0.4\textwidth,keepaspectratio]{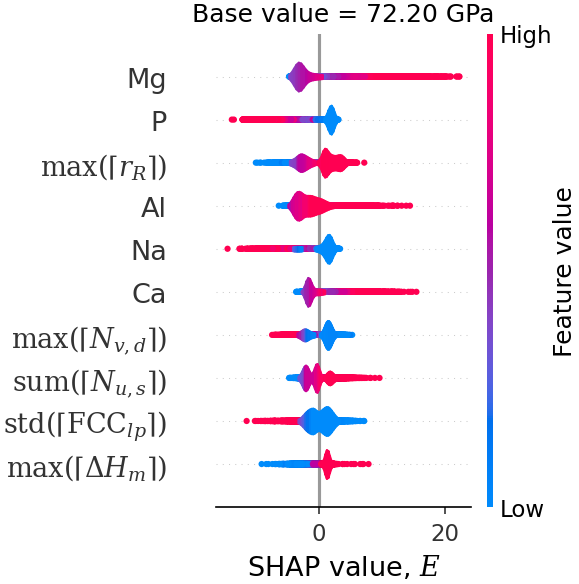}
\caption{\label{fig:shap_YoungModulus}Violin plot of SHAP values for $E$.}
\end{figure}

\begin{figure}[H]
\centering \includegraphics[width=0.4\textwidth,keepaspectratio]{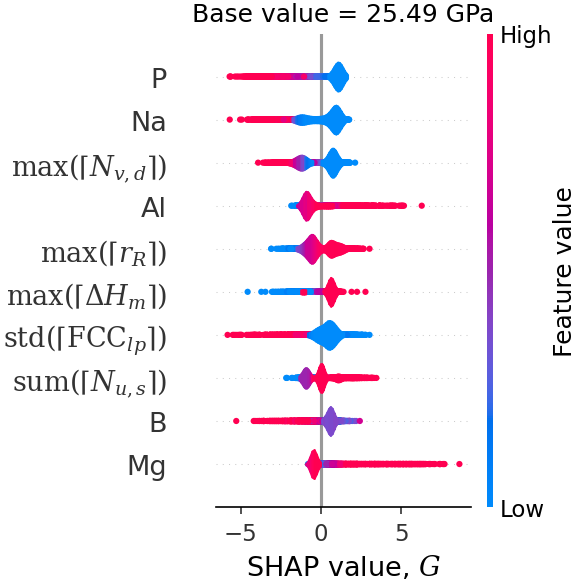}
\caption{\label{fig:shap_ShearModulus}Violin plot of SHAP values for $G$.}
\end{figure}

\begin{figure}[H]
\centering \includegraphics[width=0.4\textwidth,keepaspectratio]{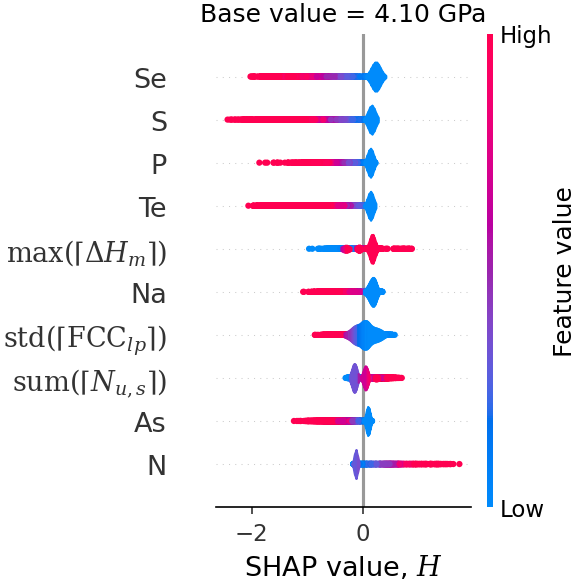}
\caption{\label{fig:shap_Microhardness}Violin plot of SHAP values for $H$.}
\end{figure}

\begin{figure}[H]
\centering \includegraphics[width=0.4\textwidth,keepaspectratio]{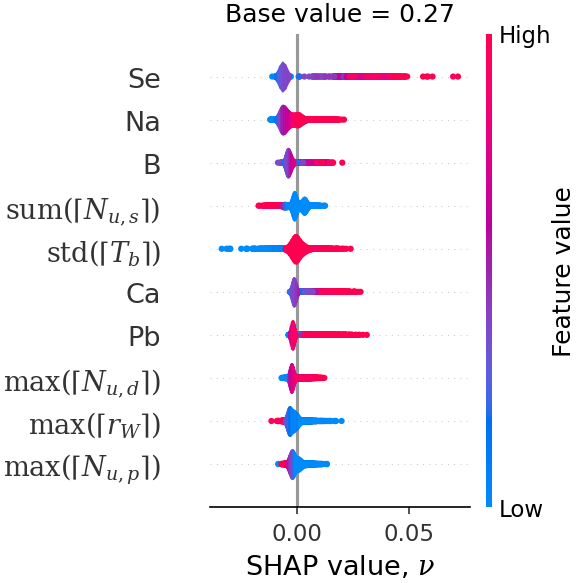}
\caption{\label{fig:shap_PoissonRatio}Violin plot of SHAP values for $\nu$.}
\end{figure}

Figures \ref{fig:shap_YoungModulus} to \ref{fig:shap_PoissonRatio}
show the violin plots of the SHAP values for the mechanical properties.
The SHAP values were calculated using the multi-headed GlassNet model.

\FloatBarrier

\subsection*{SHAP values violin plots --- density}

\label{sec:org5ab2c9b}

\begin{figure}[H]
\centering \includegraphics[width=0.4\textwidth,keepaspectratio]{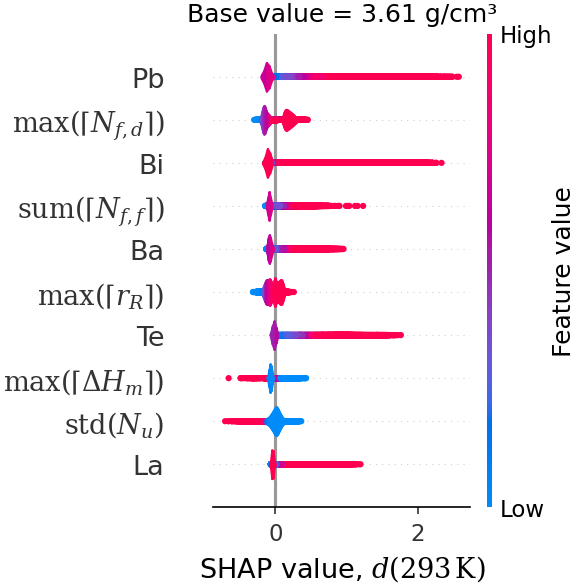}
\caption{\label{fig:shap_Density293K}Violin plot of SHAP values for $d(293\,\mathrm{K})$.}
\end{figure}

\begin{figure}[H]
\centering \includegraphics[width=0.4\textwidth,keepaspectratio]{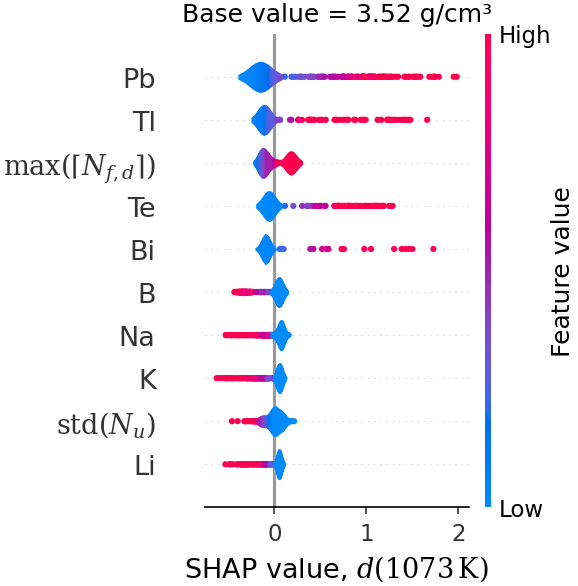}
\caption{\label{fig:shap_Density1073K}Violin plot of SHAP values for $d(1073\,\mathrm{K})$.}
\end{figure}

\begin{figure}[H]
\centering \includegraphics[width=0.4\textwidth,keepaspectratio]{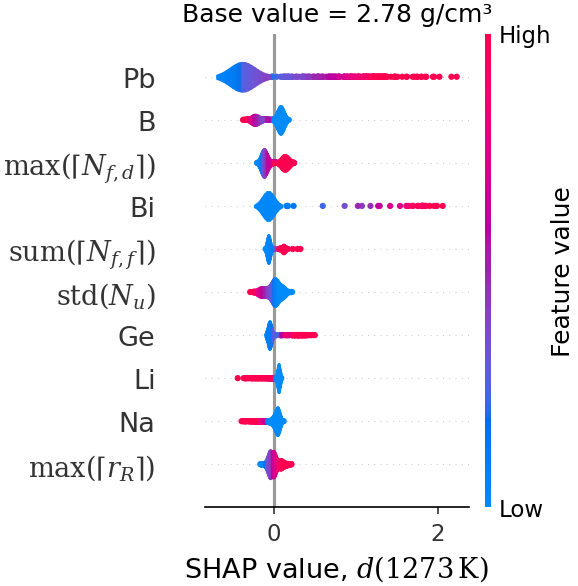}
\caption{\label{fig:shap_Density1273K}Violin plot of SHAP values for $d(1273\,\mathrm{K})$.}
\end{figure}

\begin{figure}[H]
\centering \includegraphics[width=0.4\textwidth,keepaspectratio]{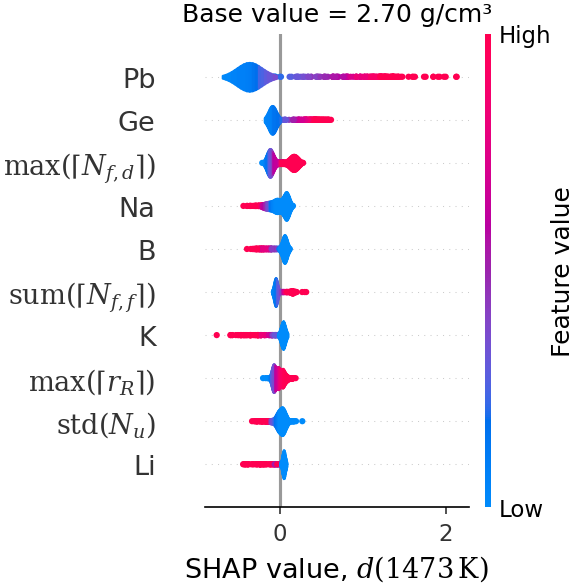}
\caption{\label{fig:shap_Density1473K}Violin plot of SHAP values for $d(1473\,\mathrm{K})$.}
\end{figure}

\begin{figure}[H]
\centering \includegraphics[width=0.4\textwidth,keepaspectratio]{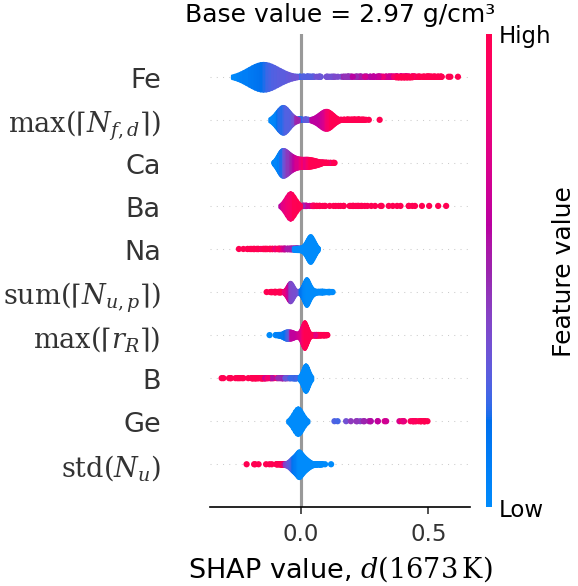}
\caption{\label{fig:shap_Density1673K}Violin plot of SHAP values for $d(1673\,\mathrm{K})$.}
\end{figure}

Figures \ref{fig:shap_Density293K} to \ref{fig:shap_Density1673K}
show the violin plots of the SHAP values for the glass density at
different temperatures. The SHAP values were calculated using the
multi-headed GlassNet model.

\FloatBarrier

\subsection*{SHAP values violin plots --- thermal properties}

\label{sec:orgd193674}

\begin{figure}[H]
\centering \includegraphics[width=0.4\textwidth,keepaspectratio]{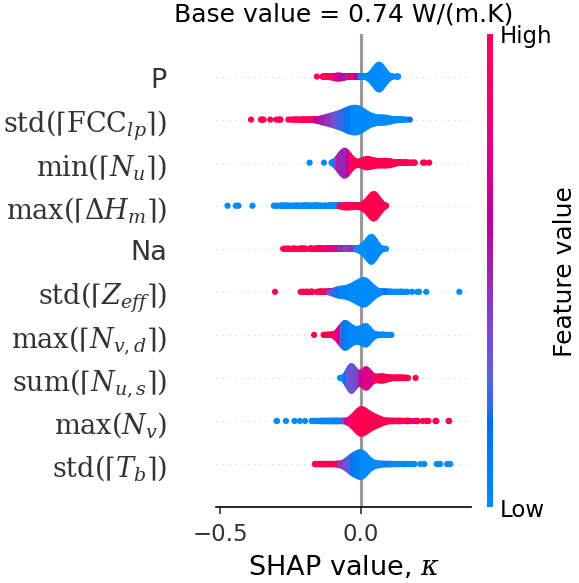}
\caption{\label{fig:shap_ThermalConductivity}Violin plot of SHAP values for
$\kappa$.}
\end{figure}

\begin{figure}[H]
\centering \includegraphics[width=0.4\textwidth,keepaspectratio]{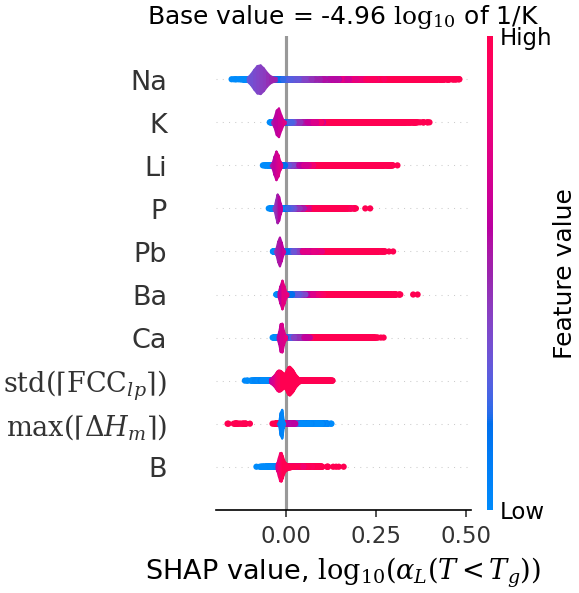}
\caption{\label{fig:shap_CTEbelowTg}Violin plot of SHAP values for $\log_{10}(\alpha_{L}(T<T_{g}))$.}
\end{figure}

\begin{figure}[H]
\centering \includegraphics[width=0.4\textwidth,keepaspectratio]{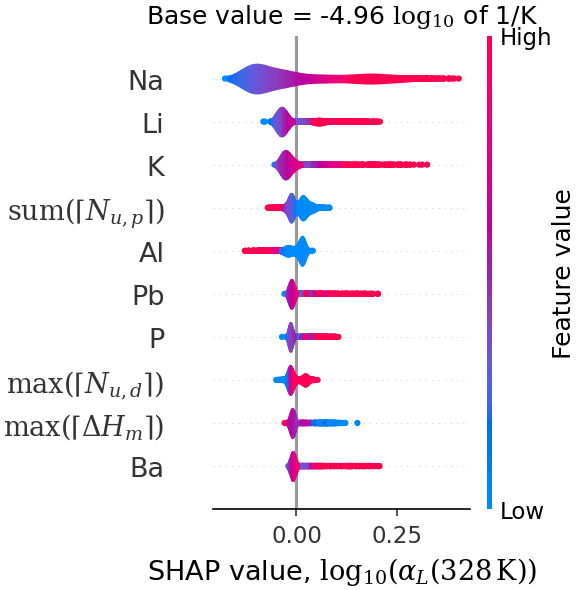}
\caption{\label{fig:shap_CTE328K}Violin plot of SHAP values for $\log_{10}(\alpha_{L}(328\,\mathrm{K}))$.}
\end{figure}

\begin{figure}[H]
\centering \includegraphics[width=0.4\textwidth,keepaspectratio]{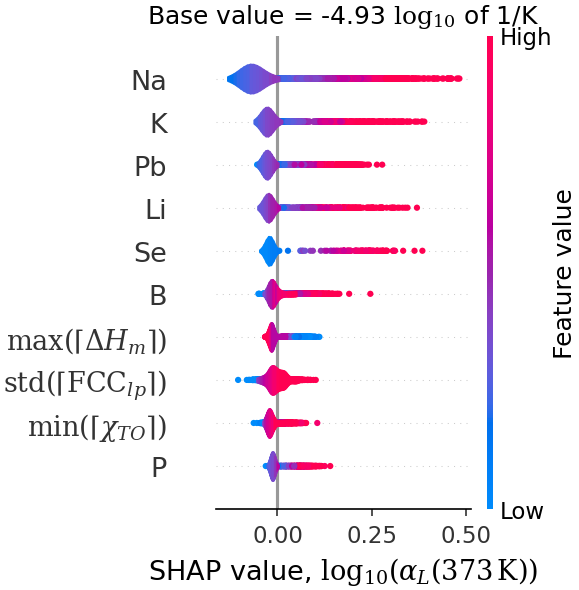}
\caption{\label{fig:shap_CTE373K}Violin plot of SHAP values for $\log_{10}(\alpha_{L}(373\,\mathrm{K}))$.}
\end{figure}

\begin{figure}[H]
\centering \includegraphics[width=0.4\textwidth,keepaspectratio]{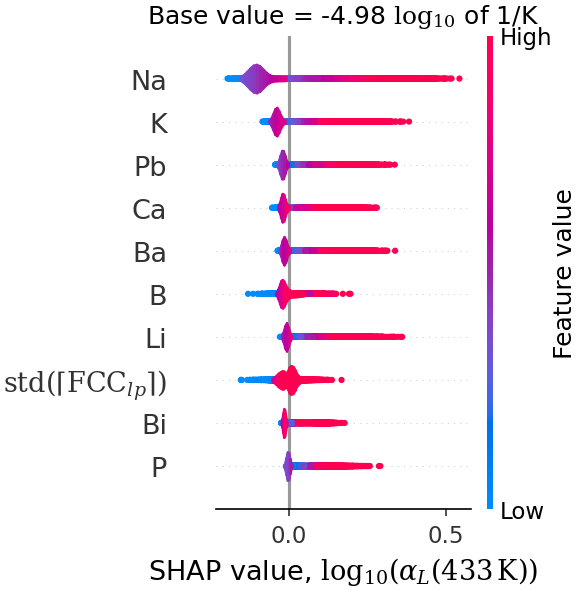}
\caption{\label{fig:shap_CTE433K}Violin plot of SHAP values for $\log_{10}(\alpha_{L}(433\,\mathrm{K}))$.}
\end{figure}

\begin{figure}[H]
\centering \includegraphics[width=0.4\textwidth,keepaspectratio]{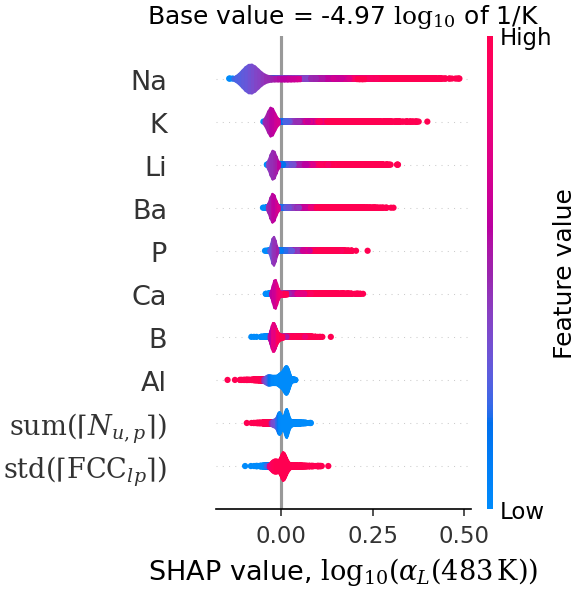}
\caption{\label{fig:shap_CTE483K}Violin plot of SHAP values for $\log_{10}(\alpha_{L}(483\,\mathrm{K}))$.}
\end{figure}

\begin{figure}[H]
\centering \includegraphics[width=0.4\textwidth,keepaspectratio]{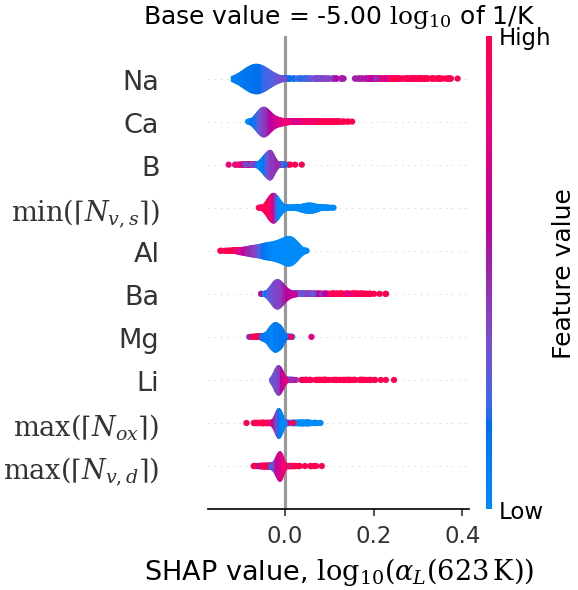}
\caption{\label{fig:shap_CTE623K}Violin plot of SHAP values for $\log_{10}(\alpha_{L}(623\,\mathrm{K}))$.}
\end{figure}

\begin{figure}[H]
\centering \includegraphics[width=0.4\textwidth,keepaspectratio]{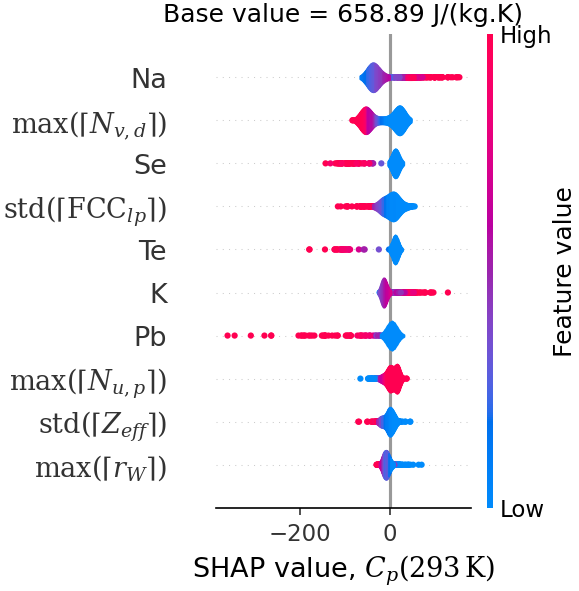}
\caption{\label{fig:shap_Cp293K}Violin plot of SHAP values for $C_{p}(293\,\mathrm{K})$.}
\end{figure}

\begin{figure}[H]
\centering \includegraphics[width=0.4\textwidth,keepaspectratio]{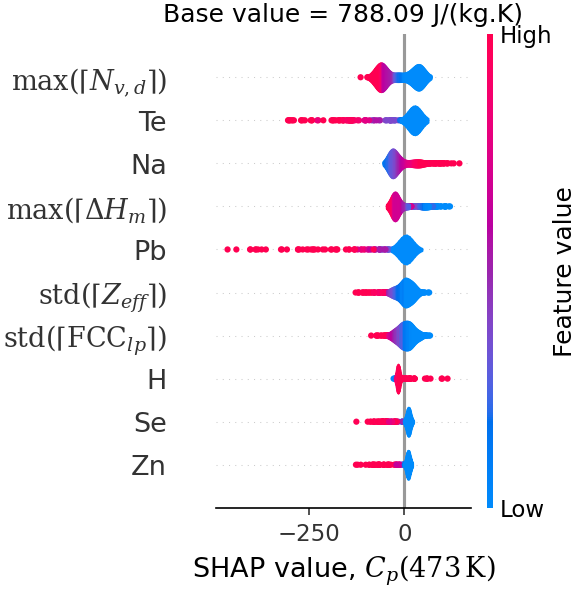}
\caption{\label{fig:shap_Cp473K}Violin plot of SHAP values for $C_{p}(473\,\mathrm{K})$.}
\end{figure}

\begin{figure}[H]
\centering \includegraphics[width=0.4\textwidth,keepaspectratio]{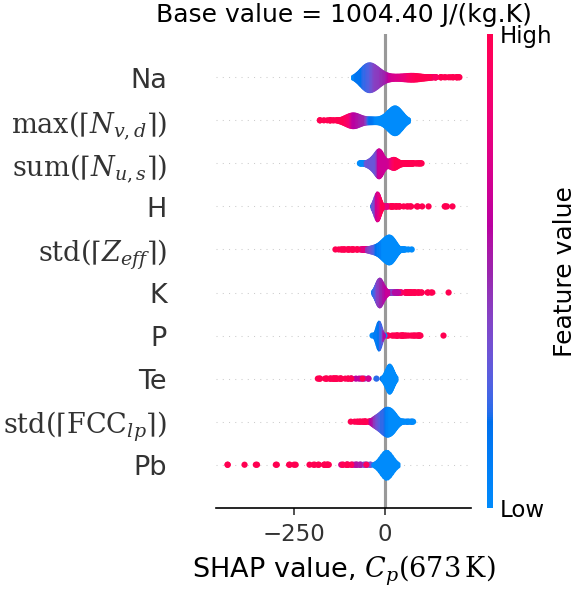}
\caption{\label{fig:shap_Cp673K}Violin plot of SHAP values for $C_{p}(673\,\mathrm{K})$.}
\end{figure}

\begin{figure}[H]
\centering \includegraphics[width=0.4\textwidth,keepaspectratio]{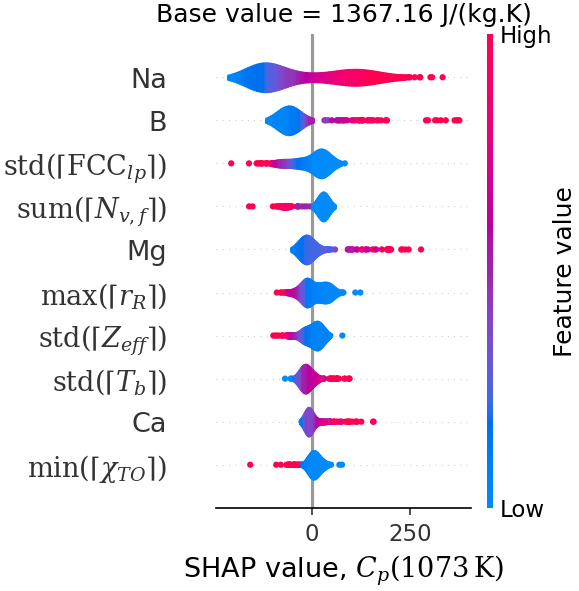}
\caption{\label{fig:shap_Cp1073K}Violin plot of SHAP values for $C_{p}(1073\,\mathrm{K})$.}
\end{figure}

\begin{figure}[H]
\centering \includegraphics[width=0.4\textwidth,keepaspectratio]{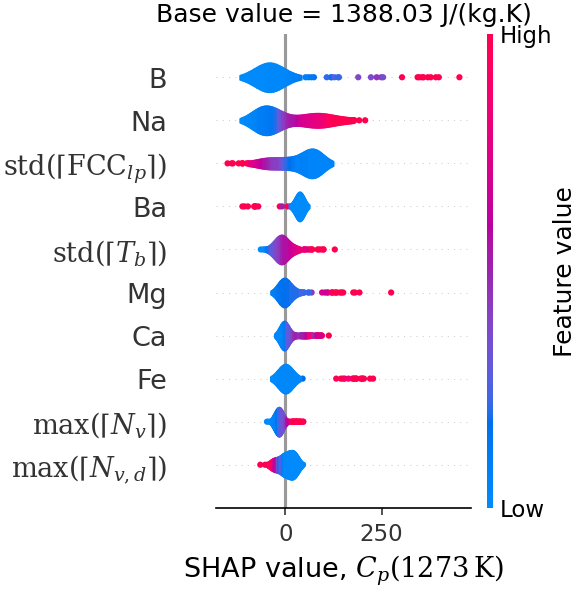}
\caption{\label{fig:shap_Cp1273K}Violin plot of SHAP values for $C_{p}(1273\,\mathrm{K})$.}
\end{figure}

\begin{figure}[H]
\centering \includegraphics[width=0.4\textwidth,keepaspectratio]{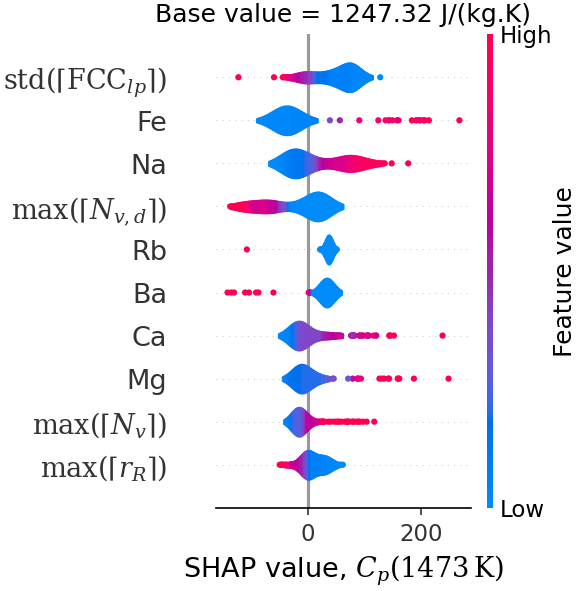}
\caption{\label{fig:shap_Cp1473K}Violin plot of SHAP values for $C_{p}(1473\,\mathrm{K})$.}
\end{figure}

\begin{figure}[H]
\centering \includegraphics[width=0.4\textwidth,keepaspectratio]{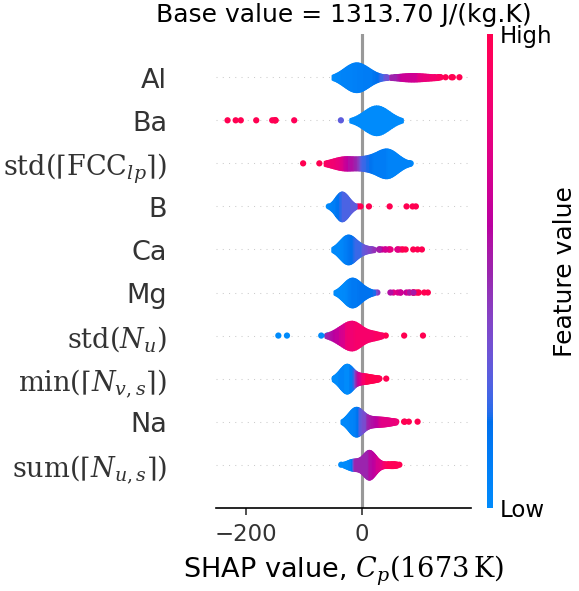}
\caption{\label{fig:shap_Cp1673K}Violin plot of SHAP values for $C_{p}(1673\,\mathrm{K})$.}
\end{figure}

Figures \ref{fig:shap_ThermalConductivity} to \ref{fig:shap_Cp1673K}
show the violin plots of the SHAP values for the thermal properties.
The SHAP values were calculated using the multi-headed GlassNet model.

\FloatBarrier

\subsection*{SHAP values violin plots --- crystallization}

\label{sec:org5a4fc4a}

\begin{figure}[H]
\centering \includegraphics[width=0.4\textwidth,keepaspectratio]{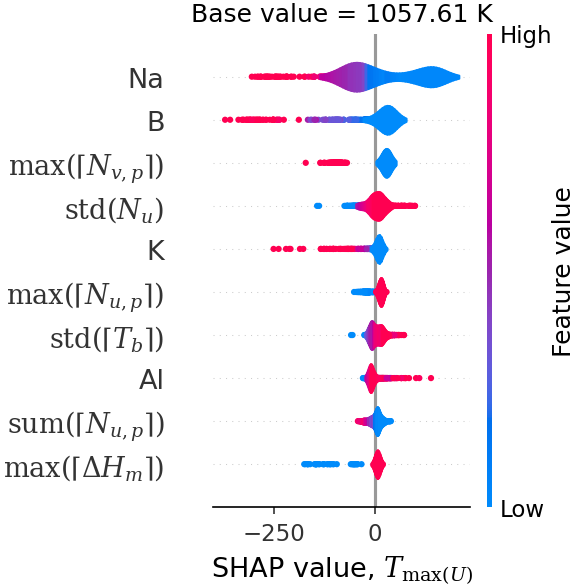}
\caption{\label{fig:shap_TMaxGrowthVelocity}Violin plot of SHAP values for
$T_{\mathrm{max}(U)}$.}
\end{figure}

\begin{figure}[H]
\centering \includegraphics[width=0.4\textwidth,keepaspectratio]{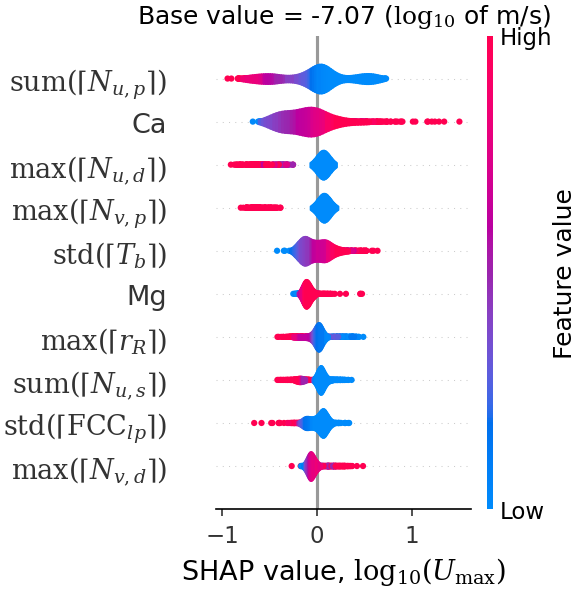}
\caption{\label{fig:shap_MaxGrowthVelocity}Violin plot of SHAP values for
$\log_{10}(U_{\mathrm{max}})$.}
\end{figure}

\begin{figure}[H]
\centering \includegraphics[width=0.4\textwidth,keepaspectratio]{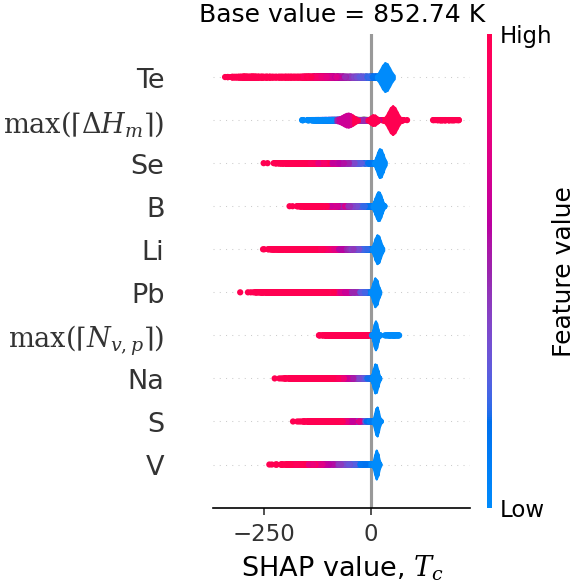}
\caption{\label{fig:shap_CrystallizationPeak}Violin plot of SHAP values for
$T_{c}$.}
\end{figure}

\begin{figure}[H]
\centering \includegraphics[width=0.4\textwidth,keepaspectratio]{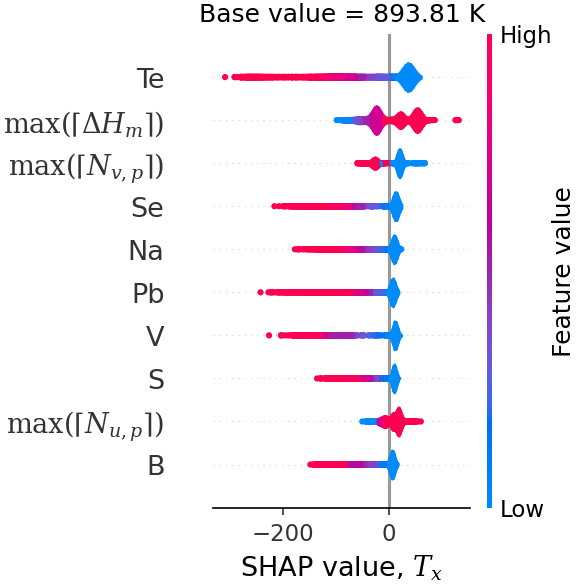}
\caption{\label{fig:shap_CrystallizationOnset}Violin plot of SHAP values for
$T_{x}$.}
\end{figure}

Figures \ref{fig:shap_TMaxGrowthVelocity} to \ref{fig:shap_CrystallizationOnset}
show the violin plots of the SHAP values for properties related to
crystallization. The SHAP values were calculated using the multi-headed
GlassNet model.

\FloatBarrier

\subsection*{SHAP values violin plots --- surface tension}

\label{sec:orga2655e5}

\begin{figure}[H]
\centering \includegraphics[width=0.4\textwidth,keepaspectratio]{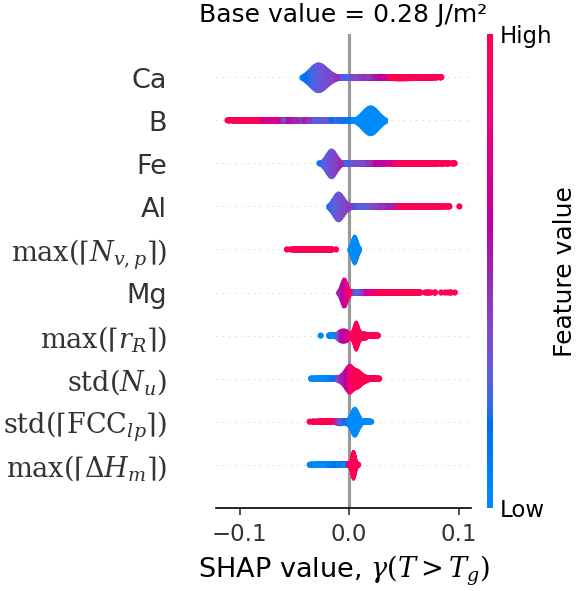}
\caption{\label{fig:shap_SurfaceTensionAboveTg}Violin plot of SHAP values
for $\gamma(T>T_{g})$.}
\end{figure}

\begin{figure}[H]
\centering \includegraphics[width=0.4\textwidth,keepaspectratio]{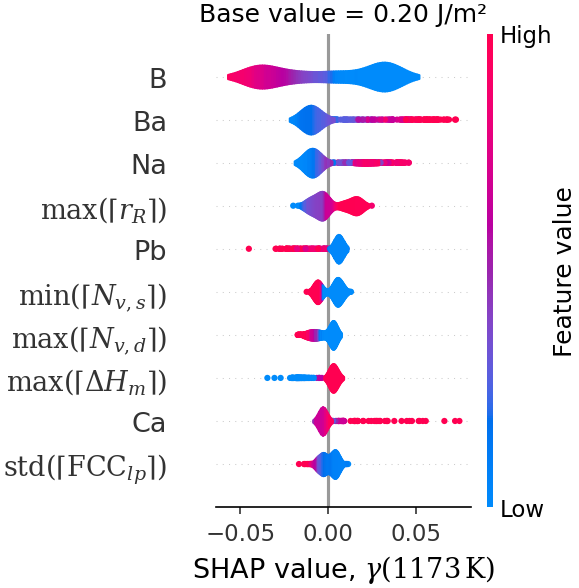}
\caption{\label{fig:shap_SurfaceTension1173K}Violin plot of SHAP values for
$\gamma(1173\,\mathrm{K})$.}
\end{figure}

\begin{figure}[H]
\centering \includegraphics[width=0.4\textwidth,keepaspectratio]{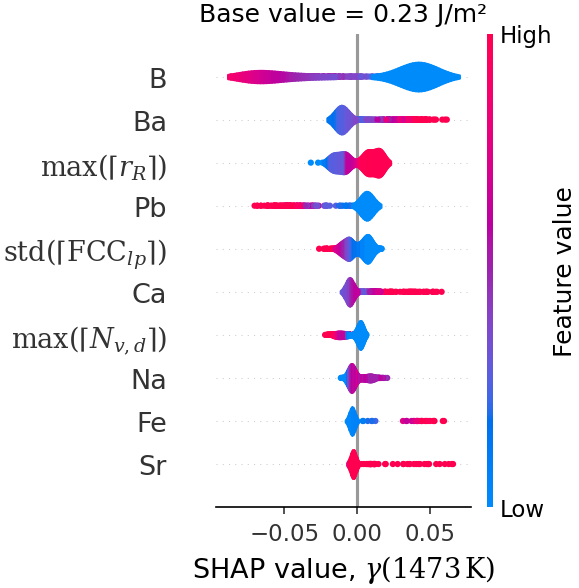}
\caption{\label{fig:shap_SurfaceTension1473K}Violin plot of SHAP values for
$\gamma(1473\,\mathrm{K})$.}
\end{figure}

\begin{figure}[H]
\centering \includegraphics[width=0.4\textwidth,keepaspectratio]{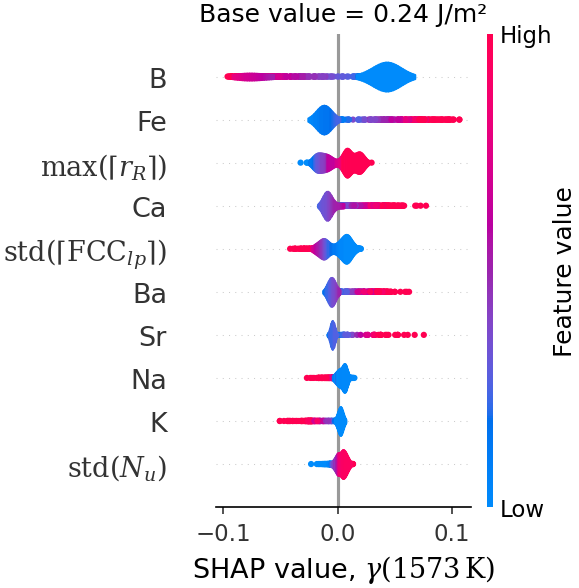}
\caption{\label{fig:shap_SurfaceTension1573K}Violin plot of SHAP values for
$\gamma(1573\,\mathrm{K})$.}
\end{figure}

\begin{figure}[H]
\centering \includegraphics[width=0.4\textwidth,keepaspectratio]{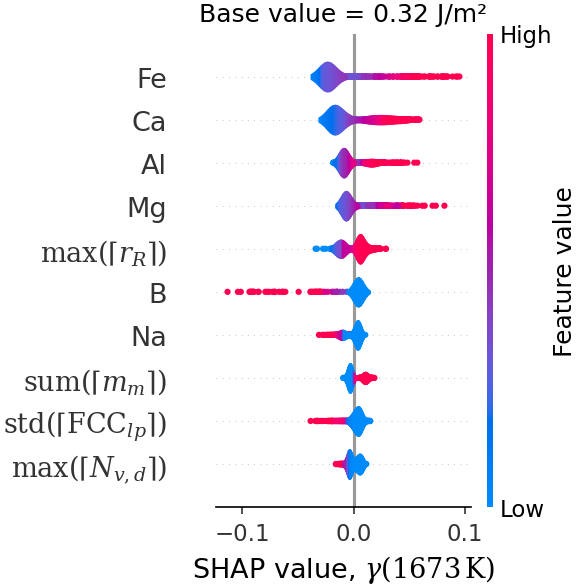}
\caption{\label{fig:shap_SurfaceTension1673K}Violin plot of SHAP values for
$\gamma(1673\,\mathrm{K})$.}
\end{figure}

Figures \ref{fig:shap_SurfaceTensionAboveTg} to \ref{fig:shap_SurfaceTension1673K}
show the violin plots of the SHAP values for the surface tension at
different temperatures. The SHAP values were calculated using the
multi-headed GlassNet model.

\FloatBarrier

\subsection*{SHAP values violin plots --- liquidus temperature and thermal shock
resistance}

\label{sec:orgb49b9ed}

\begin{figure}[H]
\centering \includegraphics[width=0.4\textwidth,keepaspectratio]{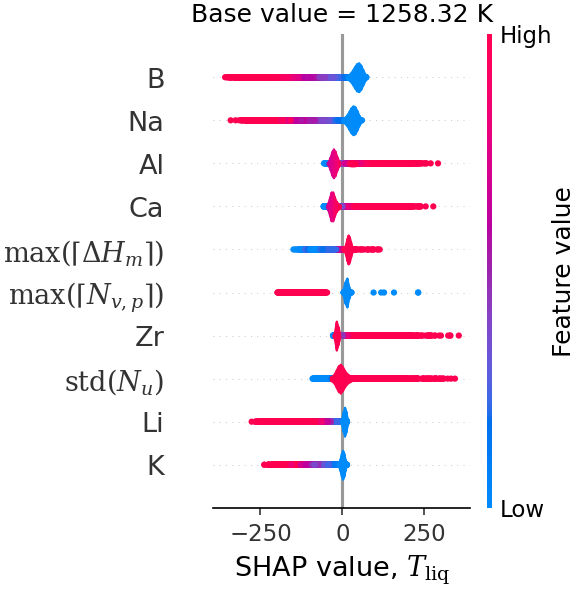}
\caption{\label{fig:shap_Tliquidus}Violin plot of SHAP values for $T_{\mathrm{liq}}$.}
\end{figure}

\begin{figure}[H]
\centering \includegraphics[width=0.4\textwidth,keepaspectratio]{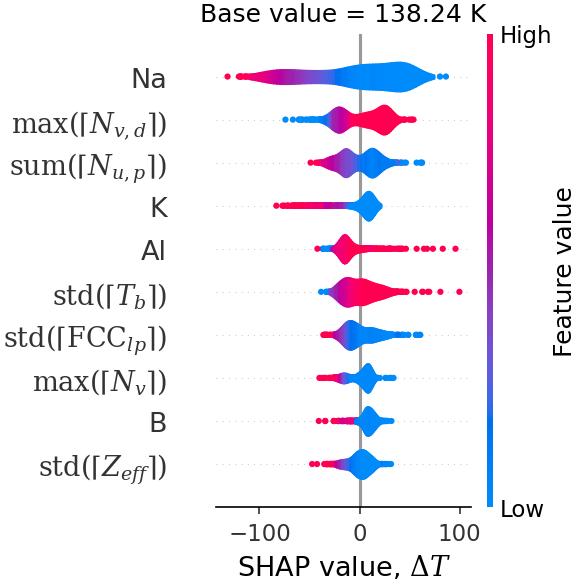}
\caption{\label{fig:shap_ThermalShockRes}Violin plot of SHAP values for $\Delta T$.}
\end{figure}

Figures \ref{fig:shap_Tliquidus} and \ref{fig:shap_ThermalShockRes}
show the violin plots of the SHAP values for the liquidus temperature
and the thermal shock resistance, respectively. The SHAP values were
calculated using the multi-headed GlassNet model.

\FloatBarrier 
\end{document}